\documentclass[a4paper,11pt]{article}

\usepackage{bm}
\usepackage{anysize}
\usepackage{amsthm}
\usepackage{amssymb}
\usepackage{amsmath}
\usepackage{fancybox}
\usepackage{bbm}
\usepackage{amsfonts}
\usepackage[T1]{fontenc}
\usepackage{latexsym}
\usepackage{makeidx}
\usepackage{cite}
\usepackage[numbers]{natbib}
\usepackage{mathrsfs}

\usepackage[pdftex]{graphicx}
\usepackage[pdftex]{color}

\usepackage[all]{xy}

\usepackage{stackengine}
\stackMath

\usepackage{lipsum}

\newtheorem{teo}{Theorem}
\newtheorem{lem}{Lemma}
\newtheorem{pro}{Proposition}
\newtheorem{cor}{Corollary}
\newtheorem{defi}{Definition} 
\newcommand*{\m}[1]{\underline{#1}}

\newcommand{\fd}{\rightarrow}

\newcommand{\inc}{\subset}

\newcommand{\iso}{\cong}

\newcommand{\al}{\alpha}
\newcommand{\be}{\beta}
\newcommand{\lan}{\lambda}
\newcommand{\fhi}{\varphi}
\newcommand{\del}{\delta}
\newcommand{\Del}{\Delta}
\newcommand{\gam}{\gamma}
\newcommand{\Gam}{\Gamma}

\newcommand{\Om}{\Omega}

\newcommand{\Z}{\mathbb{Z}}
\newcommand{\N}{\mathbb{N}}

\newcommand{\R}{\mathbb{R}}
\newcommand{\C}{\mathbb{C}}

\newcommand{\E}{\mathbb{E}}

\newcommand{\Sa}{\mathbb{S}}

\newcommand{\Sw}{{\bf \mathcal{S}}}

\newcommand{\pa}{\partial}

\newcommand{\sgn}{\mbox{sgn}}

\newcommand{\p}{\grave{}}

\newcommand{\el}{\ell}

\newtheorem{remark}{Remark}%[section]

\def\pf{\par\noindent {\em Proof.}~\par\noindent}

\def\res{\mathop{\mbox{\normalfont res}}\limits}

\def\lim{\mathop{\mbox{\normalfont lim}}\limits}

\def\pf{\par\noindent {\em Proof. }}%~\par\noindent}

\def\pa{\partial}

\begin{document}

\date{}

\title{On the Radon transform and the Dirac delta distribution in superspace}
%{Integration over $(m-k)$-surfaces using distributions and applications}
\small{
\author{ Al\'i  Guzm\'an Ad\'an$^\dagger$\thanks{Postdoctoral Fellow of the Research Foundation - Flanders (FWO)} \and Irene Sabadini$^\ddagger$ \and Frank Sommen$^\dagger$ }
%{Al\'i Guzm\'an Ad\'an$^\dagger$, Irene Sabadini$^\ddagger$, Frank Sommen$^\dagger$}
\vskip 1truecm
\date{{\footnotesize  $^\dagger$Clifford Research Group, Department of Electronics and Information Systems,\\
 Faculty of Engineering and Architecture, Ghent University, Krijgslaan 281, 9000 Gent, Belgium. \\
 {\tt ali.guzmanadan@ugent.be}, \;\;\;\;\; {\tt franciscus.sommen@ugent.be} \\[+.2cm]
$^\ddagger$Dipartimento di Matematica, Politecnico di Milano, Via Bonardi, 9, 20133 Milano, Italy.\\
 {\tt irene.sabadini@polimi.it}
}}

\maketitle

\begin{abstract} 
In this manuscript, we obtain a plane wave decomposition for the delta distribution in superspace, provided that the superdimension is not odd and negative. This decomposition allows for {explicit} inversion formulas for the {super} Radon transform in {these cases. Moreover, we prove a more general Radon inversion formula valid for all possible integer values of the superdimension. The proof of this result comes along with the study of fractional powers of the super Laplacian, their fundamental solutions, and the plane wave decompositions of super Riesz kernels.}

\vspace{0.3cm}

\small{ }
%30G35, 32A26, 46F10,  58C50 ??? , 58A10, 

\noindent
\textbf{}
\end{abstract}

\tableofcontents

\section{Introduction}
This paper inserts in the broad field of studies on superspace whose importance is {well-known} since its introduction by Berezin in the sixties, see {e.g.\ \cite{Berezin:1987:ISA:38130}},  mainly motivated by problems in theoretical physics. In the literature there are various approaches to superspace which range from differential to algebraic geometry, {see e.g.\ \cite{Berezin:1987:ISA:38130, Kostant:1975qe, MR2840967, MR2069561, MR565567, MR778559, MR574696}. In this paper we will follow a more recent approach based on an extension of harmonic and Clifford analysis to superspace,} which have been already proved to offer various advantages, among which a natural treatment of the super Dirac and super Laplace operators. 
%to harmonic analysis. The approach we will follow is the one using Clifford analysis methods, a refinement of harmonic analysis, which have been already proved to offer various advantages, among which a natural treatment of the super-Dirac and super-Laplace operators. 
 This approach started with the early paper by Sommen {\cite{sommen2000extension}} and then continued {with the works of} {Coulembier, De Bie and Sommen, see e.g.\ \cite{de2007clifford, MR2344451, MR2386499, MR3375856, MR3060765, MR2539324, MR2683546, MR2422641}.}
 
In this framework we shall consider the Radon transform, another cornerstone in theoretical and applied mathematics. In theoretical mathematics, this integral transform, originally defined on the space of lines in the plane, was then generalized to higher dimensions and also to the complex case, giving rise to the Penrose transform. The applications to practical problems such {as} tomography or image recognition are well-known. {The Radon transform in superspace was initially introduced in \cite{MR2422641} by means of the central-slice theorem, i.e.\ as the action of two consecutive Fourier transforms. In the later work \cite{MR2539324}, a more geometrical interpretation of this transform was given as an integral over the set of all hyperplanes.}

In this paper we shall study more properties of the Radon transform in the superspace setting. In particular, we will prove inversion formulas for this transform.  This poses important  differences in comparison with the purely bosonic case, i.e.\ when only commuting variables are considered. Indeed, let $R_m[\phi](\m{w},p)$ denote the Radon transform {in $\R^m$} of the function $\phi$, i.e.\ the integral of $\phi$ over the hyperplane $\langle\m{x},\m{w}\rangle=p$ where $\m{w}$ is a unit vector, $p\in\R$ and $\langle\cdot,\cdot\rangle$ {is} the Euclidean inner product in $\R^m$. Then the inversion formula of $R_m[\phi](\m{w},p)$ reads as (see e.g.\ \cite{MR754767, MR573446,MR709591})
\begin{align}
\phi({\m{x}}) &=   \frac{(-1)^{\frac{m}{2}}}{(2\pi)^m} \int_{-\infty}^\infty \frac{1}{p}\left( \int_{\Sa^{m-1}}   \pa_{p}^{m-1}\, R_{m}[\phi]({\m{w}},p+ \langle{\m{x}},{\m{w}}\rangle)  \, dS_{\m{w}}\right) \,dp, & \mbox{ for }& \;\;m \mbox{ even}, \label{CIF1}\\[+.2cm]
\phi({\m{x}}) &=  {\frac{(-1)^{\frac{m-1}{2}}}{2(2\pi)^{m-1}}}  \int_{\Sa^{m-1}}   {\pa_{p}^{m-1}}\, R_{m}[\phi]({\m{w}},p)\bigg|_{p=\langle {\m{x}},{\m{w}}\rangle}  \, dS_{\m{w}} , & \mbox{ for }& \;\;m \mbox{ odd}, \label{CIF2}
\end{align}
where $dS_{\m{w}}$ is the area element of the unit sphere $\Sa^{m-1}\inc\R^m$. The extension of these formulas to superspace requires %{in the first place} 
the replacement of the dimension $m\in\N$ by the so-called superdimension $M\in\Z$. {Clearly, the above formulas fail to preserve their classical forms} for negative values of $M$. For instance, the derivatives with respect to $p$ would {have a negative exponent, playing thus the role of an indefinite integral (or primitive function).}
%play the role of an integral (or primitive function) due to the negative exponent. 
As we shall see in our Theorem \ref{InvExpl}, when $M$ is even and negative, a novel structure for these inversion formulas is {given} in terms of a primitive function of higher order of $p^{-1}$.  

To prove these inversion formulas we are in need of a plane wave decomposition of the Dirac delta distribution in superspace, which is the {third} pillar of this paper. We show how to extend some classical formulas by adopting the point of view of hyperfunctions, namely by using the fact that the Dirac delta is {a suitable} boundary value of the super Cauchy kernel (see Theorem \ref{BVDeltaT}). This approach has been announced  in \cite{CK_Ali} where a plane wave decomposition of the super Cauchy kernel was obtained {, provided that the superdimension $M$ is not odd and negative}. {Combining these facts}
%In fact, by combining these two results, 
we obtain a plane wave expansion of the super Dirac delta {distribution}  {in these cases} (see Theorem \ref{ThmDelPW}). Again, {when the superdimension is negative and even,}
%in the cases of negative and even superdimension, 
the obtained formulas no longer resemble the structure of the classical plane wave decompositions given in \cite[Ch.1 - \S 3 ]{MR0166596}. The plane wave decomposition of the Dirac delta is a result of independent interest in the theory of distributions in superspace.  {In a forthcoming paper, we shall study the decomposition into plane waves of the super Dirac distribution and the super Cauchy kernel in the case where the superdimension $M$ is odd and negative. This shall yield explicit Radon inversion formulas in those exceptional cases.}

The inversion formulas that we prove {coincide with} the classical formulas in the purely bosonic case.  In that case, regardless of the parity of the dimension, formulas (\ref{CIF1})-(\ref{CIF2}) can be written in {a} unified way as follows
\begin{equation*}%\label{UnInvRa}
\phi({\m{x}}) = \frac{1}{2^{m}\pi^{m-1} } (-\Del_{\m{x}})^{\frac{m-1}{2}} \, \int_{\Sa^{m-1}} R_{m}[\phi](\m{w},\langle\m{x},\m{w}\rangle) \, dS_{\m{w}},
\end{equation*}
where $\Del_{\m{x}}$ is the Laplace operator in $\R^m$. The final purpose of this paper is to show how {this unified formula} can also be extended to superspace (see Theorem \ref{InvExplRed}). {Moreover, we show that this extension holds for any value of the superdimension $M\in\Z$.}
To that end, we first introduce fractional powers of the super Laplacian and construct fundamental solutions for such operators (see Theorem \ref{FundSolLap}). These results extend the work in \cite{MR2386499}, where fundamental solutions for natural powers of the super Laplace operator were obtained. Along our proof of the unified inversion formula, we also provide a plane wave decompostion for the super Riesz potential $|{\bf x}|^{-1}$. As in the case of the plane wave decomposition for the Dirac delta distribution, these two last results are of independent interest in superanalysis.

%n this paper we shall study more properties of this transform, also in relation with the central-slice theorem. In particular, we prove inversion formulas for the Radon transform. To this end we are in need of a plane wave decomposition of the Dirac delta distribution in superspace, which is the third pillar of this paper. We show how to extend some classical formulas, by adopting the point of view of hyperfunctions, namely by using the fact that the Dirac delta is the boundary value of the super Cauchy kernel.
%The plane wave decomposition of the Dirac delta is a result of independent interest in the theory of distributions in superspace.

%The inversion formulas that we prove give back the classical formulas in the purely bosonic case. Regardless of the parity of the dimension, they can be written in a unified way.

The plan of the paper is as follows. In Section \ref{S2}, we give a brief introduction  on harmonic and Clifford analysis in superspace focusing on the notions needed in the sequel. In Section \ref{S3}, we discuss some facts on distributional calculus in superspace. In particular, we introduce  important generalized superfunctions (and their classical analogues in $\R^m$) such as concentrated Dirac delta distributions and $|\bf{x}|^\lan$ {with $\lan\in\C$}, which are necessary in the subsequent sections. In Section \ref{S4}, we prove some of the main properties of the Radon transform in superspace. Section \ref{S5} is fully devoted to the plane wave decomposition of the Dirac delta distribution in superspace. First, we review the classical procedure followed to obtain some plane wave decompositions in $\R^m$, which shall also be useful for subsequent computations. Then, we proceed to obtaining plane wave decomposition formulas for the {super} Dirac delta distribution from the point of view of hyperfunctions. An alternative proof for this result, using only direct computations and the Funk-Hecke theorem, is provided in Appendix \ref{App}. This plane wave decomposition is used in Section \ref{S6} to derive explicit inversion formulas of the super Radon transform. Finally, in Section \ref{S7}, these inversion formulas are unified into a single expression, regardless of the parity and sign of the superdimension.

\section{Preliminaries}\label{S2}
%\subsection{Variables, derivatives, vector variable, Clifford algebra, inner product, norm squared}
\noindent Consider $m$ commuting (bosonic) variables $x_1,\ldots, x_m$ and $2n$ anti-commuting (fermionic) variables $x\p_1, \ldots, x\p_{2n}$ in a purely symbolic way, i.e.\ $x_jx_k = x_kx_j$, $x\p_j x\p_k = -x\p_k x\p_j$ and $x_jx\p_k = x\p_k x_j$. They give rise to the supervector variable
\[{\bf x}=(\underline{x},\underline{x\p})=\left(x_1,\ldots, x_m,x\p_1,\ldots, x\p_{2n}\right).\]
The variables $x_1,\ldots, x_m$ are generators of the polynomial algebra $\R[x_1,  \ldots, x_m]$ while $x\p_1, \ldots, x\p_{2n}$ generate a Grassmann algebra $\mathfrak{G}_{2n}$. We denote by $\mathfrak{G}^{(ev)}_{2n}$ and $\mathfrak{G}^{(odd)}_{2n}$ the subalgebras of even and odd elements of $\mathfrak{G}_{2n}$ respectively. All the variables together generate the supercommutative algebra of superpolynomials
\[ \mathcal{P}:= \mbox{Alg}_{\R}(x_1,  \ldots, x_m, x\p_1, \ldots, x\p_{2n})= \R[x_1,  \ldots, x_m] \otimes \mathfrak{G}_{2n}.\]
The bosonic and fermionic partial derivatives $\pa_{x_j}=\frac{\pa }{\pa x_j}$, $\pa_{x\p_j}=\frac{\pa }{\pa x\p_j}$ are defined as endomorphisms on $\mathcal P$ by the  relations 
\begin{equation*}%\label{ParDer1}
\begin{cases} \pa_{x_j}[1]=0,\\
 \pa_{x_j} x_k- x_k \pa_{x_j}=\del_{j,k},\\
 \pa_{x_j} x\p_k=x\p_k \pa_{x_j}, \;\; 
 \end{cases}
\hspace{.5cm} 
\begin{cases} \pa_{x\p_j}[1]=0,\\
\pa_{x\p_j} x\p_k+ x\p_k \pa_{x\p_j}=\del_{j,k},\\
\pa_{x\p_j} x_k=x_k\pa_{x\p_j}, 
\end{cases}
\end{equation*} 
{where $\del_{j,k}$ is the Kronecker symbol and $1$ denotes the constant superpolynomial $p\equiv1$. The above relations can be recursively applied for both left and right actions of the linear operators $\pa_{x_j}$ and $\pa_{x\p_j}$.}

Associated {with} these variables we consider the flat supermanifold $\R^{m|2n}=\left(\R^m, \mathcal{O}_{\R^{m|2n}}\right)$ where $\mathcal{O}_{\R^{m|2n}}$ is the structure sheaf that maps every open subset $\Om\inc \R^m$ {to} the graded algebra $C^\infty(\Om) \otimes {\mathfrak G}_{2n}$, {and $C^\infty(\Om)$ denotes the space of smooth complex-valued functions defined in $\Om$.}
%of smooth functions in $\Om$ with values in the Grassmann algebra ${\mathfrak G}_{2n}$. 
The partial derivatives $\pa_{x_j}$, $\pa_{x\p_j}$ extend from $\mathcal P$ to $C^\infty(\R^m)\otimes \mathfrak{G}_{2n}$ by density. 

\noindent Let us rewrite the supervector variable ${\bf x}$  as
\[{\bf x}=\underline{x}+\underline{x\p}=\sum_{j=1}^mx_je_j+\sum_{j=1}^{2n}x\p_j e\p_j,\]
where $e_1,\ldots,e_m,e\p_1,\ldots,e\p_{2n}$ is the standard homogeneous  basis of the graded vector space $\R^{m,2n}=\R^{m,0}\oplus \R^{0,2n}$. Here we have denoted by $\underline{x}= \sum_{j=1}^m x_je_j$ and $\underline{x\p}=\sum_{j=1}^{2n}x\p_je\p_j$ the so-called bosonic and fermionic projections of ${\bf x}$ respectively. We consider an orthosymplectic metric in $\R^{m,2n}$, giving rise to the super Clifford algebra $\mathcal C_{m,2n}:=\mbox{Alg}_\R(e_1,\ldots,e_m,e\p_1,\ldots,e\p_{2n})$ governed by the multiplication rules 
\begin{align*}%\label{CommRules}
e_je_k+e_ke_j=-2\del_{j,k}, \hspace{.3cm} e_je\p_k+e\p_ke_j=0, \hspace{.3cm} e\p_je\p_k-e\p_ke\p_j=g_{j,k},
\end{align*}
where $g_{j,k}$ is a symplectic form defined by
\[g_{2j,2k}=g_{2j-1,2k-1}=0, \hspace{.5cm} g_{2j-1,2k}=-g_{2k,2j-1}=\del_{j,k}, \hspace{.5cm} j,k=1,\ldots,n.\]

\noindent In this case the {\it inner product} of two supervectors ${\bf x}$ and ${\bf y}$ is given by
\[\langle{\bf x}, {\bf y}\rangle :=-\frac{1}{2}( {\bf x}{\bf y} + {\bf y}{\bf x})= \langle{\m x}, {\m y}\rangle + \langle{\m x\p}, {\m y\p}\rangle=\sum_{j=1}^mx_jy_j  -\frac{1}{2}\sum_{j=1}^n(x\p_{2j-1}y\p_{2j}-x\p_{2j}y\p_{2j-1}).\]

\noindent The {\it generalized norm squared} of the supervector ${\bf x}$ is thus defined by
\begin{equation}\label{NormSq}
|{\bf x}|^2=  \langle{\bf x}, {\bf x}\rangle = -{\bf x}^2=\sum_{j=1}^m x_j^2 - \sum_{j=1}^n x\p_{2j-1} x\p_{2j}.
\end{equation}
Observe that the fermionic vector variable $\m{x}\p$ is nilpotent. Indeed, its norm squared satisfies 
\[\underline{x}\p^{2n}=n! \,x\p_1x\p_2\cdots x\p_{2n-1}x\p_{2n},\]
which is the element of maximal degree in $\mathfrak{G}_{2n}$.

%\subsection{Superfunctions, Dirac and Laplace operators}
Functions in $C^\infty(\Om)\otimes \mathfrak{G}_{2n}$ {(often called superfunctions)} can be explicitly written as 
\begin{equation}\label{SupFunc}
F({\bf x})=F(\underline{x},\underline{x\p})=\sum_{A\inc \{1, \ldots, 2n\}} \, F_A(\underline{x})\, \underline{x}\p_A, 
\end{equation}
where $F_A(\underline{x})\in C^\infty(\Om)$ and  $ \underline{x}\p_A=x\p_{j_1}\ldots x\p_{j_k}$ with $A=\{j_1,\ldots, j_k\}$, $1\leq j_1< \ldots< j_k\leq 2n$.  For studying integral transforms in superspace we need of course a broader set of functions. For our purposes, it suffices to consider the function spaces $ \mathcal F(\Om) \otimes \mathfrak{G}_{2n}$ and $\mathcal F(\Om) \otimes \mathfrak{G}_{2n} \otimes \mathcal{C}_{m,2n}$ where $\mathcal F(\Om)$ stands for $C^\infty(\Om),C^\infty_0(\Om)$ {or} $\mathcal{S}(\R^m)$. {In general, the functions $F_A\in \mathcal F(\Om)$ in (\ref{SupFunc}) are complex-valued. We say that $F$ is a real superfunction when all elements $F_A$ are real-valued.}

\noindent Every superfunction can be written as the sum $F({\bf x})= F_0(\underline{x})+ {\bf F}(\underline x, \underline{x}\p)$ where the  {complex}-valued function $F_0(\underline{x}):=F_\emptyset(\underline{x})$ is called the {\it body} $F$, and ${\bf F}:=\sum_{|A|\geq 1} \, F_A(\underline{x})\,\underline{x}\p_A$ is the {\it nilpotent part} of $F$. Indeed, it is clearly seen that  ${\bf F}^{2n+1}=0$. 

%\noindent For studying integral transforms in superspace we need of course a broader set of functions. For our purposes, it suffices to consider the function spaces $ \mathcal F(\Om) \otimes \mathfrak{G}_{2n}$ and $\mathcal F(\Om) \otimes \mathfrak{G}_{2n} \otimes \mathcal{C}_{m,2n}$ where $\mathcal F(\Om)$ stands for $C^\infty(\Om),C^\infty_0(\Om)$ and $\mathcal{S}(\R^m)$.

\noindent The  bosonic and fermionic Dirac operators are defined by
\[\pa_{\underline x}=\sum_{j=1}^m e_j\pa_{x_j}, \hspace{1cm} \pa_{\underline x\p}=2\sum_{j=1}^n \left(e\p_{2j}\pa_{x\p_{2j-1}}-e\p_{2j-1}\pa_{x\p_{2j}}\right),\]
giving rise to the left and right super Dirac operators (super-gradient)
$\pa_{\bf x} \cdot =\pa_{\underline x\p}\cdot -\pa_{\underline x}\cdot$ and $\cdot\pa_{\bf x}  =- \cdot\pa_{\underline x\p} - \cdot \pa_{\underline x}$ respectively. As in the classical setting, the action of $\pa_{\bf x}$ on the vector variable ${\bf x}$ results in the superdimension
\[M:=\pa_{\bf x}[{\bf x}]=[{\bf x}]\pa_{{\bf x}}=\pa_{\underline{x\p}}[\underline{x\p}]-\pa_{\underline{x}}[\underline{x}]=m-2n.\]

Given an open set $\Om\inc \R^m$, a superfunction $F\in C^\infty(\Om)\otimes \mathfrak{G}_{2n}\otimes \mathcal{C}_{m,2n}$ is said to be (left) {\it monogenic} if $\pa_{\bf x}[F]=0$. As the super Dirac operator factorizes the super Laplace operator:
\[\Del_{\bf x}=-\pa_{\bf x}^2=\sum_{j=1}^{m} \pa^2_{x_j}-4\sum_{j=1}^n \pa_{x\p_{2j-1}}\pa_{x\p_{2j}},\]
monogenicity also constitutes  a refinement of harmonicity in superanalysis. {The super Laplace operator can be decomposed as $\Del_{\bf x}=\Del_{\m{x}}+\Del_{\m{x}\p}$ where $\Del_{\m{x}}=\sum_{j=1}^{m} \pa^2_{x_j}$ and $\Del_{\m{x}\p}=-4\sum_{j=1}^n \pa_{x\p_{2j-1}}\pa_{x\p_{2j}}$ are the bosonic and fermionic Laplacians  with respect to ${\bf x}$, respectively.}

The super Euler operator is defined by
\[
\E=\sum_{j=1}^m x_j\pa_{x_j} + \sum_{j=1}^{2n} x\p_j\pa_{x\p_j}.
\]
We denote by $\N_0:=\{0\}\cup\N$ the set of non-negative integers. Homogeneous superpolynomials of degree $j\in \N_0$ are eigenfunctions of the super Euler operator with eigenvalue $j$. We denote the space of homogeneous superpolynomials of degree $j\in \N_0$ as $\mathcal{P}_j= \{R\in \mathcal{P} : \E[R]=j\,R\}$, which allows for the decomposition
\[
\mathcal{P} = \bigoplus_{j=0}^\infty \mathcal{P}_j.
\]
%An element $R\in \mathcal P$ is called a {spherical harmonic of degree $j$} if it satisfies 
%\[
%\Del_{\bf x}[R]=0,  \;\;\;\; \mbox{ and }  \;\;\;\; \E[R]=jR,  \;\; \mbox{(i.e. $R\in \mathcal{P}_j$)}.
%\]
%The space of all spherical harmonics of degree $j$ is denoted by $\mathcal{H}_j$. 

The operators $\Del_{\bf x}$, ${\bf x}^2$ and $\E$ satisfy the canonical commutation relations of {the special linear Lie algebra} $\mathfrak{sl}_2$   (see e.g.\ \cite{MR2344451})
\begin{align}\label{sl2}
\left[\frac{\Del_{\bf x}}{2}, \frac{-{\bf x}^2}{2}\right] &= \E + \frac{M}{2}, & \left[\frac{\Del_{\bf x}}{2}, \E + \frac{M}{2}\right] &= {\Del_{\bf x}}, & \left[\frac{-{\bf x}^2}{2}, \E + \frac{M}{2}\right] &= {{\bf x}^2}, 
\end{align}
{where $[a,b]:=ab-ba$. The following calculation also extends the bosonic case
\[
\pa_{\bf x} {\bf x} + {\bf x}\pa_{\bf x} = 2\left(\E+\frac{M}{2}\right).
\]
This means that the same computation rules of classical Clifford and harmonic analysis can be transferred to the superspace setting by substituting the Euclidean dimension $m$ by the superdimension $M$. In particular, the following identity can be proved using formulae (\ref{sl2}) iteratively,
see e.g. \cite{MR2344451, MR2386499},
\begin{equation}\label{LapPowX}
\Del_{\bf x}^j \left[{\bf x}^{2{{\el}}}\right] = \begin{cases} (-4)^j \frac{{{\el}}!}{({{\el}}-j)!} \frac{\Gam\left(\frac{M}{2}+{{\el}}\right)}{\Gam\left(\frac{M}{2}+{{\el}}-j\right)} {\bf x}^{2{{\el}}-2j}, & j\leq {{\el}},\\ 0, & {{\el}}<j.\end{cases}
\end{equation}
More details on the theory of monogenic and harmonic superfunctions can be found for instance in \cite{MR2344451, MR2386499, MR3375856}.

\section{Distributional calculus in superspace}\label{S3}
In this section we discuss the notion of generalized functions in the superspace setting, study some important examples, and introduce integration over the supersphere by means of the concentrated Dirac distribution. These ideas shall be useful when dealing with the super Radon transform in the subsequent {sections}.

The analogue in superspace of the classical integral $\int_{\R^m}\; dV_{\underline x}$ in $\R^m$ is given by
\[{\int_{\R^{m|2n}_{\bf x}} }= \int_{\R^m} dV_{\underline{x}} \int_{B,{\m{x}\p}}=\int_{B,\m{x}\p} \int_{\R^m} dV_{\underline{x}},\]
where $dV_{\underline{x}}=d{x_1}\cdots d{x_m}$ {is} the classical volume element in $\R^m$ and the integral over fermionic variables is given by the Berezin integral (see \cite{Berezin:1987:ISA:38130}), defined by 
\[\int_{B,\m{x}\p} {:=} \pi^{-n} \, \pa_{x\p_{2n}}\cdots \pa_{x\p_{1}}=\frac{(-1)^n \pi^{-n}}{4^n n!} \pa_{\underline{x}\p}^{2n}.\]
The subscript $\m{x}\p$ means that we are integrating with respect to the $\m{x}\p$ variable. 

Similarly to the function spaces, we define the spaces of generalized functions $ \mathcal F'(\Om) \otimes \mathfrak{G}_{2n}$ and $ \mathcal F'(\Om) \otimes \mathfrak{G}_{2n} \otimes \mathcal{C}_{m,2n}$,
%$\mathcal F'(\Om)|_{m|2n} = \mathcal F'(\Om) \otimes \mathfrak{G}_{2n} \otimes \mathcal{C}_{m,2n}$, 
where $\mathcal F'(\Om)$ stands for the spaces of generalized functions $\mathcal{E}'(\Om),\mathcal{D}'(\Om)$ and $ \mathcal{S}'(\R^m)$ respectively. The action of a superdistribution $\al\in\mathcal F'(\Om) \otimes \mathfrak{G}_{2n}$ given by
\begin{equation*}%\label{SupDistForm}
\al=\sum_{A\inc\{1,\ldots, 2n\}} \al_A \underline{x}\p_A, \;\;\;\;\;\;\; \;\;\; \al_A \in \mathcal F'(\Om),
\end{equation*}
on a test superfunction $F\in \mathcal F(\Om) \otimes \mathfrak{G}_{2n}$ {of the form (\ref{SupFunc})} is defined as
\[
{\int_{\R^{m|2n}_{\bf x}}} \al F =\sum_{A,{C}\inc\{1,\ldots, 2n\}} \langle \al_A, {f_C} \rangle {\int_{B,\m{x}\p}}   \underline{x}\p_A\, {\underline{x}\p_C}, \]
where, as usual, the notation
\[
\langle \al_A, {f_C} \rangle= \int_{\R^m} \al_A(\m{x}) {f_C}(\m{x}) \; dV_{\underline x},
\]
is used for the action of the real distribution $\al_A\in \mathcal{F}^\prime(\Om)$ on the test function ${f_C}\in \mathcal{F}(\Om)$.

\subsection{Useful generalized functions in $\R^m$}
Before introducing the specific generalized superfunctions needed {in} this {paper}, we {recall} some basic facts about the {generalized} functions $t_+^\lan$, $t_-^\lan$, $|t|^\lan$, $\sgn(t)|t|^\lan$, $|\m{x}|^\lan$ with $\lan\in\C$. We refer the reader to \cite{MR0166596} for a complete study of these generalized functions.

When Re$(\lan) >-1$, the functionals $t_+^\lan$ and $t_-^\lan$ are defined on  $\Sw(\R)$ by means of the integrals
\begin{align*}
\langle t_+^\lan, \phi(t)\rangle = \int_0^\infty t^\lan \phi(t)\, dt, \;\;\;\;\; \mbox{ and }  \;\;\;\;\; \langle t_-^\lan, \phi(t)\rangle = \langle t_+^\lan, \phi(-t)\rangle= \int_{-\infty}^0 |t|^\lan \phi(t)\, dt,
\end{align*}
respectively. The mappings $\lan\mapsto t_+^\lan$ and $\lan\mapsto t_-^\lan$ extend from the complex region $\{\textup{Re}(\lan)>-1\}$ to an analytic mapping on $\C\setminus\{-1, -2,\ldots\}$ with values in $\Sw'(\R)$, i.e.\ $\lan\mapsto \langle t_+^\lan, \phi\rangle$ and $\lan\mapsto \langle t_-^\lan, \phi\rangle$ are analytic functions on $\C\setminus\{-1, -2,\ldots\}$ for all $\phi \in\Sw(\R)$. These analytic continuations of $t_\pm^\lan$ can be written in the strip $-{\el}-1<\textup{Re}(\lan)<-{\el}$,  {with $\el\in\N$,} as %follows
%These integrals can be extended by analytic continuation (on the parameter $\lan$) to all $\lan\neq -1, -2, \ldots $ yielding to the following definitions in the strip $-{\el}-1<\textup{Re}(\lan)<-{\el}$
\begin{align}\label{t+defi}
\langle t_+^\lan, \phi\rangle &= \int_0^\infty t^\lan \left(\phi(t) - \sum_{j=0}^{{\el}-1} \frac{t^j}{j!}\phi^{(j)}(0)\right)\, dt,   & 
\;\;\;\;\langle t_-^\lan, \phi\rangle &= \int_0^\infty t^\lan \left(\phi(-t) - \sum_{j=0}^{{\el}-1} (-1)^j\frac{t^j}{j!}\phi^{(j)}(0)\right)\, dt. 
\end{align}
From these formulas, it  follows that $t_+^\lan$ and $t_-^\lan$ have simple poles at $\lan= -1, -2, \ldots $ and their residues at $\lan=-{\el}$ are given by
\begin{equation}\label{t+-Res}
\frac{(-1)^{{\el}-1}}{({\el}-1)!} \del^{({\el}-1)}(t),  \;\;\;\;\; \mbox{ and }  \;\;\;\;\; \frac{\del^{({\el}-1)}(t)}{({\el}-1)!},
\end{equation}
respectively.

We will also use the generalized functions 
\[|t|^\lan := t_+^\lan +  t_-^\lan,  \;\;\; \mbox{ and } \;\;\;  \sgn(t)\, |t|^\lan := t_+^\lan - t_-^\lan,\]
where $\sgn(t)$ stands for the sign of $t$. By virtue of (\ref{t+-Res}), it follows that $|t|^\lan$ has poles only at $\lan=-1,-3,-5\ldots$, while $\sgn(t)\, |t|^\lan$ has poles only at $\lan=-2,-4,-6\ldots$. Moreover, their residues are given by 
\begin{align*}
\res_{\lan=-2{\el}-1} |t|^\lan = 2 \frac{\del^{(2{\el})}(t)}{(2{\el})!},   \;\;\;\;\; \mbox{ and }  \;\;\;\;\;  \res_{\lan=-2{\el}} \sgn(t)\, |t|^\lan = -2 \frac{\del^{(2{\el}-1)}(t)}{(2{\el}-1)!}.
\end{align*}
%Direct computations yield the following explicit formulas for $|t|^\lan$ and $\sgn(t)\, |t|^\lan$
%\begin{align}
%\langle |t|^\lan, \phi\rangle &= \int_0^\infty t^\lan \left(\phi(t) + \phi(-t) -2\sum_{j=0}^{{\el}-1} \frac{t^{2j}}{(2j)!}\phi^{(2j)}(0)\right)\, dt, & -2{\el}-1&<\textup{Re}(\lan)<-2{\el}+1,\label{ModLan}\\
%\langle |t|^\lan \sgn(t), \phi\rangle &= \int_0^\infty t^\lan \left(\phi(t) - \phi(-t) -2\sum_{j=1}^{{\el}} \frac{t^{2j-1}}{(2j-1)!}\phi^{(2j-1)}(0)\right)\, dt, &-2{\el}-2&<\textup{Re}(\lan)<-2{\el}. \label{ModLanSig}
%\end{align}
%This allows to define the functionals $t^{-\el}$ for $\el\in\N$. Indeed, for $\lan=-2{\el}$ in (\ref{ModLan}) we get
In particular, this means that $|t|^\lan$ is defined for $\lan=-2{\el}$, while $\sgn(t)\, |t|^\lan$ is defined for $\lan=-2{\el}-1$. This yields the following definitions for $t^{-2{\el}}$ and $t^{-2{\el}-1}$,
\begin{equation}\label{10}
\langle t^{-2{\el}}, \phi(t)\rangle = \int_0^\infty t^{-2{\el}} \left(\phi(t) + \phi(-t) -2\sum_{j=0}^{{\el}-1} \frac{t^{2j}}{(2j)!}\phi^{(2j)}(0)\right)\, dt,
\end{equation}
%and for $\lan=-2{\el}-1$ in (\ref{ModLanSig}) we get 
\begin{equation}\label{11}
\langle t^{-2{\el}-1}, \phi(t)\rangle = \int_0^\infty t^{-2{\el}-1} \left(\phi(t) - \phi(-t) -2\sum_{j=1}^{{\el}} \frac{t^{2j-1}}{(2j-1)!}\phi^{(2j-1)}(0)\right)\, dt.
\end{equation}

Given $\underline{x}= \sum_{j=1}^m x_je_j \in \R^m$, consider its Euclidean norm $|\m{x}|=\left(\sum_{j=1}^m x_j^2\right)^{1/2}$. If Re$(\lan)>-m$, the generalized function $|\m{x}|^\lan$ is defined by
\[
\langle |\m{x}|^\lan, \phi\rangle = \int_{\R^m} |\m{x}|^\lan  \phi(\m{x}) \, dV_{\m{x}}, \;\;\;\;\; \phi\in\Sw(\R^m).
\]
{Using} spherical coordinates in the above integral, i.e.\ $x=r\m{w}$ with $r=|\m{x}|$ and $\m{w}\in\Sa^{m-1}$, we get
\begin{equation}\label{Sphr+}
\langle |\m{x}|^\lan, \phi\rangle = \int_0^\infty r^{\lan+m-1} \left(\int_{\Sa^{m-1}} \phi(r\m{w}) \, dS_{\m{w}}\right) dr = \sigma_m \, \langle r_+^{\lan+m-1}, \Sigma[\phi](r)\rangle,
\end{equation}
where $dS_{\m{w}}$ is the Lebesgue measure on the unit sphere $\Sa^{m-1}=\{\m{w}\in\R^m: |\m{w}|=1\}$, $\sigma_{m}=\frac{2\pi^{\frac{M}{2}}}{\Gam\left(\frac{M}{2}\right)}$ is the surface area of $\Sa^{m-1}$, and 
\[
\displaystyle\Sigma[\phi](r) = \frac{1}{\sigma_m} \int_{\Sa^{m-1}} \phi(r\m{w}) \, dS_{\m{w}}\] is the so-called spherical mean of $\phi$. Clearly, $\phi\in\Sw(\R^m)$ implies that $\Sigma[\phi]\in\Sw(\R_+)$  where  $\R_+$ denotes the set of non-negative real numbers. Similarly, $\phi\in C^\infty_0(\R^m)$ implies that $\Sigma[\phi]\in C^\infty_0(\R_+)$.

From (\ref{Sphr+}) we see that $|\m{x}|^\lan$ represents the {action} of $ r_+^{\lan+m-1}$ {on} $\Sigma[\phi](r)$. As a function of the complex parameter $\lan$, this action is an analytic function in the region Re$(\lan)>-m$. Using (\ref{t+defi}), it can be {extended  to} the whole $\lan$-plane except for the points $\lan=-m,-m-1,-m-2, \ldots$, where it has simple poles. The residue of $\langle r_+^{\lan+m-1}, \Sigma[\phi](r)\rangle$ at $\lan=-m-{\el}+1$ {$(\el\in\N)$} can be computed using (\ref{t+-Res}) as follows
\[
\sigma_m \;  \res_{{\lan=-m-{\el}+1}} \langle r_+^{\lan+m-1}, \Sigma[\phi](r)\rangle  \,=\, \sigma_m \;  \res_{{\lan=-{\el}}} \langle r_+^{\lan}, \Sigma[\phi](r)\rangle  \,=\, \frac{\sigma_m}{({\el}-1)!} \, \frac{d^{{\el}-1}}{dr^{{\el}-1}} \Sigma[\phi](0).
\]
But it is known that the derivatives of odd order of {the spherical mean} $\Sigma[\phi]$ vanish at $r=0$, see \cite[Ch.1 - \S3.9]{MR0166596}. Thus the poles corresponding to even values of ${\el}$ do not exist. This leaves us with the poles corresponding to ${\el}=1,3,5\ldots$ or equivalently $\lan=-m-2{\el}$, ${\el}\in\N_0$. In \cite{MR0166596}, the residues of $|\m{x}|^\lan$ for $\lan=-m-2{\el}$ were computed to be
\begin{equation}\label{ResRLan}
\res_{\lan=-m-2{\el}} {{|\m{x}|}}^\lan = \frac{2\pi^{\frac{m}{2}}\Del_{\m{x}}^{{\el}} \del(\m{x})}{2^{2{\el}} {\el}! \Gam\left(\frac{m}{2}+{\el}\right)}, 
\end{equation}
{where $\del(\underline{x})=\del(x_1)\cdots \del(x_m)$ is the $m$-dimensional real Dirac distribution.} These simple poles can be eliminated if we divide $|\m{x}|^\lan$ by an ordinary function of $\lan$ with exactly the same simple poles. A good candidate for such a function is $\Gam\left(\frac{\lan+m}{2}\right)$ which has simple poles at $\lan=-m-2{\el}$ with residue $\res_{\lan=-m-2{\el}} \Gam\left(\frac{\lan+m}{2}\right) = \frac{2 (-1)^{{\el}}}{{\el}!}$. Thus the generalized function $\frac{{{|\m{x}|}}^\lan}{\Gam\left(\frac{\lan+m}{2}\right)}$ is an entire analytic mapping of $\lan$, and its values at the singular points of ${{|\m{x}|}}^\lan$ can be computed as 
\begin{equation}\label{ValSingP}
\frac{{{|\m{x}|}}^\lan}{\Gam\left(\frac{\lan+m}{2}\right)}\Bigg|_{\lan=-m-2{\el}} =\frac{\res_{\lan=-m-2{\el}} {{|\m{x}|}}^\lan}{\res_{\lan=-m-2{\el}} \Gam\left(\frac{\lan+m}{2}\right)} = \frac{(-1)^{{\el}} \pi^{\frac{m}{2}}\Del_{\m{x}}^{{\el}} \del(\m{x})}{2^{2{\el}} \Gam\left(\frac{m}{2}+{\el}\right)}.
\end{equation}

\subsection{Taylor expansions to construct superdistributions}
We can extend the definitions of the above {generalized} functions to superspace by considering their finite Taylor expansions. This {also} is  an important technique to produce interesting even superfunctions {(i.e.\ elements of $C^\infty(\Om)\otimes \mathfrak{G}_{2n}^{(ev)}$)}
%(and also generalized superfunctions) 
out of real-valued functions 
%(and also generalized functions), 
see e.g.\ \cite{Berezin:1987:ISA:38130}.

\begin{defi}\label{CompSupFunct}
Consider a function $F \in C^\infty(E)$ where $E$ is an open region of $\R^\ell$, and $\el$ {even real} {superfunctions} $a_j({\bf x})\in C^\infty(\Om)\otimes {\mathfrak G}^{(ev)}_{2n}$, $j=1,\ldots, \ell$. We expand every $a_j$ as the sum of its body and its nilpotent part, i.e.\ $a_j({\bf x})=[a_j]_0(\underline{x})+{\bf a_j}({\bf x})$. If the domain $E\subseteq \R^\ell$ contains the image of the function $([a_1]_0, \ldots, [a_\el]_0)$, we define the composed superfunction $F\left(a_1({\bf x}), \ldots,a_\ell({\bf x})\right) \in C^\infty(\Om)\otimes {\mathfrak G}_{2n}$ by means of the Taylor expansion as
%\begin{equation}
%F_A(y_1({\bf x}), \ldots, y_\ell({\bf x}))= \sum_{k_1,\ldots, k_\ell\geq 0} \frac{F_A^{(k_1,\ldots, k_\ell)}\left([y_1]_0(\underline{x}), \ldots, [y_\ell]_0(\underline{x})\right)}{k_1! \cdots k_\ell!}  {\bf y_1}({\bf x})^{k_1} \cdots {\bf y_\ell}({\bf x})^{k_\ell},
%\end{equation}
\begin{equation}\label{Tay_Ser}
F(a_1, \ldots, a_\ell)= \sum_{k_1,\ldots, k_\ell\geq 0} \frac{F^{(k_1,\ldots, k_\ell)}\left([a_1]_0, \ldots, [a_\ell]_0\right)}{k_1! \cdots k_\ell!}  {\bf a_1}^{k_1} \cdots {\bf a_\ell}^{k_\ell}.
\end{equation}
%{A superfunction $F$ is said to be an even superfunction if it takes values in the even subalgebra ${\mathfrak G}^{(ev)}_{2n}$ of ${\mathfrak G}_{2n}$.} % i.e.\ $F\in C^\infty(\Om)\otimes {\mathfrak G}^{(ev)}_{2n}$.
\end{defi}

\begin{remark}
Note that the series in the above definition of $F(a_1, \ldots, a_\ell)$ is finite in view of the nilpotency of ${\bf a_j}({\bf x})$. Moreover, it is clear that Definition \ref{CompSupFunct} can be used also for (generalized) functions that are not $C^\infty$ as long as all the derivatives appearing in the formula exist.
\end{remark}
%
%A form of producing interesting even superfunctions is by considering finite Taylor expansions of real-valued functions, see e.g.\ \cite{Berezin:1987:ISA:38130}. For example, consider a smooth function $F\in C^\infty(E)$, where $E\inc \R$ is an open domain, and an even superfunction $a=a_0+{\bf a}\in C^\infty(\Om)\otimes \mathfrak{G}^{(ev)}_{2n}$ where $a_0$ and ${\bf a}$ are the body and nilpotent part of $a$, respectively. If the domain $E\subseteq \R^\ell$ contains the image of the function $a_0$, we define the composed superfunction $F(a({\bf x}))\in C^\infty(\Om)\otimes \mathfrak{G}^{(ev)}_{2n}$ as
%\begin{equation}\label{BosSuFun}
%F(a)=F(a_0+{\bf a})=\sum_{j=0}^{n} \;\frac{{\bf a}^j}{j!}\;F^{(j)}(a_0).
%\end{equation}
%Note that the above series is finite in view of the nilpotency of ${\bf a}({\bf x})$. Moreover, it is clear that (\ref{BosSuFun}) can be used for functions that are not $C^\infty$ as long as all the derivatives appearing in the formula exist. The same procedure is also valid to define generalized functions in superspace out of real-valued generalized functions.

The expansion (\ref{Tay_Ser}) is used to define arbitrary real powers of even superfunctions. Let $\lan\in \R$ and $a=a_0+{\bf a}\in C^\infty(\Om)\otimes \mathfrak{G}_{2n}^{(ev)}$, then for $a_0>0$ we define  
\begin{equation}\label{GenPow}
a^\lan {:=} \sum_{j=0}^{n}\;\frac{{\bf a}^j}{j!}\; \frac{\Gam(\lan+1)}{\Gam(\lan-j+1)} \,a_0^{\lan-j}. 
\end{equation}
If $m\neq 0$, we use this idea to define the {norm} of the supervector variable ${\bf x}$. Indeed, its norm squared $-{\bf x}^2$ is an even smooth superfunction with non-negative body $|\m{x}|^2=\sum_{j=1}^m x_j^2$, see (\ref{NormSq}). {Hence, the norm of ${\bf x}$ is defined as}
%we define the {\it norm} of ${\bf x}$ by
\[|{\bf x}| {:=} (-{\bf x}^2)^{1/2}=\left(|\underline{x}|^2- \underline{x\p}^{\, 2} \right)^{1/2}=\sum_{j=0}^n \frac{(-1)^j \underline{x\p}^{\,2j}}{j!} \, \frac{\Gam\left(\frac{3}{2}\right)}{\Gam\left(\frac{3}{2}-j\right)} |\underline{x}|^{1-2j}.\]
Similarly, the analogue of the generalized function $|\underline{x}|^{\lan}$ in superspace is given by
\begin{align}\label{PowSup}
|{\bf x}|^\lan &{:=} \left(|\underline{x}|^2- \underline{x\p}^{\, 2} \right)^{\lan/2} = \sum_{j=0}^n \frac{(-1)^j \underline{x\p}^{\,2j}}{j!} \, \frac{\Gam\left(\frac{\lan}{2}+1\right)}{\Gam\left(\frac{\lan}{2}-j+1\right)} |\underline{x}|^{\lan-2j} = \sum_{j=0}^n \frac{ \underline{x\p}^{\,2j}}{j!} \, \frac{\Gam\left(-\frac{\lan}{2}+j\right)}{\Gam\left(-\frac{\lan}{2}\right)} |\underline{x}|^{\lan-2j},
\end{align}
{where, in the last equality,} we have used the identity $(-1)^j \frac{\Gam\left(\frac{\lan}{2}+1\right)}{\Gam\left(\frac{\lan}{2}-j+1\right)} =  \frac{\Gam\left(-\frac{\lan}{2}+j\right)}{\Gam\left(-\frac{\lan}{2}\right)}$. It is easily seen that $|{\bf x}|^\lan$ has {simple} poles at {$\lan=-M-2{\el}$ with ${\el}\in\N_0$.}
%$\lan=-M, -M-2, -M-4, \ldots$. 

{To compute the corresponding residues of $|{\bf x}|^\lan$, we need to introduce first the Dirac delta distribution in the supervector variable ${\bf x}$
\[\del({\bf x}):=\del(\underline{x}) \frac{\pi^n}{n!} \underline{x}\p^{\, 2n} = \pi^n \del(\m{x}) x\p_1 \cdots x\p_{2n},\]
where the product $\del(\underline{x}\p):=  \pi^n x\p_1 \cdots x\p_{2n}$ defines the Dirac distribution with respect to the fermionic variables. Indeed, it can be verified that
\[\langle \del({\bf x}), G({\bf x})\rangle= \int_{\R^{m}} \int_{B,{\m{x}\p}} \del({\bf x})G({\bf x}) \, dV_{\m{x}}=G(0),
\]
where $G\in C^\infty(U)\otimes \mathfrak{G}_{2n}$ and $U\inc \R^{m}$ is a neighborhood of the origin. We may now extend formulas (\ref{ResRLan})-(\ref{ValSingP}) to superspace.
}
{
\begin{teo}\label{NorSup}
Let $m\neq 0$, $M=m-2n$ and ${\el}\in \N_0$. Then the following properties hold.
\begin{itemize}
\item[$i)$] $\displaystyle \Del_{\bf x}^{{\el}} \del({\bf x}) = \pi^n {{\el}}! \sum_{{j}=0}^{min({{\el}},n)} \frac{4^{{j}} \, \m{x}\p^{\,2n-2{j}} }{({{\el}}-{j})!(n-{j})!} \Del_{\m{x}}^{{{\el}}-{j}}\del(\m{x})$.
\item[$ii)$] $\displaystyle \res_{\lan=-M-2{{\el}}} |{\bf x}|^\lan = \frac{2\pi^{\frac{M}{2}}}{2^{2{{\el}}}\, {{\el}}!\, \Gam\left(\frac{M}{2}+{{\el}}\right)} \Del_{\bf x}^{{\el}} \del({\bf x})$.
\item[$iii)$] The normalization $\dfrac{|{\bf x}|^\lan}{\Gam(\frac{\lan+M}{2})}$ defines an analytic mapping on the entire $\lan$-plane with values in $\Sw'(\R^m)\otimes \mathfrak{G}_{2n}$. Moreover,
\begin{equation*}
\frac{|{\bf x}|^\lan}{\Gam(\frac{\lan+M}{2})}\Bigg|_{\lan=-M-2{{\el}}} = \frac{(-1)^{{\el}} \pi^{\frac{M}{2}}}{2^{2{{\el}}} \Gam\left(\frac{M}{2}+{{\el}}\right)} \Del_{\bf x}^{{\el}} \del({\bf x}).
\end{equation*}
\end{itemize}
\end{teo}
}
{
\pf
\begin{itemize}
\item[$i)$] It is easily seen that
\[
\Del_{\bf x}^{{\el}} \del({\bf x}) =\frac{\pi^n}{n!} \, \left(\Del_{\m{x}}+\Del_{\m{x}\p}\right)^{{\el}}\left[ \del(\underline{x})  \underline{x}\p^{\, 2n} \right] = \frac{\pi^n}{n!}  \sum_{{j}=0}^{{\el}} \binom{{{\el}}}{{j}} \, \Del_{\m{x}\p}^{{j}} \left[\underline{x}\p^{\, 2n} \right] \,  \Del_{\m{x}}^{{{\el}}-{j}}\left[ \del(\underline{x}) \right]. 
\]
Formula (\ref{LapPowX}) now gives $\displaystyle \Del_{\m{x}\p}^{{j}} \left[\underline{x}\p^{\, 2n} \right] = \begin{cases} 4^{{j}} \frac{n!\, {j}!}{(n-{j})!} \m{x}\p^{2n-2{j}} & {j}\leq n,\\ 0, & {j}>n. \end{cases}$ Substituting this expression into the above sum yields the desired conclusion.
\item[$ii)$] From formula (\ref{ResRLan}) we obtain
\begin{align*}
\res_{\lan=-M-2{{\el}}} |\m{x}|^{\lan-2j} = \res_{\lan=-m-2({{\el}}+j-n)} |\m{x}|^{\lan} = \begin{cases} \dfrac{2\pi^{\frac{m}{2}}\Del_{\m{x}}^{{{\el}}+j-n} \del(\m{x})}{2^{2{{\el}}+2j-2n} \, ({{\el}}+j-n)! \, \Gam\left(\frac{m}{2}+{{\el}}+j-n\right)}& {{\el}}\geq n-j,\\[+.4cm]  0, & {{\el}}<n-j.\end{cases}
\end{align*}
Then, using formula (\ref{PowSup}) we obtain
\begin{align*}
 \res_{\lan=-M-2{{\el}}} |{\bf x}|^\lan &= %\sum_{j=0}^n \frac{ \underline{x\p}^{\,2j}}{j!} \, \left(\frac{\Gam\left(-\frac{\lan}{2}+j\right)}{\Gam\left(-\frac{\lan}{2}\right)}\Bigg|_{\lan=-M-2{{\el}}}\right) \left(\res_{\lan=-M-2{{\el}}} |\underline{x}|^{\lan-2j} \right) \\
\sum_{j=0}^n \frac{ \underline{x\p}^{\,2j}}{j!} \, \frac{\Gam\left(\frac{M}{2}+{{\el}}+j\right)}{\Gam\left(\frac{M}{2}+{{\el}}\right)} \left(\res_{\lan=-M-2{{\el}}} |\underline{x}|^{\lan-2j} \right) \\
&= \sum_{{j}=0}^{min(n,{{\el}})} \frac{ \underline{x\p}^{\,2n-2{j}}}{(n-{j})!} \, \frac{\Gam\left(\frac{M}{2}+{{\el}}+n-{j}\right)}{\Gam\left(\frac{M}{2}+{{\el}}\right)} \dfrac{2\pi^{\frac{m}{2}}\Del_{\m{x}}^{{{\el}}-{j}} \del(\m{x})}{2^{2{{\el}}-2{j}} \, ({{\el}}-{j})! \, \Gam\left(\frac{m}{2}+{{\el}}-{j}\right)}\\
&= \frac{2\pi^{\frac{m}{2}}}{2^{2{{\el}}} \Gam\left(\frac{M}{2}+{{\el}}\right)} \sum_{{j}=0}^{min(n,{{\el}})} \frac{4^{{j}}\,  \underline{x\p}^{\,2n-2{j}}}{(n-{j})!\, ({{\el}}-{j})!} \Del_{\m{x}}^{{{\el}}-{j}} \del(\m{x}),
\end{align*}
 where we have replaced the index $j$ by $n-{j}$ in the second equality. Comparison of the above equality with $i)$ completes the proof.
\item[$iii)$] We recall that $\res_{\lan=-M-2{{\el}}} \Gam\left(\frac{\lan+M}{2}\right) = \frac{2 (-1)^{{\el}}}{{{\el}}!}$. Hence, it immediately follows from $ii)$ that 
 \[
 \frac{|{\bf x}|^\lan}{\Gam(\frac{\lan+M}{2})}\Bigg|_{\lan=-M-2{{\el}}} = \dfrac{\res_{\lan=-M-2{{\el}}} |{\bf x}|^\lan}{\res_{\lan=-M-2{{\el}}} \Gam\left(\frac{\lan+M}{2}\right)} =  \frac{(-1)^{{\el}} \pi^{\frac{M}{2}}}{2^{2{{\el}}} \Gam\left(\frac{M}{2}+{{\el}}\right)} \Del_{\bf x}^{{\el}} \del({\bf x}). 
 \]
$ \hfill\square$
\end{itemize}
}
%, and the corresponding residues can be computed using formula (\ref{ResRLan}).  Thus, the normalization  

%\noindent {Note to self: Should we explicitly compute the residues of $|{\bf x}|^\lan$ and the values of $\frac{|{\bf x}|^\lan}{\Gam(\frac{\lan+M}{2})}$ at the singular points of $|{\bf x}|^\lan$?}

\subsection{Concentrated delta distributions and integral over the supersphere}
{Throughout this paper, we will integrate over supermanifolds of co-dimension 1 in $\R^{m|2n}$. To that end, we need the following definition of concentrated Dirac delta distribution.}
\begin{defi}\label{ConDelSS}
Consider {an even real} superfunction $g=g_0+{\bf g}\in C^\infty(\R^m)\otimes \mathfrak{G}_{2n}^{(ev)}$, where $g_0$ and ${\bf g}$ are the {body} and the nilpotent part of $g$ respectively, and such that $\pa_{\underline x}[g_0]\neq0$ on the surface $g_0^{-1}(0):=\{\underline{w}\in\R^{m}: g_0(\underline{w})=0\}$. The distribution $\del^{(\el)}(g)$, $\el\in\N_0$, is defined as the Taylor series
\begin{equation}\label{delk}
\del^{(\el)}(g) {:=} \sum_{j=0}^{n} \frac{{\bf g}^j}{j!} \, \del^{(\el+j)}(g_0).
\end{equation}
When $\el=0$, the above distribution is the {\it concentrated delta distribution} on the supermanifold defined by the equation $g({\bf x})=0$. 
\end{defi}

{
%We refer the reader to \cite[Ch.3 - \S1]{MR0166596} for a detailed treatment of concentrated delta distribution in the classical case.
}

In \cite{Guz_Somm5}, integration over general supermanifolds of codimension $1$ was introduced by means of the action of  concentrated Dirac distributions. In particular, the supersphere $\Sa^{m-1,2n}$ is algebraically defined by the relation ${\bf x}^2+1=0$ if
$m\neq 0$. Thus the classical integral over the unit sphere in $\R^m$ is extended to $\Sa^{m-1,2n}$ as (see also \cite{MR2344451, MR2539324})
\begin{equation}\label{IntSupSph}
 \int_{\Sa^{m-1,2n}} F({\bf x}) \, dS_{\bf x}= 2\int_{\R^m} \int_{B,{\m{x}\p}} \del({\bf x}^2+1) \,F({\bf x})\; dV_{\underline x},
 \end{equation}
 where $\del({\bf x}^2+1)=\sum_{j=0}^n \frac{\underline{x}\p^{\,2j}}{j!} \;\del^{(j)}(1-|\underline{x}|^2)$ is the concentrated delta distribution on the supersphere. {For a superfunction $F$ of the form (\ref{SupFunc}), the above integral reads as follows
\[
 \int_{\Sa^{m-1,2n}} F({\bf x}) \, dS_{\bf x} = 2 \sum_{j = 0,\ldots, n \atop A\inc \{1, \ldots, 2n\}} \frac{1}{j!}\left(\int_{B,{\m{x}\p}} \m{x}\p^{\, 2j} \m{x}\p_A \right) \left(\int_{\R^m} \del^{(j)}(1-|\underline{x}|^2) F_A(\m{x})\, V_{\m{x}}\right),
\]
where  $\del^{(j)}(1-|\underline{x}|^2)$ is the $j$-th derivative of the concentrated delta distribution (or $(j+1)$-fold layer) on the unit sphere $\Sa^{m-1}=\{\m{w}\in \R^m: |\m{w}|=1\}$. As usual, the notation $\int_{\R^m} \del^{(j)}(1-|\underline{x}|^2) F_A(\m{x})\, V_{\m{x}}$ is used for the evaluation of the distribution $\del^{(j)}(1-|\underline{x}|^2)$ on the real function $F_A(\m{x})$, which can be computed as
\[
\int_{\R^m} \del^{(j)}(1-|\underline{x}|^2) F_A(\m{x})\, V_{\m{x}} =\frac{1}{2} \int_{\Sa^{m-1}} \left(\frac{\pa}{\pa{r^2}}\right)^j \left[r^{m-2}F_A(r\m{w})\right]\Big|_{r=1} \; dS_{\m{w}}.
\]
 We refer the reader to \cite{MR0166596, MR1996773} for a complete treatment on concentrated delta distributions
and $j$-fold layer integrals in $\R^m$, 
and to \cite{MR2539324, Guz_Somm5} for concrete examples of their use in integration over the supersphere.}

{ The integral (\ref{IntSupSph})} is (up to a multiplicative constant) the unique $\mathfrak{osp}(m|2n)$-invariant integration operator over the supersphere that satisfies (see \cite{MR3060765, MR2539324})
\begin{equation}\label{IntSphMod1}
 \int_{\Sa^{m-1,2n}} f(|{\bf x}|) F({\bf x})\, dS_{\bf x} = f(1) \int_{\Sa^{m-1,2n}}  F({\bf x})\, dS_{\bf x},
\end{equation}
for any $f:\R\fd \R$ {smooth} in a neighborhood of the point $x=1$.

{In \cite{MR2539324}, it was proven that} the integral (\ref{IntSupSph}) over $\Sa^{m-1,2n}$ reduces to the following Pizzetti formula when integrating  superpolynomials %(see e.g.\ \cite{MR2539324})
\begin{equation}\label{PizzSS}
\int_{\Sa^{m-1,2n}} R({\bf x})\, dS_{\bf x}=\sum_{j=0}^\infty  \frac{2\pi^{M/2}}{2^{2j}\, j!\, \Gamma(j+M/2)} \Del_{\bf x}^j [R]\Big|_{{\bf x} =0}.
\end{equation}
In particular, one obtains that the surface area $ \sigma_{M}$ of  $\Sa^{m-1,2n}$ is given by $\sigma_{M}= \frac{2\pi^{\frac{M}{2}}}{\Gam\left(\frac{M}{2}\right)}$.

%\begin{remark}\label{NorInt}
In the case $M=-2k$, the {first} ($k+1$) terms of the Pizzetti sum vanish, i.e.
\[\int_{\Sa^{m-1,2n}} R({\bf x})\, dS_{\bf x}=\sum_{j=k+1}^\infty  \frac{2\pi^{M/2}}{2^{2j}\, j!\, \Gamma(j+M/2)} \Del_{\bf x}^j [R]\Big|_{{\bf x} =0}.\]
{This implies that} the integral of any polynomial of degree $\leq 2k+1$ vanishes. In particular, the area of the supersphere equals zero in this case, i.e. 
{\[\sigma_{-2k}={\int_{\Sa^{m-1,2n}} 1\, dS_{\bf x}=} \frac{2\pi^{-k}}{\Gam(-k)}=0.\] }
Thus the normalized integral $\frac{1}{\sigma_{-2k}}\int_{\Sa^{m-1,2n}} F({\bf x})\, dS_{\bf x}$ is in general not well-defined. However, for certain functions $F$ with a vanishing integral over the supersphere, it is possible to define a (non-vanishing) normalized integral. {We now recall a few important facts about this normalized integral for even and negative superdimensions. We refer the reader to \cite{MR2344451, CK_Ali}  for more details.}

{The idea behind the definition of the normalized integral is as follows. Consider $M$ as a formal complex parameter,  the Pizzetti formula (\ref{PizzSS}) for the normalized integral of a polynomial $R({\bf x})\in \bigoplus_{j=0}^{2k+1}\mathcal{P}_j$ reads %as
\[
\frac{1}{\sigma_{M}} \int_{\Sa^{m-1,2n}} R({\bf x})\, dS_{\bf x} = \sum_{j=0}^k  \frac{\Gam(M/2)}{2^{2j}\, j!\, \Gamma(j+M/2)} \Del_{\bf x}^j [R]\Big|_{{\bf x}=0}.
\]}
Taking the limit of this expression for $M\fd -2k$, and using the fact that the Gamma function has simple poles at $-k, -k+1, \ldots, 0$, we obtain the following definition
% Consider $M$ as a formal complex parameter and divide the Pizzetti formula (\ref{PizzSS}) for $R({\bf x})$  by the area of the supersphere $\sigma_{M}= \frac{2\pi^{\frac{M}{2}}}{\Gam\left(\frac{M}{2}\right)}$. Taking the limit of this expression for $M\fd -2k$, and using the fact that the Gamma function has simple poles, we obtain the following definition
\begin{align}\label{NormInt}
\frac{1}{\sigma_{-2k}} \int_{\Sa^{m-1,2n}} R({\bf x})\, dS_{\bf x} &:= \lim_{M\fd-2k } \sum_{j=0}^k  \frac{\Gam\left(\frac{M}{2}\right)}{2^{2j}\, j!\, \Gamma(j+M/2)} \Del_{\bf x}^j [R]\Big|_{{\bf x} =0}\nonumber\\
&= \sum_{j=0}^k  \frac{(k-j)!}{2^{2j}\, j!\, k!} \left(-\Del_{\bf x}\right)^j [R]\Big|_{{\bf x} =0}.
\end{align}
This definition is particularly interesting in the purely fermionic case $m=0$, in which the Pizzetti formula (\ref{PizzSS}) yields the trivial functional $\int_{\Sa^{-1,2n}} \cdot\, dS_{\bf x}\equiv 0$. The normalized integral (\ref{NormInt}) still satisfies (\ref{IntSphMod1}) (see \cite{CK_Ali}), namely if $R_{2j}\in\mathcal{P}_{2j}$ {and  $j+\el\leq k$, then
\begin{align}\label{NormIntMod1}
\frac{1}{\sigma_{-2k}} \int_{\Sa^{m-1,2n}} {\bf x}^{\,2\el} R_{2j}({\bf x})\, dS_{{\bf x}} &=(-1)^\el \; \frac{1}{\sigma_{-2k}} \int_{\Sa^{m-1,2n}}  R_{2j}({\bf x})\, dS_{{\bf x}}.
% \begin{cases} \displaystyle (-1)^\el \; \frac{1}{\sigma_{-2k}} \int_{\Sa^{-1,2n}}  R_{2j}({\bf x})\, dS_{{\bf x}} & \el+j\leq k, \\[+.3cm]
%0 & \el+j>k. \end{cases}
\end{align}}
%For more details about this normalized integral and its applications we refer the reader to \cite{MR2344451, CK_Ali}.
%\end{remark}

\section{Radon transform in superspace}\label{S4}
In this section we discuss some of the main properties of the Radon transform in superspace. Let us start by providing a brief overview about the classical Radon transform in $\R^m$. In the purely bosonic case, the Radon transform of a function $\phi\in \Sw(\R^m)$ is defined as
\[
R_m[\phi](\m{w},p) = \int_{\R^m} \del(\langle\m{x}, \m{w}\rangle-p) \, \phi(\m{x})\, dV_{\m{x}} = \frac{1}{|\m{w}|} \int_{\langle\m{x}, \m{w}\rangle=p}  \phi(\m{x})\, dS_{\m{x}},
\]
where $dS_{\m{x}}$ denotes the Lebesgue measure on the hyperplane $\langle\m{x}, \m{w}\rangle=p$. This transform maps functions in $\Sw(\R^m)$ into functions in $\Sw(\mathscr{P}^m)$, where $\mathscr{P}^m$ is the space of all hyperplanes in $\R^m$. See e.g.\ \cite{MR754767, MR709591, MR573446} for {a more complete} discussion of the properties of the Radon transform.

We recall that each hyperplane in $\mathscr{P}^m$ can be written as $\{\m{x}\in\R^m: \langle\m{x}, \m{w}\rangle=p\}$ with $\m{w}\in\Sa^{m-1}$ and $p\in \R$. Note that the pairs $(\m{w},p)$ and $(-\m{w},-p)$ define the same hyperplane in $\mathscr{P}^m$. Thus, the mapping  
\[(\m{w},p) \mapsto {\{\m{x}\in\R^m: \langle\m{x}, \m{w}\rangle=p\}} \in \mathscr{P}^m\] is a double covering of $\Sa^{m-1}\times\R$ onto $ \mathscr{P}^m$.

The Radon transform is closely related to the Fourier transform in $\R^m$ 
\[
\mathcal{F}_m[\phi](\m{\xi}) = \frac{1}{(2\pi)^{m/2}} \int_{\R^m} e^{-i \langle \m{x},\m{\xi}\rangle} \phi (\m{x})\, dV_{\m{x}}.
\]
In fact, from the {central-slice} theorem we have (see \cite{MR754767, MR573446})
\begin{align}\label{FouVsRad}
\mathcal{F}_m[\phi](r\m{w}) &= \frac{1}{(2\pi)^{m/2}} \int_{-\infty}^{\infty} e^{-irp} R_m[\phi](\m{w},p)\, dp, & (\m{w},p)&\in \Sa^{m-1}\times\R.
\end{align}
It is easily seen that $\phi\in \Sw(\R^m)$ implies that $\mathcal{F}_m[\phi](r\m{w}) \in \Sw(\R)$ for each fixed $\m{w}$. Thus it is clear from (\ref{FouVsRad}) that the function $p\mapsto R_m[\phi](\m{w},p)$  {belongs to} $ \Sw(\R)$. Property (\ref{FouVsRad}) can be further extended as follows.

\begin{pro}\label{CSLTGEN}
Let $g\in\Sw'(\R)$, $\phi\in \Sw(\R^m)$ and $\m{\xi}\in\R^m\setminus\{0\}$. Then
\[
\int_{\R^m} g(\langle \m{\xi},\m{x}\rangle) \, \phi(\m{x})\, dV_{\m{x}} = \int_{-\infty}^{\infty} g(p) \,R_m[\phi](\m{\xi},p)\, dp.
\]
\end{pro}
\pf
Let us write $\m{\xi}=r\m{w}$ with $r>0$ and $\m{w}\in\Sa^{m-1}$. Consider the coordinate transformation $\m{y}=M\m{x}$ where $M\in\textup{SO}(m)$ is a rotation matrix whose first row is given by the unit vector $\m{w}$, i.e.\ the first component of $\m{y}$ is $y_1=\langle \m{w},\m{x}\rangle$. Then,
\begin{align}\label{P1}
\int_{\R^m} g(\langle \m{\xi},\m{x}\rangle) \, \phi(\m{x})\, dV_{\m{x}} &= \int_{\R^m} g\left(ry_1\right) \, \phi\left(M^{-1}\m{y}\right) \, dy_1\ldots dy_m \nonumber \\
&= \int_{-\infty}^{\infty} g(p) \left(\frac{1}{r} \int_{\R^{m-1}} \psi\left(\frac{p}{r}, y_2, \ldots, y_m\right) \, dy_2\ldots dy_m\right) \, dp,
\end{align}
where we have used the substitutions $p=r y_1$ and $\psi(\m{y})= \phi\left(M^{-1}\m{y}\right)$. Using the same coordinate transformation, we obtain
\begin{align}\label{P2}
R_m[\phi](\m{\xi},p)&=  \int_{\R^m} \del(\langle\m{x}, \m{\xi}\rangle-p) \, \phi(\m{x})\, dV_{\m{x}} \nonumber\\
&=  \int_{\R^m} \del(ry_1-p) \,\psi(\m{y})\, dy_1\ldots dy_m \nonumber\\
&=\frac{1}{r} \int_{\R^{m-1}} \psi\left(\frac{p}{r}, y_2, \ldots, y_m\right) \, dy_2\ldots dy_m.
\end{align}
Combining (\ref{P1}) and (\ref{P2}) we obtain the desired result. $\hfill\square$ 

{We now have the following definition of the Radon transform in the superspace setting.}
\begin{defi}\label{DefRadonTrans}
Let $m\neq 0$. Given a commuting variable $p$ and a supervector variable ${\bf w}$, both independent of ${\bf x}$, we define the Radon transform of a superfunction $\phi\in \mathcal S(\R^m) \otimes \mathfrak{G}_{2n}$ as
\begin{equation}\label{RadonTrans}
R_{m|2n}[\phi]({\bf w}, p) = \int_{\R^{m|2n}_{\bf x}} \del(\langle{\bf x},{\bf w}\rangle -p) \phi({\bf x}), \;\;\; \mbox{ with }  \;\;\;\del(\langle{\bf x},{\bf w}\rangle -p)=\sum_{j=0}^{2n}\frac{\langle\m{x}\p,\m{w}\p\rangle^j}{j!}  \, \del^{(j)}(\langle\m{x},\m{w}\rangle -p).
\end{equation}
\end{defi}

\begin{remark}\label{C-STheoSS}
{The Radon transform in superspace was introduced in the works \cite{MR2422641, MR2539324} where some basic properties were studied.} Initially, the super Radon transform was introduced in \cite{MR2422641} in terms of the super Fourier transform using the central-slice property (\ref{FouVsRad}) as definition. In the later work  \cite{MR2539324}, the above definition in terms of the Dirac distribution was introduced. It has been proven that this definition indeed satisfies the central-slice theorem, i.e.\
\begin{equation}\label{CenSliSS}
\mathcal{F}_{m|2n}[\phi](r {\bf w}) = \frac{1}{(2\pi)^{M/2}} \int_{-\infty}^{\infty} e^{-irp} R_{m|2n}[\phi]({\bf w},p)\, dp, 
\end{equation}
where 
\[
\mathcal{F}_{m|2n}[\phi]({\bf y}) = \frac{1}{(2\pi)^{M/2}} \int_{\R^{m|2n}_{\bf x}} e^{-i \langle {\bf x},{\bf y}\rangle} \phi ({\bf x}),
\]
is the Fourier transform in superspace. The transform $\mathcal{F}_{m|2n}$ defines an isomorphism of $\mathcal S(\R^m) \otimes \mathfrak{G}_{2n}$, {see \cite[Theorem 7]{MR2422641}. Then the condition} 
%It thus follows that 
$\phi\in \mathcal S(\R^m) \otimes \mathfrak{G}_{2n}$ implies that, {given a supervector parameter ${\bf w}$, the function $r\mapsto \mathcal{F}_{m|2n}[\phi](r{\bf w})$} is a rapidly decreasing function of the real variable $r$.
%$\mathcal{F}_{m|2n}[\phi](r{\bf w})$ is a rapidly decreasing function of the real-variable $r>0$.
\end{remark}

Now, we show some additional properties of the super Radon transform which shall be useful in what follows. 

\begin{pro}\label{RTPP}
The Radon transform in superspace satisfy the following properties:
\begin{itemize}
\item[$i)$]\underline{{\it Homogeneity:}} Consider an even {real} superfunction $h=h_0+{\bf h}\in C^\infty(\R^m)\otimes \mathfrak{G}_{2n}^{(ev)}$, where $h_0$ and ${\bf h}$ are the {body} and the nilpotent part of $h$ respectively. If $h_0>0$ in $\R^m$, then
\[R_{m|2n}[\phi](h{\bf w}, hp) = \frac{1}{h}R_{m|2n}[\phi]({\bf w}, p).\]
\item[$ii)$] \underline{{\it Shifting property:}} Consider the translation $\phi_{\bf y}({\bf x})=\phi({\bf x}-{\bf y})$. Then

 \[R_{m|2n}[\phi_{\bf y}]({\bf w}, p) = R_{m|2n}[\phi]({\bf w}, p-\langle{\bf y}, {\bf w}\rangle).\]
 
 \item[$iii)$] \underline{{\it Derivatives of the transform:}} 
 \begin{align*}
 \pa_{w_j}R_{m|2n}[\phi]({\bf w}, p) &= -\pa_p \, R_{m|2n}[x_j \phi]({\bf w}, p),\\
 \pa_{w\p_{2j-1}}R_{m|2n}[\phi]({\bf w}, p) &= \frac{1}{2}\pa_p \, R_{m|2n}[x\p_{2j} \phi]({\bf w}, p),\\
  \pa_{w\p_{2j}}R_{m|2n}[\phi]({\bf w}, p) &= \frac{-1}{2}\pa_p \, R_{m|2n}[x\p_{2j-1} \phi]({\bf w}, p).
 \end{align*}
 
 \item[$iv)$] \underline{{\it Action of the Dirac and Laplace operators:}} 
  \begin{align*}
 \pa_{{\bf w}}R_{m|2n}[\phi]({\bf w}, p) &= \pa_p \, R_{m|2n}[{\bf x} \phi]({\bf w}, p),\\
 \Del_{{\bf w}}R_{m|2n}[\phi]({\bf w}, p) &=\pa^2_p \, R_{m|2n}[|{\bf x}|^2 \phi]({\bf w}, p).
 \end{align*}
\end{itemize}
\end{pro}
\pf
\begin{itemize}
\item[$i)$] {Given two real superfunctions $g,h\in C^\infty(\R^m)\otimes \mathfrak{G}_{2n}^{(ev)}$, it is known that $\del(hg)=\frac{\del(g)}{h}$, provided that the body $h_0$ of $h$ is positive in $\R^m$, and that the body $g_0$ of $g$ has a non-vanishing gradient on the surface $g_0^{-1}(0)$. This was proved in \cite[Proposition 7]{Guz_Somm5}.}
%for any $g\in C^\infty(\R^m)\otimes \mathfrak{G}_{2n}^{(ev)}$ such that its real part $g_0$ has a non-vanishing gradient on the surface $g_0^{-1}(0)$. 
 Taking $g({\bf x})=\langle{\bf x},{\bf w}\rangle -p$, we obtain
\[R_{m|2n}[\phi](h{\bf w}, hp) ={\int_{\R^{m|2n}_{\bf x}}} \del\left(h(\langle{\bf x},{\bf w}\rangle -p)\right) \,\phi({\bf x})  = \frac{1}{h} R_{m|2n}[\phi]({\bf w}, p). \]

\item[$ii)$] In virtue of the translation invariance of the Berezin integral (see \cite[Ch.2 - \S2]{Berezin:1987:ISA:38130}), the change of variable ${\bf u}={\bf x}-{\bf y}$ yields
\begin{align*}
R_{m|2n}[\phi_{\bf y}]({\bf w}, p) &= {\int_{\R^{m|2n}_{\bf x}}} \del(\langle{\bf x},{\bf w}\rangle -p) \,\phi({\bf x}-{\bf y})\\ 
&= {\int_{\R^{m|2n}_{\bf u}}} \del\left(\langle{\bf u},{\bf w}\rangle + \langle{\bf y},{\bf w}\rangle -p\right) \,\phi({\bf u})\\ 
&= R_{m|2n}[\phi]({\bf w}, p-\langle{\bf y}, {\bf w}\rangle).
\end{align*}
\item[$iii)$] By the chain rule in superspace (see \cite[Ch.2 - \S1]{Berezin:1987:ISA:38130}), we obtain
\begin{align*}
 \pa_{w\p_{2j-1}}R_{m|2n}[\phi]({\bf w}, p) &= {\int_{\R^{m|2n}_{\bf x}}}  \pa_{w\p_{2j-1}} \,\del(\langle{\bf x},{\bf w}\rangle -p) \,\phi({\bf x})\\ 
 &={\int_{\R^{m|2n}_{\bf x}}}  -\frac{x\p_{2j}}{2} \; \del^\prime(\langle{\bf x},{\bf w}\rangle -p) \,\phi({\bf x}) \\
 &= \frac{1}{2} \pa_{p} \, R_{m|2n}[x\p_{2j} \phi]({\bf w}, p).
\end{align*}
The formulas for $ \pa_{w_j}R_{m|2n}[\phi]({\bf w}, p)$ and $\pa_{w\p_{2j}}R_{m|2n}[\phi]({\bf w}, p)$ can be proven in a similar way. 
\item[$iv)$]  It follows from direct computations using $iii)$. $\hfill\square$ \\
\end{itemize}

Proposition \ref{CSLTGEN} extends to the superspace setting as follows.

\begin{pro}\label{P3}
Let $g\in\mathcal{S}'(\R)$ and $\phi\in\mathcal{S}(\R^m)\otimes \mathfrak{G}_{2n}$ . Then
\begin{equation}\label{SLT}
\int_{\R^{m|2n}_{\bf x}}  g(\langle{\bf x},{\bf w}\rangle) \phi({\bf x}) = \int_{-\infty}^\infty g(p) \, R_{m|2n}[\phi]({\bf w}, p) \, dp.
\end{equation}
\end{pro}
\pf
From {formula (\ref{CenSliSS})}
%the central-slice theorem 
and the fact that $\mathcal{F}_{m|2n}[\phi](r {\bf w})$ is a $\Sw(\R)$ function of {$r\in \R$}, it follows that 
{
\[
R_{m|2n}[\phi]({\bf w},p) = (2\pi)^{\frac{M}{2}-1} \int_{-\infty}^{\infty} e^{irp}\mathcal{F}_{m|2n}[\phi](r {\bf w})\, dr,
%\mathcal{F}_{m|2n}[\phi](r {\bf w}) = \frac{1}{(2\pi)^{M/2}} \int_{-\infty}^{\infty} e^{-irp} R_{m|2n}[\phi]({\bf w},p)\, dp, 
\]
which is of class $\Sw(\R)$ when considered as a function of the real variable $p$.
}
%$R_{m|2n}[\phi]({\bf w}, p)$ is a $\Sw(\R)$ function of the variable $p$ (see Remark \ref{C-STheoSS}).
Thus, the action of the tempered distribution $g(p)$ on $R_{m|2n}[\phi]({\bf w}, p)$ in (\ref{SLT}) is well-defined. Using Proposition \ref{CSLTGEN} and the finite Taylor expansion of $g(\langle{\bf x},{\bf w}\rangle)$ we obtain
\begin{align*}
\int_{\R^{m|2n}_{\bf x}}  g(\langle{\bf x},{\bf w}\rangle) \phi({\bf x}) &= \sum_{j=0}^{2n} \int_{B, {\m{x}\p}} \frac{\langle \m{x}\p,\m{w}\p\rangle^j}{j!} \int_{\R^m}g^{(j)}(\langle \m{x},\m{w}\rangle) \phi({\bf x})\, dV_{\m{x}} \\
&= \sum_{j=0}^{2n} \int_{B,  {\m{x}\p}} \frac{\langle \m{x}\p,\m{w}\p\rangle^j}{j!}  \int_{-\infty}^\infty g^{(j)}(p)\,  R_{m}[\phi](\m{w}, p) \, dp \\
&= \sum_{j=0}^{2n} \int_{B,  {\m{x}\p}} \frac{\langle \m{x}\p,\m{w}\p\rangle^j}{j!} (-1)^j \int_{-\infty}^\infty g(p)\,  \pa_p^j R_{m}[\phi](\m{w}, p) \, dp\\
&= \int_{-\infty}^\infty g(p) \left(\int_{B,  {\m{x}\p}} \sum_{j=0}^{2n} \frac{\langle \m{x}\p,\m{w}\p\rangle^j}{j!}  \int_{\R^m} \del^{(j)}(\langle\m{x}, \m{w}\rangle-p) \, \phi({\bf x})\, dV_{\m{x}}\right) \, dp \\
&= \int_{-\infty}^\infty g(p) \left( \int_{\R^{m|2n}_{\bf x}} \del(\langle{\bf x},{\bf w}\rangle -p) \phi({\bf x})\right) \, dp \\
&=\int_{-\infty}^\infty g(p) \, R_{m|2n}[\phi]({\bf w}, p) \, dp,
\end{align*}
which proves the result. $\hfill\square$

\begin{cor}\label{Cor1PWtoRad}
Let $g\in\mathcal{S}'(\R)$ and $\phi\in\mathcal{S}(\R^m)\otimes \mathfrak{G}_{2n}$. Then
\begin{enumerate}
\item[$i)$] If $g\equiv 1$, {it follows that}
 \[ \int_{\R^{m|2n}_{\bf x}}  \phi({\bf x}) = \int_{-\infty}^\infty R_{m|2n}[\phi]({\bf w}, p) \, dp.\]
 \item[$ii)$] Given {a real superfunction} $a({\bf w})\in C^\infty(\R^m)\otimes \mathfrak{G}_{2n}^{(ev)}$, {we have}
    \[ \int_{\R^{m|2n}_{\bf x}} g(\langle{\bf x},{\bf w}\rangle + a({\bf w})) \,\phi({\bf x}) = \int_{-\infty}^\infty g(p) \, R_{m|2n}[\phi]({\bf w}, p-  a({\bf w})) \, dp.\]
\end{enumerate}
\end{cor}
\pf
{We only prove} $ii)$ since $i)$ is a direct consequence of Proposition \ref{P3}. Let us write $a({\bf w})= a_0(\m{w})+{\bf a}({\bf w})$ where $a_0$ and ${\bf a}$ are the {body} and nilpotent parts of $a$ respectively. Then
\[
\int_{\R^{m|2n}_{\bf x}} g(\langle{\bf x},{\bf w}\rangle + a({\bf w})) \phi({\bf x}) =  \sum_{j=0}^n \frac{{\bf a}({\bf w})^j}{j!} \int_{\R^{m|2n}_{\bf x}} g^{(j)}\left(\langle{\bf x},{\bf w}\rangle + a_0(\m{w})\right)  \phi({\bf x}).
\]
Using Proposition \ref{P3} for the tempered distributions $g^{(j)}\left(t + a_0(\m{w})\right)$ {on the real line ($t\in \R$)}, we obtain 
\begin{align*}
\int_{\R^{m|2n}_{\bf x}} g(\langle{\bf x},{\bf w}\rangle + a({\bf w})) \phi({\bf x}) &=  \sum_{j=0}^n \frac{{\bf a}({\bf w})^j}{j!} \int_{-\infty}^\infty g^{(j)}\left(p + a_0(\m{w})\right) R_{m|2n}[\phi]({\bf w}, p) \, dp \\
&=  \sum_{j=0}^n \frac{{\bf a}({\bf w})^j}{j!} \int_{-\infty}^\infty g^{(j)}\left(p \right) R_{m|2n}[\phi]({\bf w}, p-  a_0(\m{w})) \, dp \\
&= \int_{-\infty}^\infty g(p) \left(  \sum_{j=0}^n (-1)^j \frac{{\bf a}({\bf w})^j}{j!} \, \pa_p^jR_{m|2n}[\phi]({\bf w}, p-  a_0(\m{w})) \right)\, dp \\
&= \int_{-\infty}^\infty g(p) \, R_{m|2n}[\phi]({\bf w}, p-  a({\bf w})) \, dp,
\end{align*}
which proves the result. $\hfill\square$

\section{Plane wave decomposition of the delta distribution}\label{S5}
The main goal of this section is to obtain a plane wave decomposition of the Dirac delta distribution in superspace. This is an important step towards the inversion formulas for the Radon transform, which will be obtained in the next section. {Before studying the superspace case,} let us recall first some {useful} plane wave decompositions in $\R^m$.

\subsection{Plane wave decompositions in $\R^m$}\label{5.1}
Let $\m{w}\in \Sa^{m-1}$, $\lan\in\C$ with Re$(\lan)>-1$ and let $F_\lan(\langle\m{x},\m{w}\rangle)$ be the generalized function defined by
\[
\left\langle F_\lan(\langle\m{x},\m{w}\rangle), \phi(\m{x}) \right\rangle {:=} \int_{\R^m} \frac{\left|\langle\m{x},\m{w}\rangle\right|^\lan}{\Gam\left(\frac{\lan+1}{2}\right)} \, \phi(\m{x})\, dV_{\m{x}}, \;\;\;\;\; {\phi \in \mathcal{S}(\R^m).}
\]
{By} Proposition \ref{CSLTGEN}, this generalized function in $\R^m$ can be written as the following {1-dimensional} functional  
\[
\left\langle F_\lan(\langle\m{x},\m{w}\rangle), \phi(\m{x}) \right\rangle = \int_{-\infty}^{+\infty} \frac{|p|^\lan}{\Gam\left(\frac{\lan+1}{2}\right)} R_m[\phi](\m{w},p)\, dp.
\]
{Then, on account of (\ref{ValSingP}), $F_\lan(\langle\m{x},\m{w}\rangle)$ can be analytically continued to the entire $\lan$-plane.} Since $F_\lan(\langle\m{x},\m{w}\rangle)$ is a functional depending continuously on the parameter $\m{w}$, we can integrate $F_\lan(\langle\m{x},\m{w}\rangle)$ over $\Sa^{m-1}$. In this way, we obtain a new functional $G_\lan$ given by
\[
\left\langle G_\lan, \phi(\m{x}) \right\rangle {:=} \int_{\Sa^{m-1}} \left\langle F_\lan(\langle\m{x},\m{w}\rangle), \phi(\m{x}) \right\rangle \, dS_{\m{w}}.
\]
{This integral can be explicitly computed for $\textup{Re}(\lan)>-1$ yielding (see e.g.\ \cite[Ch.1 - \S 3.10 ]{MR0166596})}
%In \cite[Ch.1 - \S 3.10 ]{MR0166596}, the following explicit formula for the generalized function  $G_\lan$ was computed
\begin{equation}\label{PWRLan}
\frac{1}{\pi^{\frac{m-1}{2}} \Gam\left(\frac{\lan+1}{2}\right)}  \int_{\Sa^{m-1}} \left|\langle\m{x},\m{w}\rangle\right|^\lan \, dS_{\m{w}} = \frac{2{{|\m{x}|}}^\lan }{\Gam\left(\frac{\lan+m}{2}\right)}.
\end{equation}
{Again,} the validity of this formula can be extended by analytic continuation from $\{\textup{Re}(\lan)>-1\}$ to the rest of the $\lan$-plane. Formula (\ref{PWRLan}) provides {the so-called} plane wave decomposition {for} %the functional 
\[
\frac{2{{|\m{x}|}}^\lan }{\Gam\left(\frac{\lan+m}{2}\right)}.
\]
Evaluating formula (\ref{PWRLan}) at $\lan=-m-2{{\el}}$, and making use of (\ref{ValSingP}), we {obtain}
\begin{align}\label{Princ}
\frac{1}{\pi^{\frac{m-1}{2}}}  \int_{\Sa^{m-1}} \frac{\left|\langle\m{x},\m{w}\rangle\right|^\lan}{ \Gam\left(\frac{\lan+1}{2}\right)}\Bigg|_{\lan=-m-2{{\el}}} \, dS_{\m{w}} %\;= \; \frac{2r^\lan }{\Gam\left(\frac{\lan+m}{2}\right)} \Bigg|_{\lan=-m-2{{\el}}} 
 \;=\; \frac{(-1)^{{\el}} 2\pi^{\frac{m}{2}}\Del_{\m{x}}^{{\el}} \del(\m{x})}{2^{2{{\el}}} \Gam\left(\frac{m}{2}+{{\el}}\right)}.
\end{align}
Thus, if $m$ is even, this formula reduces to 
\begin{equation}\label{PrincEven}
\frac{1}{\pi^{\frac{m-1}{2}} \Gam\left(\frac{1-m}{2}-{{\el}}\right)}   \int_{\Sa^{m-1}} \langle\m{x},\m{w}\rangle^{-m-2{{\el}}} \, dS_{\m{w}} = \frac{(-1)^{{\el}} 2\pi^{\frac{m}{2}}\Del_{\m{x}}^{{\el}} \del(\m{x})}{2^{2{{\el}}} \Gam\left(\frac{m}{2}+{{\el}}\right)}.
\end{equation}
On the other hand, if $m$ is odd, the integrand in formula (\ref{Princ}) is evaluated at one of the simple poles of $|t|^\lan$. {Using formula (\ref{ValSingP}) on the real line, i.e.\ {when} $m=1$, we obtain}
\begin{equation}\label{PinValDim1}
\frac{\left|t\right|^\lan}{ \Gam\left(\frac{\lan+1}{2}\right)}\Bigg|_{\lan=-2j-1} = \; \frac{(-1)^j j!}{(2j)!} \del^{(2j)}(t).
\end{equation}
{Then, taking $t=\langle\m{x},\m{w}\rangle$ and $j=\frac{m-1}{2}+{{\el}}$, formula (\ref{Princ}) yields}
%In this case, formula  (\ref{Princ}) yields
\begin{equation}\label{PrincOdd}
\frac{(-1)^{\frac{m-1}{2}} \left(\frac{m-1}{2}+{{\el}}\right)!}{\pi^{\frac{m-1}{2}} (m-1+2{{\el}})!} \int_{\Sa^{m-1}} \del^{(m-1+2{{\el}})}(\langle\m{x},\m{w}\rangle)\,  dS_{\m{w}} =  \frac{2\pi^{\frac{m}{2}}\Del_{\m{x}}^{{\el}} \del(\m{x})}{2^{2{{\el}}} \Gam\left(\frac{m}{2}+{{\el}}\right)}.
\end{equation}
{Summarizing}, taking ${{\el}}=0$ in (\ref{PrincEven}) and (\ref{PrincOdd}) we obtain the following plane wave decompositions of $\del(\m{x})$
\begin{equation}\label{PWDelRm}
\del({\m{x}}) =  
\begin{cases}
\displaystyle \frac{(-1)^{\frac{m}{2}}(m-1)!}{(2\pi)^m} \int_{\Sa^{m-1}}  \langle{\m{x}},{\m{\omega}}\rangle^{-m} \, dS_{\m{\omega}}, & \mbox{ for } \;\;m \mbox{ even},\\[+.5cm]
\displaystyle {\frac{(-1)^{\frac{m-1}{2}}}{2 (2\pi)^{m-1}}} \int_{\Sa^{m-1}} \del^{(m-1)}(\langle{\m{x}},{\m{\omega}}\rangle) \, dS_{\m{\omega}}, & \mbox{ for } \;\;m \mbox{ odd}.
 \end{cases}
\end{equation}
For a {more complete} study of these plane wave decompositions we refer the reader to \cite[Ch.1 - \S 3 ]{MR0166596}.

\subsection{Plane wave decompositions in superspace}
%The delta distribution on the supervector variable ${\bf x}$ is given by
%\[\del({\bf x})=\del(\underline{x}) \frac{\pi^n}{n!} \underline{x}\p^{\, 2n} = \pi^n \del(\m{x}) x\p_1 \cdots x\p_{2n},\]
%where $\del(\underline{x})=\del(x_1)\cdots \del(x_m)$ is the $m$-dimensional real Dirac distribution, and $\del(\underline{x}\p)=  \pi^n x\p_1 \cdots x\p_{2n}$ is the Dirac distribution with respect to the fermionic variables. Indeed, it can be verified that
%\[\langle \del({\bf x}), G({\bf x})\rangle= \int_{\R^{m}} \int_{B,{\m{x}\p}} \del({\bf x})G({\bf x}) \, dV_{\m{x}}=G(0),
%\]
%where $G\in C^\infty(U)\otimes \mathfrak{G}_{2n}$ with $U\inc \R^{m}$ being a neighborhood of the origin.

In this section, we {obtain a decomposition of the distribution $\del({\bf x})=\del(\underline{x}) \frac{\pi^n}{n!} \underline{x}\p^{\, 2n}$ into plane waves integrated over the supersphere}, thus extending formulas (\ref{PWDelRm}) to superspace. To that end we will not follow the classical approach from \cite{MR0166596}, which was briefly described in Section \ref{5.1}. Instead, we will look at this problem from the perspective of hyperfunctions. Indeed, we will first show that $\del({\bf x})$ is {a suitable} %the 
boundary value of the super Cauchy kernel, i.e.\ the fundamental solution of the {generalized Cauchy-Riemann} operator $\pa_{\bf x}-\pa_{x_0}$, {where $x_0$ is an extra real variable}. Then, we combine this result with the plane wave decomposition obtained in \cite{CK_Ali} for this Cauchy kernel. 

% To that end, we show first that $\del({\bf x})$ can be approximated by the boundary value of the super Cauchy kernel, i.e.\ the fundamental solution of the operator $\pa_{\bf x}-\pa_{x_0}$. Then, we combine this result with the plane wave decomposition for this Cauchy kernel obtained in \cite{CK_Ali}. 

A fundamental solution $\fhi_1^{m+1|2n} (x_0,{\bf x})$  of $\pa_{\bf x}-\pa_{x_0}$ must satisfy the condition
\[(\pa_{\bf x}-\pa_{x_0})\, \fhi_1^{m+1|2n} (x_0,{\bf x}) = \del(x_0)\del({\bf x}).\]
The following explicit expressions for $\fhi_1^{m+1|2n} (x_0,{\bf x})$ were computed in \cite{CK_Ali}. For a detailed account on fundamental solutions of the super Dirac operator $\pa_{\bf x}$ and super Laplace operator $\Del_{\bf x}$ we refer the reader to \cite{MR2386499}.
\begin{lem}\label{Lem5}
A fundamental solution of $\pa_{\bf x}-\pa_{x_0}$ is given by 
\begin{equation}\label{OrigFor}
\fhi_1^{m+1|2n} (x_0,{\bf x}) =  \pi^n \sum_{{{j}}=0}^n \frac{(-1)^{{j}} 2^{2{{j}}} {{j}}!}{(n-{{j}})!}\, \fhi_{2{{j}}+1}^{m+1|0} \underline{x}\p^{\, 2n-2{{j}}} - \pi^n \sum_{{{j}}=0}^{n-1} \frac{(-1)^{{j}} 2^{2{{j}}+1} {{j}}!}{(n-{{j}}-1)!}\, \nu_{2{{j}}+2}^{m+1|0} \underline{x}\p^{\, 2n-2{{j}}-1},
\end{equation}
where  {$ \nu_{2{{j}}+2}^{m+1|0}$ is the fundamental solution of %$(\pa_{x_0}^2+\Del_{\m{x}})^{{{j}}+1}$, where $\pa_{x_0}^2+\Del_{\m{x}}$ is the classical bosonic Laplacian in $(m+1)$ dimensions,}
$\Del_{m+1|0}^{{{j}}+1}$, being $\Del_{m+1|0} := \pa_{x_0}^2+\Del_{\m{x}}$ the bosonic Laplacian in $(m+1)$ dimensions,} 
and $\fhi_{2{{j}}+1}^{m+1|0}:= (\pa_{\m{x}}-\pa_{x_0}) \nu_{2{{j}}+2}^{m+1|0}$ is a fundamental solution of %the operator 
$(-\pa_{\m{x}}-\pa_{x_0})\Del_{m+1|0}^{{{j}}}$.

Moreover, if $M+1\notin -2\N_0$, the fundamental solution $\fhi_1^{m+1|2n}$ can also be written as 
\begin{align*}
\fhi_1^{m+1|2n}(x_0,{\bf x}) &= \frac{-1}{\sigma_{M+1}}  \frac{x_0-{\bf x}}{|x_0-{\bf x}|^{M+1}}%\\
 %&= \frac{-\sgn(x_0)}{\sigma_{M+1}} \left[ \sum_{j=0}^\infty \frac{{\bf x}^{2j}}{j!} \frac{\Gam\left(\frac{M+1}{2}+j\right)}{\Gam\left(\frac{M+1}{2}\right)} |x_0|^{-M-2j} -\sgn(x_0) \sum_{j=0}^\infty \frac{{\bf x}^{2j+1}}{j!} \frac{\Gam\left(\frac{M+1}{2}+j\right)}{\Gam\left(\frac{M+1}{2}\right)} |x_0|^{-M-2j-1}\right],
\end{align*} 
where $|x_0-{\bf x}|=|x_0+{\bf x}|=\left(x_0^2+|\m{x}|^2-\m{x}\p^{\, 2}\right)^\frac{1}{2}$.% and $\sigma_{M+1} = \frac{2\pi^{\frac{M+1}{2}}}{\Gam\left(\frac{M+1}{2}\right)}$ is the surface area of the supersphere $\Sa^{m|2n}$. %(see e.g. \cite{Guz_Somm5}).% and  $\sgn(x_0)$ is the sign of $x_0$.
\end{lem}

Before stating the plane wave decomposition theorem for the super  Cauchy kernel, we recall how to construct monogenic plane waves out of  holomorphic functions, see e.g.\ \cite{CK_Ali, MR985370}. Let $g(z)=g_1(a,b)+ig_2(a,b)$ be a holomorphic $\C$-valued function of the complex variable $z=a+ib$ in an open domain $\Om\subseteq \R^2\iso \C$. If $m\neq 0$, given a supervector parameter ${\bf w}=\m{w}+\m{w}\p$, we define 
\begin{equation}\label{HolSF}
g(\langle{\bf x},{\bf w}\rangle - x_0 {\bf w} ) = g_1(\langle{\bf x},{\bf w}\rangle, x_0 |{\bf w}|) - \frac{{\bf w}}{|{\bf w}|} g_2(\langle{\bf x},{\bf w}\rangle, x_0 |{\bf w}|),
\end{equation}
as an element of $C^\infty(\Om_{\m{w}})\otimes \mathfrak{G}_{2n}\otimes \mathcal{C}_{m,2n}$ where $\Om_{\m{w}}=\left\{{(\m{x}, x_0)}\in \R^{m+1}: \left(\langle\m{x},\m{w}\rangle, \, x_0|\m{w}|\right)  \in \Om \right\}$. Here the functions $g_\el(\langle{\bf x},{\bf w}\rangle, x_0 |{\bf w}|)$, $\el=1,2$, are defined as in (\ref{Tay_Ser}). {A function of the type (\ref{HolSF}) is  called a {\it monogenic plane wave}.  The monogenicity of $g(\langle{\bf x},{\bf w}\rangle - x_0 {\bf w} )$ with respect to the operator $\left(\pa_{\bf x} - \pa_{x_0}\right)$ was indeed established in \cite{CK_Ali}, i.e.\ $\left(\pa_{\bf x} - \pa_{x_0}\right) \, g(\langle{\bf x},{\bf w}\rangle - x_0 {\bf w} )=0$.}
%Since $g_1,g_2$ are real analytic functions in $\Om$, the definition of $g_\el(\langle{\bf x},{\bf w}\rangle, x_0 |{\bf w}|)$, $\el=1,2$, is independent of the splitting of $\langle{\bf x},{\bf w}\rangle$ and $x_0 |{\bf w}|$, {provided that the conditions of Lemma \ref{Lem1} are satisfied}. Therefore, if {$(\underline{0},x_0|\m{w}|) \in \Om$} and $\left(\langle\m{x},\m{w}\rangle, \, x_0|\m{w}|\right)$ belongs to the region of  convergence of the Taylor series of $g_\el$ around $(\underline{0},x_0|\m{w}|)$,   we obtain %the expression
%\begin{align}\label{Taylor_PW}
%g(\langle{\bf x},{\bf w}\rangle - x_0 {\bf w})&= \sum_{j=0}^\infty \frac{\langle{\bf x},{\bf w}\rangle^j}{j!} \, \pa_a^j g_1 \left(0, x_0 |{\bf w}|\right) - \frac{{\bf w}}{|{\bf w}|} \sum_{j=0}^\infty \frac{\langle{\bf x},{\bf w}\rangle^j}{j!} \, \pa_a^j g_2\left(0, x_0 |{\bf w}|\right) \\
%&= \sum_{j=0}^\infty \frac{\langle{\bf x},{\bf w}\rangle^j}{j!} \,  g^{(j)}\left(- x_0 {\bf w}\right),\nonumber
%\end{align}
%which coincides with the Taylor expansion of $g$ as a function of one complex variable.
\begin{remark}
{In (\ref{HolSF})},  we have replaced the role of the complex imaginary unit $i$ by {the} supervector $-\frac{{\bf w}}{|{\bf w}|}$. We recall that this correspondence does not exist if $m=0$ since ${\bf w}=\m{w}\p$ is nilpotent. However, it is still possible to find an analogue of definition  (\ref{HolSF}) in this context given by
\begin{equation*}%\label{HolPF}
g(\langle\m{x}\p,\m{w}\p\rangle - x_0 \m{w}\p) = \sum_{j=0}^{2n} \frac{(\langle\m{x}\p,\m{w}\p\rangle - x_0 \m{w}\p)^j}{j!} g^{(j)}(0),
\end{equation*}
for any real-valued function $g$ of class $C^{2n}$ in a neighborhood of $z=0$. Therefore, in the case $m=0$, it suffices to consider only the generators $(\langle\m{x}\p,\m{w}\p\rangle - x_0 \m{w}\p)^j$ for $j=0, 1, \ldots, 2n$.
\end{remark}

We also introduce the following sequence of complex functions
\begin{equation}\label{RecFunc}
G_\el(z) = \frac{z^\el}{\el!} \ln(z) - a_\el z^\el, \;\;\;\; \mbox{ with } \;\;\;  a_{\el+1}=\frac{1}{\el+1} \left(a_\el + \frac{1}{(\el+1)!}\right), \;\; a_0=0,
\end{equation}
%\begin{remark}
where $\ln(z)$ is the the principle branch of the complex logarithm, i.e.\ we consider $-\pi < Arg(z) \leq \pi$. 
%defined in the principle branch $-\pi < Arg(z) \leq \pi$ of the logarithm function in the complex plane. 
The sequence $\{a_\el\}$ can also be explicitly defined as $a_\el = \frac{\Psi(\el+1) - \Psi(1)}{\el!}$ where $\Psi(z) = \frac{\Gam'(z)}{\Gam(z)}$ is the digamma function. 
%The functions in the sequence $\{G_\el\}_{\el\in\N_0}$ appear in the plane wave decomposition of the Cauchy kernel in even negative superdimensions. 
The functions in the sequence $\{G_\el\}_{\el\in\N_0}$ are primitives of the complex logarithm function. Indeed, it is easy to check that $G'_{\el+1}=G_\el$ while $G_0(z)=\ln(z)$.

We can now formulate the decomposition into plane waves of the super Cauchy kernel $\fhi_1^{m+1|2n} (x_0,{\bf x})$ obtained in \cite{CK_Ali}.

\begin{teo}{\bf [Plane wave decomposition of Cauchy kernel]}\label{PWCK}
\noindent Let $x_0\neq 0$ and $M+1\notin -2\N_0$, {with $M=m-2n$}. Then
\begin{itemize}
\item[$i)$] If {$M\geq1$},
\begin{align*}
\frac{-1}{\sigma_{M+1}}  \frac{x_0-{\bf x}}{|x_0-{\bf x}|^{M+1}} &= -\sgn(x_0) \frac{(-1)^{\frac{M}{2}}(M-1)!}{2 (2\pi)^M} \int_{\Sa^{m-1,2n}} (\langle{\bf x},{\bf w}\rangle - x_0 {\bf w})^{-M}\, dS_{\bf w}, & \mbox{ for }& \;\;M \mbox{ even},\\[+.2cm]
\frac{-1}{\sigma_{M+1}}  \frac{x_0-{\bf x}}{|x_0-{\bf x}|^{M+1}}&= - \frac{(-1)^{\frac{M+1}{2}}(M-1)!}{2 (2\pi)^M} \int_{\Sa^{m-1,2n}} (\langle{\bf x},{\bf w}\rangle - x_0 {\bf w})^{-M} {\bf w}\, dS_{\bf w}, & \mbox{ for }& \;\;M \mbox{ odd}.
\end{align*}

\item[$ii)$] If $M =-2k$ ($m\neq 0$),
\begin{multline*}
\frac{-1}{\sigma_{M+1}}  \frac{x_0-{\bf x}}{|x_0-{\bf x}|^{M+1}} = \frac{- \sgn(x_0)}{ 4^{k} (k!)^2 \;\sigma_{-2k+1}}\; \Del_{\bf w}^k \left[(\langle{\bf x},{\bf w}\rangle - x_0 {\bf w})^{2k}\right]\\
 + \frac{(-1)^k (4\pi^2)^k}{2} \sgn(x_0) \int_{\Sa^{m-1,2n}} G_{2k}\left(\langle{\bf x},{\bf w}\rangle - x_0 {\bf w}\right) \, dS_{\bf w},
\end{multline*}
with $G_{2k}$ defined as in (\ref{RecFunc}).
\item[$iii)$] If $M =-2n$ ($m= 0$),
\begin{align*}
\frac{-1}{\sigma_{-2n+1}}  \frac{x_0-{\m{x}\p}}{|x_0-\m{x}\p|^{-2n+1}} &= \frac{-\sgn(x_0)}{4^n \, (n!)^2 \, \sigma_{-2n+1}}\, \Del_{\m{w}\p}^n \left[(\langle\m{x}\p,\m{w}\p\rangle - x_0 \m{w}\p)^{2n}\right].
\end{align*}
\end{itemize}
% If $M=-2k$ with $k\in \N-1$, the above plane wave decomposition of $\fhi_1^{m+1|2n}$ does not hold. In this case one has 
% \[\int_{\Sa^{m-1,2n}} (\langle{\bf x},{\bf w}\rangle - x_0 {\bf w})^{-M}\, dS_{\bf w} =0.\]
\end{teo}

\begin{remark}
The actions of the Laplace operators in $ii)$  and $iii)$ on the corresponding plane wave polynomials can be seen as normalized integrals over the supersphere as defined in (\ref{NormInt}).
%Remark \ref{NorInt}. 
In particular, 
 \begin{align*}
 \frac{- \sgn(x_0)}{ 4^{k} (k!)^2 \;\sigma_{-2k+1}}\; \Del_{\bf w}^k (\langle{\bf x},{\bf w}\rangle - x_0 {\bf w})^{2k}   &= \frac{- (-1)^k \sgn(x_0)}{\sigma_{-2k+1}} \frac{1}{\sigma_{-2k}} \int_{\Sa^{m-1,2n}} (\langle{\bf x},{\bf w}\rangle - x_0 {\bf w})^{2k} \, dS_{\bf w}. 
 \end{align*}
 and 
 \begin{align*}
 \frac{-\sgn(x_0)}{4^n \, (n!)^2 \, \sigma_{-2n+1}}\, \Del_{\m{w}\p}^n \left[(\langle\m{x}\p,\m{w}\p\rangle - x_0 \m{w}\p)^{2s}\right] &= \frac{-(-1)^n \, \sgn(x_0)}{\sigma_{-2n+1}} \;  \frac{1}{\sigma_{-2n}} \int_{\Sa^{-1,2n}} (\langle\m{x}\p,\m{w}\p\rangle - x_0 \m{w}\p)^{2n} dS_{\m{w}\p}.
\end{align*}
\end{remark}

%\subsection{Radon decomposition of the Delta distribution}
{We now show that,} similarly to the classical case, the delta distribution in superspace can be written as the boundary value of the Cauchy kernel. We define the boundary value of a generalized  {superfunction} $f(x_0, {\bf x})$ at $x_0=0$ by
\[
\textup{B.V.} \left[f\right]({\bf x}) := \lim_{x_0\fd0^+}\left[f(x_0, {\bf x}) - f(-x_0, {\bf x})\right].
\] 

\begin{teo}\label{BVDeltaT}
Let $M+1\notin -2\N_0$,  {with $M=m-2n$}. Then
\begin{equation}\label{BVDelta}
-\del({\bf x}) = \textup{B.V.} \left[\fhi_1^{m+1|2n}\right] ({\bf x})= \lim_{x_0\fd0^+}\left[\fhi_1^{m+1|2n}(x_0, {\bf x}) - \fhi_1^{m+1|2n}(-x_0, {\bf x})\right].
\end{equation}
This is equivalent to the following set of equalities (see Lemma \ref{Lem5}),
%\begin{itemize}
%\item[$i)$] $\textup{B.V.} \left[\fhi_1^{m+1|0} \right] ({\bf x}) = -\del(\m{x})$,
%\item[$ii)$] $\textup{B.V.} \left[\fhi_{2j+1}^{m+1|0} \right] ({\bf x})=0 $, for $j=1, \ldots, n$,
%\item[$iii)$] $\textup{B.V.} \left[ \nu_{2j+2}^{m+1|0} \right] ({\bf x}) = 0$, for $j=0, \ldots, n-1$.
%\end{itemize}
\begin{align}%\label{BVDeltaCond}
\textup{B.V.} \left[\fhi_1^{m+1|0} \right] (\m{x}) &= -\del(\m{x}), \label{BVDeltaCond1}\\
\textup{B.V.} \left[\fhi_{2j+1}^{m+1|0} \right] (\m{x}) &= 0, & j&=1, \ldots, n, \label{BVDeltaCond2}\\
\textup{B.V.} \left[ \nu_{2j+2}^{m+1|0} \right] (\m{x}) &= 0, & j&=0, \ldots, n-1. \label{BVDeltaCond3}
\end{align}
\end{teo}
\pf 
The equivalence between (\ref{BVDelta}) and (\ref{BVDeltaCond1})-(\ref{BVDeltaCond3}) immediately follows from {the definition of $\del({\bf x})$} and formula (\ref{OrigFor}). We proceed to prove each of the formulas (\ref{BVDeltaCond1})-(\ref{BVDeltaCond3}).

\noindent In order to prove (\ref{BVDeltaCond1}) we need to introduce the notion of Cauchy transform of a distribution in $\R^m${, see e.g.\ \cite{MR697564}}. Let $T\in\mathcal{E}'(\R^m)$, the Cauchy transform of $T$ is defined as 
\begin{align*}
\widehat{T}(x_0+\m{x}) = 
%\frac{1}{\sigma_{m+1}} \left\langle T_{\m{u}}, \frac{\m{u}+x_0-\m{x}}{|\m{u}+x_0-\m{x}|^{m+1}}\right\rangle = 
-\left\langle T_{\m{u}},\, \fhi_1^{m+1|0}(x_0,\m{x}-\m{u})\right\rangle =   \frac{1}{\sigma_{m+1}} \int_{\R^m}  \frac{\m{u}+x_0-\m{x}}{|\m{u}+x_0-\m{x}|^{m+1}} T(\m{u})\, dV_{\m{u}}, \;\;\;\;\; x_0\neq 0,
\end{align*}
where $\fhi_1^{m+1|0}(x_0,\m{x}) = {\dfrac{-1}{\sigma_{m+1}} \dfrac{x_0-\m{x}}{|x_0-\m{x}|^{m+1}}}$ is the Cauchy kernel in the purely bosonic case, i.e.\ the fundamental solution of $-(\pa_{x_0}+\pa_{\m{x}})$. Since $x_0\neq 0$, it is clear that  $\fhi_1^{m+1|0}(x_0,\m{x}-\m{u})\in C^\infty(\R^m)$ in the variable $\m{u}$. Thus the above action of the distribution $T_{\m{u}}$ is well-defined and $\widehat{T}(x_0+\m{x})$ is a monogenic generalized function. {Here monogenicity is understood with respect to the operator $\pa_{x_0}+\pa_{\m{x}}$.}

\noindent From \cite[Theorem 27.7]{MR697564}, any distribution $T\in\mathcal{E}'(\R^m)$ can be written as the boundary value of its Cauchy transform, i.e.\ $T=\textup{B.V.} [\widehat{T}]$, or equivalently, %for every $\phi\in C_0^\infty(\R^m)$ we have
\begin{equation}\label{TeoCau}
\left\langle T,\, \phi \right\rangle = \lim_{x_0\fd 0^+} \int_{\R^m} \left(\widehat{T}(x_0+\m{x})-\widehat{T}(-x_0+\m{x})\right) \phi(\m{x}) dV_{\m{x}}, \;\;\;\;\; \phi\in C_0^\infty(\R^m).
\end{equation}
Now, if we take $T=\del(\m{x})=\del(x_1)\ldots\del(x_m)$, we have %that
\[
\widehat{\del}(x_0+\m{x}) =-\left\langle \del({\m{u}}),\, \fhi_1^{m+1|0}(x_0,\m{x}-\m{u})\right\rangle = -\fhi_1^{m+1|0}(x_0,\m{x}) =\frac{1}{\sigma_{m+1}} \frac{x_0-\m{x}}{|x_0-\m{x}|^{m+1}}.
\]
Then by formula (\ref{TeoCau}) we obtain $\del(\m{x})=-\textup{B.V.} \left[\fhi_1^{m+1|0}(x_0,\m{x}) \right]$, which proves (\ref{BVDeltaCond1}).

\noindent Now, we proceed to prove (\ref{BVDeltaCond2}). {We first} recall that the  fundamental solution $\nu_{2j}^{m+1|0}$ of  $\Del_{m+1|0}^{j}$, with $m+1-2j\notin -2\N_0$, is given by (see \cite{MR745128})
\begin{align}\label{PolyMonFundSolEven}
\nu_{2j}^{m+1|0} &= c(m+1,j) |x_0+\m{x}|^{2j-m-1}, &{\mbox{ where}\;\;\;\;\;} c(m+1,j)&= \frac{(-1)^j \Gam\left(\frac{m+1}{2}-j\right)}{2^{2j} \pi^{\frac{m+1}{2}} \Gam(j)}.
% \frac{(-1)^j \Gam\left(\frac{m+1}{2}-j\right)}{2^{2j} \pi^{\frac{m+1}{2}} \Gam(j)} \frac{|x_0+\m{x}|^{2j}}{|x_0+\m{x}|^{m+1}}, \;\;\;\;\; \mbox{ if } \;\; m+1-2j\notin -2\N_0.
\end{align}
Thus the fundamental solution $\fhi_{2j+1}^{m+1|0}$ of the operator $(-\pa_{\m{x}}-\pa_{x_0})\Del_{m+1|0}^{{j}}$ can be computed as (see \cite{CK_Ali})
\begin{align*}%\label{PolyMonFundSolOdd}
\fhi_{2j+1}^{m+1|0}&= (\pa_{\m{x}}-\pa_{x_0}) \nu_{2j+2}^{m+1|0} =  d(m+1,j) \frac{(x_0-\m{x}) }{|x_0+\m{x}|^{m+1-2j}},  & j&={1,\ldots, n,}
\end{align*}
{where $d(m+1,j) =\frac{(-1)^{j+1} \Gam\left(\frac{m+1}{2}-j\right)}{2^{2j+1} \pi^{\frac{m+1}{2}} \Gam(j+1)}$.}
%\begin{align}\label{PolyMonFundSolOdd}
%\fhi_{2j+1}^{m+1|0}&= (\pa_{\m{x}}-\pa_{x_0}) \nu_{2j+2}^{m+1|0} =  d(m+1,j) \frac{(x_0-\m{x}) }{|x_0+\m{x}|^{m+1-2j}},  &d(m+1,j) =&\frac{(-1)^{j+1} \Gam\left(\frac{m+1}{2}-j\right)}{2^{2j+1} \pi^{\frac{m+1}{2}} \Gam(j+1)},
%\end{align}
%which is valid for $j=0,\ldots, n$. 
Then, for any test function $\phi\in C_0^\infty(\R^m)$  we obtain
\begin{align*}
{I_j}(x_0) &:= \int_{\R^m} \left(\fhi_{2j+1}^{m+1|0}(x_0,\m{x})-\fhi_{2j+1}^{m+1|0}(-x_0,\m{x})\right) \phi(\m{x})\, dV_{\m{x}} = 2\, d(m+1,j) \, x_0\int_{\R^m} \frac{\phi(\m{x})}{|x_0+\m{x}|^{m+1-2j}} \,dV_{\m{x}}.
\end{align*}
{If $m+1-2j<0$, it is immediately seen that $\lim_{x_0\fd 0^+} I_j(x_0) =0$. On the other hand, if $m+1-2j\geq 0$, we use the identity $\frac{1}{|x_0+\m{x}|}\leq \frac{1}{|\m{x}|}$ to show that}
%Observe that $|x_0-\m{x}|^2=x_0^2+|\m{x}|^2\geq|\m{x}|^2$,  which implies that $\frac{1}{|x_0-\m{x}|}\leq \frac{1}{|\m{x}|}$. Thus, if $m+1-2j\geq 0$, we obtain by means of the change of coordinates $\m{x}=r\m{w}$, $r>0$, $\m{w}\in\Sa^{m-1}$, that
\begin{align}\label{Appr}
\frac{{|I_j(x_0)|}}{2\, d(m+1,j) } \leq x_0 \int_{\R^m} \frac{{|\phi(\m{x})|}}{|\m{x}|^{m+1-2j}} \,dV_{\m{x}} 
%=x_0 \int_0^\infty r^{2j-2} \left(\int_{\Sa^{m-1}} \phi(r\m{w})  \,dS_{\m{w}}\right) dr 
=   \sigma_m\,  x_0 \int_0^\infty r^{2j-2}\;  \Sigma\left[{|\phi|}\right](r) \, dr, 
\end{align}
{Since} $\Sigma\left[{|\phi|}\right](r)=\frac{1}{\sigma_{m}}\int_{\Sa^{m-1}} {|\phi(r\m{w})|}  \,dS_{\m{w}} $ is 
%the so-called spherical mean of the function $\phi$. It is 
clearly compactly supported, {we have} %Therefore 
\[
\int_0^\infty r^{2j-2}\;  \Sigma\left[{|\phi|}\right](r) \, dr<\infty, \;\;\;\;\;\; \mbox{ for }\;\;\;\;\; j=1,\ldots, n.
\]
%$\int_0^\infty r^{2j-2} \Sigma[\phi](r) dr<\infty$ for $j=1,\ldots, n$. 
Hence, formula (\ref{Appr}) yields $\lim_{x_0\fd {0^+}} I_j(x_0)=0$,
%. On the other hand, if $m+1-2j<0$, it is immediately seen that $\lim_{x_0\fd 0^+} I(x_0) =0$
which proves (\ref{BVDeltaCond2}).

\noindent The proof of (\ref{BVDeltaCond3}) follows immediately from (\ref{PolyMonFundSolEven}). Indeed,
\[
\textup{B.V.} \left[ \nu_{2j+2}^{m+1|0} \right] (\m{x}) =  c(m+1,j+1) \lim_{x_0\fd 0^+}\left(|x_0+\m{x}|^{2j-m-1} - |-x_0+\m{x}|^{2j-m-1}\right)=0. \]
$\hfill\square$

\begin{remark}
{It is known that $C_0^\infty(\R^m)$ is dense in $\Sw(\R^m)$, see e.g. \cite[Lemma~7.1.8]{MR1996773}. Thus,}
although formula (\ref{BVDelta}) was proven for actions {on} test functions in $C_0^\infty(\R^m)$, it can be extended by density to actions on {test functions} in $\Sw(\R^m)$.
\end{remark}

Combining {Theorems} \ref{PWCK} and \ref{BVDeltaT} we obtain the following plane wave decomposition for the delta distribution in superspace. {In appendix \ref{App} we provide an alternative proof for this result, which follows from the Funk-Hecke Theorem in superspace and that can be of independent interest.}

\begin{teo}\label{ThmDelPW}
Let $M+1\notin -2\N_0$, {with $M=m-2n$}. Then
\begin{itemize}
\item[$i)$] If $M {\geq}1$,
\begin{align*}
\del({\bf x}) &=  \frac{(-1)^{\frac{M}{2}}(M-1)!}{(2\pi)^M} \int_{\Sa^{m-1,2n}}  \langle{\bf x},{\bf w}\rangle^{-M} \, dS_{\bf w}, & \mbox{ for }& \;\;M \mbox{ even},\\[+.2cm]
\del({\bf x}) &=  {\frac{(-1)^{\frac{M-1}{2}}}{2 (2\pi)^{M-1}} }\int_{\Sa^{m-1,2n}} \del^{(M-1)}(\langle{\bf x},{\bf w}\rangle) \, dS_{\bf w}, & \mbox{ for }& \;\;M \mbox{ odd}.
\end{align*}
%In the above formulas we have made use of the following superdistributions
%\begin{align*}
%\textup{P.V.} \left[\langle{\bf x},{\bf w}\rangle^{-M}\right]  &= \sum_{j=0}^{2n} \frac{\langle\m{x}\p,\m{w}\p\rangle^j}{j!} \frac{(-1)^j(M+j-1)!}{(M-1)!} \; \textup{P.V.} \left[\langle\m{x},\m{w}\rangle^{-M-j} \right], \\
% \del^{(M-1)}(\langle{\bf x},{\bf w}\rangle) &= \sum_{j=0}^{2n} \frac{\langle\m{x}\p,\m{w}\p\rangle^j}{j!}\, \del^{(M-1+j)} (\langle\m{x},\m{w}\rangle)
%\end{align*}

\item[$ii)$] If $M =-2k$ ($m\neq 0$),
\[
\del({\bf x})= \frac{1}{ 2^{2k-1} (k!)^2 \;\sigma_{-2k+1}}\, \Del_{\bf w}^k \left[\langle{\bf x},{\bf w}\rangle^{2k}\right] -  (-1)^k (4\pi^2)^k  \int_{\Sa^{m-1,2n}}  G_{2k}\left(|\langle{\bf x},{\bf w}\rangle| \right)\, dS_{\bf w}.
\]
%with $\textup{P.V.} \left[G_{2k}\left(|\langle{\bf x},{\bf w}\rangle| \right)\right]$ defined, in a similar way as $\textup{P.V.} \left[\langle{\bf x},{\bf w}\rangle^{-M}\right]$, by means of its Taylor expansion.
\item[$iii)$] If $M =-2n$ ($m= 0$),
\begin{align*}
\del(\m{x}\p)&=\frac{1}{ 2^{2n-1} (n!)^2 \;\sigma_{-2n+1}}\, \Del_{\m{w}\p}^n \left[\langle\m{x}\p,\m{w}\p\rangle^{2n}\right] .
\end{align*}
\end{itemize}
\end{teo}
\pf The theorem is proved by combining formula (\ref{BVDelta}) with the plane wave decompositions of the Cauchy kernel in Theorem \ref{PWCK}.

\paragraph{Case $i)$ $M {\geq}1$.} If $M$ is even, we obtain 
\[
-\del({\bf x}) = - \frac{(-1)^{\frac{M}{2}}(M-1)!}{2 (2\pi)^M} \int_{\Sa^{m-1,2n}} \textup{B.V.}\left[\sgn(x_0) (\langle{\bf x},{\bf w}\rangle - x_0 {\bf w})^{-M}\right]\, dS_{\bf w}.
\]
{It} is easily seen that 
\[\textup{B.V.}\left[\sgn(x_0) (\langle{\bf x},{\bf w}\rangle - x_0 {\bf w})^{-M}\right] = \lim_{x_0\fd 0^+}\Big(\langle{\bf x},{\bf w}\rangle - x_0 {\bf w}\Big)^{-M} + \Big(\langle{\bf x},{\bf w}\rangle + x_0 {\bf w}\Big)^{-M} =2\langle{\bf x},{\bf w}\rangle^{-M},\]
{where 
\[\langle{\bf x},{\bf w}\rangle^{-M} = \sum_{j=0}^{2n} (-1)^j \frac{\langle{\m{x}\p},\m{w}\p\rangle^j}{j!} \; \frac{(M+j-1)!}{(M-1)!} \,\langle{\m{x}},\m{w}\rangle^{-M-j},\] 
and} %where 
the functionals $\langle{\m{x}},\m{w}\rangle^{-M-j}$ are defined as in (\ref{10})-(\ref{11}). We thus conclude that
\[
\del({\bf x}) =  \frac{(-1)^{\frac{M}{2}}(M-1)!}{(2\pi)^M} \int_{\Sa^{m-1,2n}}  \langle{\bf x},{\bf w}\rangle^{-M} \, dS_{\bf w}.
\]
On the other hand, if $M$ is odd we obtain
\begin{equation}\label{BVDel1}
-\del({\bf x}) = - \frac{(-1)^{\frac{M+1}{2}}(M-1)!}{2 (2\pi)^M} \int_{\Sa^{m-1,2n}} \textup{B.V.} \left[(\langle{\bf x},{\bf w}\rangle - x_0 {\bf w})^{-M} {\bf w} \right]\, dS_{\bf w}.
\end{equation}
We recall that the Dirac delta distribution on the real line can be written as the boundary value of the complex Cauchy kernel, i.e.
\[
\del(a) = \frac{-1}{2\pi i}\;  \textup{B.V.} \left[ \frac{1}{z}\right] =  \lim_{b\fd 0^+} \, \frac{-1}{2\pi i} \left[ \frac{1}{a+ib} - \frac{1}{a-ib}\right].
\]
Differentiating $(M-1)$ times yields the identity
%\begin{equation}\label{DelCompId}
\[
\del^{(M-1)}(a) =\frac{-(M-1)!}{2\pi i} \, \textup{B.V.} \left[ z^{-M}\right].
\]
%\end{equation}
We may now make the identifications $a\mapsto \langle{\bf x},{\bf w}\rangle$, $b\mapsto x_0$, $i\mapsto \frac{-{\bf w}}{|{\bf w}|}$ and $\frac{1}{i} \mapsto  \frac{{\bf w}}{|{\bf w}|}$ and obtain
\begin{equation}\label{BVDel2}
\del^{(M-1)}(\langle{\bf x},{\bf w}\rangle) = \frac{-(M-1)!}{2\pi} \, \textup{B.V.} \left[\frac{{\bf w}}{|{\bf w}|} \left(\langle{\bf x},{\bf w}\rangle - x_0 \frac{{\bf w}}{|{\bf w}|}\right)^{-M}  \right].
\end{equation}
Here $\del^{(M-1)}(\langle{\bf x},{\bf w}\rangle)$ is defined by means of the Taylor expansion (\ref{delk}) and $\left(\langle{\bf x},{\bf w}\rangle - x_0 \frac{{\bf w}}{|{\bf w}|}\right)^{-M}$ is defined by (\ref{HolSF})  after taking $g(z)=z^{-M}$ and {replacing} $x_0$ by {$\dfrac{x_0}{|{\bf w}|}$}, which yields the plane wave 
\begin{equation}\label{MonPWMod}
g\left(\langle{\bf x},{\bf w}\rangle - x_0 \frac{{\bf w}}{|{\bf w}|}\right) = g_1(\langle{\bf x},{\bf w}\rangle, x_0 ) - \frac{{\bf w}}{|{\bf w}|} g_2(\langle{\bf x},{\bf w}\rangle, x_0 ).
\end{equation}
On account of (\ref{IntSphMod1}), and combining (\ref{BVDel1}) with (\ref{BVDel2}), we obtain
\begin{align*}
\del({\bf x}) &= \frac{(-1)^{\frac{M+1}{2}}(M-1)!}{2 (2\pi)^M} \int_{\Sa^{m-1,2n}} \textup{B.V.} \left[\frac{{\bf w}}{|{\bf w}|} \left(\langle{\bf x},{\bf w}\rangle - x_0 \frac{{\bf w}}{|{\bf w}|}\right)^{-M}  \right] dS_{\bf w}\\
& =\frac{(-1)^{\frac{M-1}{2}}}{2 (2\pi)^{M-1}} \int_{\Sa^{m-1,2n}} \del^{(M-1)}(\langle{\bf x},{\bf w}\rangle) \, dS_{\bf w},
\end{align*}
which proves the theorem for this case.

\paragraph{Case $ii)$ $M=-2k$, $m\neq 0$.} In this case, {combining} Theorem \ref{PWCK} with formula (\ref{BVDelta}) 
{yields}
\begin{multline}\label{DelPWInit}
-\del({\bf x})= \frac{- 1}{ 4^{k} (k!)^2 \;\sigma_{-2k+1}}\; \Del_{\bf w}^k \Big[ \textup{B.V.} \left[\sgn(x_0) (\langle{\bf x},{\bf w}\rangle - x_0 {\bf w})^{2k}\right]\Big]\\
 + \frac{(-1)^k (4\pi^2)^k}{2}  \int_{\Sa^{m-1,2n}}  \textup{B.V.} \Big[\sgn(x_0)\,G_{2k}\left(\langle{\bf x},{\bf w}\rangle - x_0 {\bf w}\right)\Big] \, dS_{\bf w}.
\end{multline}
The first boundary value can be easily computed as $\textup{B.V.} \left[\sgn(x_0) (\langle{\bf x},{\bf w}\rangle - x_0 {\bf w})^{2k}\right]=2\langle{\bf x},{\bf w}\rangle^{2k}$. {In order to} compute the other boundary value, we first recall that 
\begin{align*}
\lim_{b\fd 0^+} G_{2k}(a\pm ib) %&=  \lim_{b\fd 0^+} \left(\frac{(a\pm ib)^{2k}}{(2k)!} \ln(a\pm ib) -a_{2k}(a\pm ib)^{2k}\right) \\
\;=\; 
\begin{cases} 
\frac{a^{2k}}{(2k)!} \ln(a) -a_{2k}a^{2k}, & a>0\\[+.2cm]
\frac{a^{2k}}{(2k)!} \left(\ln(|a|)\pm i\pi\right) -a_{2k}a^{2k}, & a < 0
\end{cases}
\;=\; G_{2k}(|a|) \pm  \frac{a^{2k}}{(2k)!} i \pi H(-a),
\end{align*}
where $H(a)=\begin{cases} 1, & a>0\\ 0, &a\leq 0,\end{cases}$ is the Heaviside distribution. Then \[\lim_{b\fd 0^+} G_{2k}(a+ ib) +G_{2k}(a- ib)=2G_{2k}(|a|),\] 
and therefore, $2G_{2k}(|a|) = \textup{B.V.} \left[ \sgn(b)\, G_{2k}(z)\right]$. %Hence, 
Under the same identifications as before, we {thus} obtain
\[
\textup{B.V.} \left[\sgn(x_0)\,G_{2k}\left(\langle{\bf x},{\bf w}\rangle - x_0 \frac{{\bf w}}{|{\bf w}|}\right)\right] = 2 G_{2k}\left(|\langle{\bf x},{\bf w}\rangle|\right),
\]
which according to (\ref{IntSphMod1}) yields 
\[
\int_{\Sa^{m-1,2n}}  \textup{B.V.} \Big[\sgn(x_0)\,G_{2k}\left(\langle{\bf x},{\bf w}\rangle - x_0 {\bf w}\right)\Big] \, dS_{\bf w} = 2 \int_{\Sa^{m-1,2n}}  G_{2k}\left(|\langle{\bf x},{\bf w}\rangle|\right) \, dS_{\bf w}.
\]
Finally, substituting the above expressions for {the two boundary} values into (\ref{DelPWInit}), we obtain
\[
\del({\bf x})= \frac{1}{ 2^{2k-1} (k!)^2 \;\sigma_{-2k+1}}\, \Del_{\bf w}^k \left[\langle{\bf x},{\bf w}\rangle^{2k}\right] -  (-1)^k (4\pi^2)^k  \int_{\Sa^{m-1,2n}}  G_{2k}\left(|\langle{\bf x},{\bf w}\rangle| \right)\, dS_{\bf w},
\]
which completes the proof in this case.

\paragraph{Case $iii)$ $M=-2n$.} In this case, we have
\[
-\del(\m{x}\p)= \frac{- 1}{ 4^{n} (n!)^2 \;\sigma_{-2n+1}}\; \Del_{\bf w}^n \Big[ \textup{B.V.} \left[\sgn(x_0) (\langle\m{x}\p,\m{w}\p\rangle - x_0 \m{x}\p)^{2n}\right]\Big], \;\;\;\; 
\]
and  $\textup{B.V.} \left[\sgn(x_0) (\langle\m{x}\p,\m{w}\p\rangle - x_0 \m{x}\p)^{2n}\right] = 2 \langle\m{x}\p,\m{w}\p\rangle^{2n}$. Combining these two equalities, we obtain the desired result. $\hfill\square$\\

%\noindent {In appendix \ref{App} we provide an alternative proof of Theorem \ref{ThmDelPW} by direct computation.}

\section{Inversion formulas for the Radon transform}\label{S6}
In this section, we use the previous plane wave decomposition of $\del({\bf x})$ to obtain explicit inversion formulas for the Radon transform (\ref{RadonTrans}).

\begin{teo}\label{InvExpl}
Let $M+1\notin -2\N_0$, {with $M=m-2n$,} and $\phi\in \Sw(\R^m)\otimes \mathfrak{G}_{2n}$. Then
\begin{itemize}
\item[$i)$] If $M {\geq} 1$,
\begin{align*}
\phi({\bf y}) &=   \frac{(-1)^{\frac{M}{2}}}{(2\pi)^M} \int_{-\infty}^\infty \frac{1}{p}\left( \int_{\Sa^{m-1,2n}}   \pa_{p}^{M-1}\, R_{m|2n}[\phi]({\bf w},p+ \langle{\bf y},{\bf w}\rangle)  \, dS_{\bf w}\right) \,dp, & \mbox{ for }& \;\;M \mbox{ even},\\[+.2cm]
\phi({\bf y}) &=   {\frac{(-1)^{\frac{M-1}{2}}}{2(2\pi)^{M-1}}}\int_{\Sa^{m-1,2n}}   {\pa_{p}^{M-1}}\, R_{m|2n}[\phi]({\bf w},p)\bigg|_{p=\langle{\bf y},{\bf w}\rangle}  \, dS_{\bf w} , & \mbox{ for }& \;\;M \mbox{ odd}.
\end{align*}

\item[$ii)$] If $M =-2k$ ($m\neq 0$),
\begin{multline}\label{45}
\phi({\bf y})= \frac{1}{ 2^{2k-1} (k!)^2 \;\sigma_{-2k+1}}\, \int_{-\infty}^\infty p^{2k} \, \Del_{\bf w}^k R_{m|2n}[\phi]({\bf w},p+ \langle{\bf y},{\bf w}\rangle) \, dp\\
 -  (-1)^k (4\pi^2)^k  \int_{-\infty}^\infty  G_{2k}(|p|) \left(\int_{\Sa^{m-1,2n}}   R_{m|2n}[\phi]({\bf w},p+ \langle{\bf y},{\bf w}\rangle)\, dS_{\bf w} \right) dp.
\end{multline}

\item[$iii)$] If $M =-2n$ ($m= 0$),
\begin{align*}
\phi(\m{y}\p)&=\frac{1}{ 2^{2n-1} (n!)^2 \;\sigma_{-2n+1}}\, \Del_{\m{w}\p}^n \int_{B,\m{x}\p}\langle\m{x}\p-\m{y}\p,\m{w}\p\rangle^{2n} \phi(\m{x}\p) .
\end{align*}
\end{itemize}
\end{teo}
{
\begin{remark}
The formula provided in $iii)$ is an inversion formula for the fermionic integral transform 
\[
R^{\dagger} \,[\phi](\m{w}\p,p) = \int_{B,\m{x}\p} \left(\langle\m{x}\p,\m{w}\p\rangle - p\right)^{2n}  \phi(\m{x}\p).
\]
We recall that the Radon transform $R_{m|2n}$ is not well-defined in the purely fermionic case $m=0$, see Definition \ref{DefRadonTrans}. In this case, the plane wave decomposition of the %purely fermionic 
Dirac delta distribution $\del(\m{x}\p)$(Theorem \ref{ThmDelPW}) yields an inversion formula for the transform $R^{\dagger}$ instead, {i.e.\ $\displaystyle\phi(\m{y}\p)= \frac{1}{ 2^{2n-1} (n!)^2 \;\sigma_{-2n+1}}\, \Del_{\m{w}\p}^n \; R^{\dagger} \,[\phi](\m{w}\p,\langle\m{y}\p,\m{w}\p\rangle)$.}
\end{remark}
%Note to self: Can we define a Radon transform in the purely fermionic case? Is there an interpretation for this last formula?
}

\pf
\paragraph{Case $i)$ $M>1$.} If $M$ is even, Theorem \ref{ThmDelPW} $i)$ yields
\begin{equation}\label{DeltoRad1}
\phi({\bf y}) = \int_{\R^{m|2n}_{\bf x}} \del({\bf x}-{\bf y})\,  \phi({\bf x}) =  \frac{(-1)^{\frac{M}{2}}(M-1)!}{(2\pi)^M} \int_{\Sa^{m-1,2n}} \left( \int_{\R^{m|2n}_{\bf x}} \langle{\bf x}-{\bf y},{\bf w}\rangle^{-M} \phi({\bf x})  \right) dS_{\bf w}.
\end{equation}
From Corollary \ref{Cor1PWtoRad} $ii)$ we obtain
\[
\int_{\R^{m|2n}_{\bf x}} \left(\langle{\bf x},{\bf w}\rangle - \langle{\bf y},{\bf w}\rangle\right)^{-M} \phi({\bf x}) = \int_{-\infty}^\infty p^{-M} \, R_{m|2n}[\phi]({\bf w}, p  +\langle{\bf y},{\bf w}\rangle) \, dp,
\]
{while} $p^{-M} = \frac{(-1)^{M-1}}{(M-1)!} \; \pa_p^{M-1}\left[p^{-1}\right]$ in the distributional sense, see {e.g.\ \cite[Ch.1 - \S3.3]{MR0166596}}. Hence,
\begin{align*}
\int_{\R^{m|2n}_{\bf x}} \langle{\bf x}-{\bf y},{\bf w}\rangle^{-M} \phi({\bf x}) &=  \frac{(-1)^{M-1}}{(M-1)!} \int_{-\infty}^\infty \pa_p^{M-1}\left[p^{-1}\right] \, R_{m|2n}[\phi]({\bf w}, p+\langle{\bf y},{\bf w}\rangle) \, dp\\
&=  \frac{1}{(M-1)!} \int_{-\infty}^\infty p^{-1} \; \pa_p^{M-1} R_{m|2n}[\phi]({\bf w}, p+\langle{\bf y},{\bf w}\rangle) \, dp.
\end{align*}
Substituting this last formula into (\ref{DeltoRad1}) yields the desired result.

\noindent On the other hand, if $M$ is odd we obtain
\begin{equation}\label{DeltoRad2}
\phi({\bf y}) = \int_{\R^{m|2n}_{\bf x}} \del({\bf x}-{\bf y})\,  \phi({\bf x}) =  {\frac{(-1)^{\frac{M-1}{2}}}{2 (2\pi)^{M-1}}} \int_{\Sa^{m-1,2n}} \left( \int_{\R^{m|2n}_{\bf x}}\del^{(M-1)}(\langle{\bf x}-{\bf y},{\bf w}\rangle) \, \phi({\bf x})  \right) dS_{\bf w}.
\end{equation}
Again, by Corollary \ref{Cor1PWtoRad} $ii)$, we get
\begin{align*}
\int_{\R^{m|2n}_{\bf x}}\del^{(M-1)} \left(\langle{\bf x},{\bf w}\rangle - \langle{\bf y},{\bf w}\rangle\right) \, \phi({\bf x}) &= \int_{-\infty}^\infty \del^{(M-1)}(p) \; R_{m|2n}[\phi]({\bf w}, p  +\langle{\bf y},{\bf w}\rangle) \, dp \\
&= (-1)^{M-1}\; \pa_p^{M-1} R_{m|2n}[\phi]({\bf w}, p)\Big|_{p=\langle{\bf y},{\bf w}\rangle}.
\end{align*}
Substituting this last expression into (\ref{DeltoRad2}) we obtain the {assertion}.

\paragraph{Case $ii)$ $M=-2k$, $m\neq 0$.} In this case we have from Theorem \ref{ThmDelPW} $ii)$ that
\begin{align}\label{DeltoRad3}
\phi({\bf y}) &= \int_{\R^{m|2n}_{\bf x}} \del({\bf x}-{\bf y})\,  \phi({\bf x})\nonumber \\
 &= \frac{1}{ 2^{2k-1} (k!)^2 \;\sigma_{-2k+1}}\, \Del_{\bf w}^k \int_{\R^{m|2n}_{\bf x}}  \left(\langle{\bf x},{\bf w}\rangle-\langle{\bf y},{\bf w}\rangle\right)^{2k}\phi({\bf x}) \\
 &\phantom{=}  -  (-1)^k (4\pi^2)^k  \int_{\Sa^{m-1,2n}} \left(\int_{\R^{m|2n}_{\bf x}}  G_{2k}\left(|\langle{\bf x},{\bf w}\rangle - \langle{\bf y},{\bf w}\rangle | \right) \phi({\bf x})  \right)\, dS_{\bf w}.\nonumber
\end{align}
Using Corollary \ref{Cor1PWtoRad} $ii)$, we obtain
\begin{align*}
\Del_{\bf w}^k \int_{\R^{m|2n}_{\bf x}}  \left(\langle{\bf x},{\bf w}\rangle-\langle{\bf y},{\bf w}\rangle\right)^{2k}\phi({\bf x}) &= \int_{-\infty}^{+\infty} p^{2k} \, \Del_{\bf w}^k R_{m|2n}[\phi]({\bf w},p+ \langle{\bf y},{\bf w}\rangle) \, dp,
\end{align*}
and 
\begin{align*}
\int_{\R^{m|2n}_{\bf x}}  G_{2k}\left(|\langle{\bf x},{\bf w}\rangle - \langle{\bf y},{\bf w}\rangle | \right) \phi({\bf x}) &= \int_{-\infty}^{+\infty} G_{2k}(|p|) \, R_{m|2n}[\phi]({\bf w},p+ \langle{\bf y},{\bf w}\rangle)\; dp.
\end{align*}
Substituting these two formulas into (\ref{DeltoRad3}), we obtain (\ref{45}).

{
\noindent Note that the first summand in the right-hand side of (\ref{45}) must be independent of the variables $p$ and ${\bf w}$. The independence of $p$ is clear since this term is an integral on the variable $p$. However, the independence of ${\bf w}$ is not directly seen, but it can be verified using some properties of the Radon transform. Indeed, from the shifting property and the action of the super Laplace operator (Proposition \ref{RTPP} $ii)$ and $iv)$), we obtain
\begin{align*}
\int_{-\infty}^\infty p^{2k} \, \Del_{\bf w}^k R_{m|2n}[\phi]({\bf w},p+ \langle{\bf y},{\bf w}\rangle) \, dp &= \int_{-\infty}^\infty p^{2k} \, \Del_{\bf w}^k R_{m|2n}\left[\phi({\bf x}{+}{\bf y})\right]({\bf w},p) \, dp \\
&= \int_{-\infty}^\infty p^{2k} \, \pa_p^{2k} R_{m|2n}\left[|{\bf x}|^{2k}\phi({\bf x}{+}{\bf y})\right]({\bf w},p) \, dp
\end{align*}
Finally, using integration by parts and Corollary \ref{Cor1PWtoRad} $i)$ we obtain
\begin{align*}
\int_{-\infty}^\infty p^{2k} \, \Del_{\bf w}^k R_{m|2n}[\phi]({\bf w},p+ \langle{\bf y},{\bf w}\rangle) \, dp &= (2k)! \int_{-\infty}^\infty  R_{m|2n}\left[|{\bf x}|^{2k}\phi({\bf x}{+}{\bf y})\right]({\bf w},p) \, dp \\
&= \int_{\R_{\bf x}^{m|2n}} |{\bf x}|^{2k}\phi({\bf x}{+}{\bf y}),
\end{align*}
which is clearly independent of ${\bf w}$.}

\paragraph{Case $iii)$ $M=-2k$, $m\neq 0$.} The result in this case follows directly from Theorem \ref{ThmDelPW} $iii)$. $\hfill\square$

\section{Inversion formulas unified}\label{S7}
In the purely bosonic case, Theorem \ref{InvExpl} $i)$ yields the inversion formulas for the classical Radon transform in $M=m$ dimensions. These formulas can be rewritten in an unified way, regardless of the parity of $m$, as follows (see e.g.\ \cite{MR754767, MR573446}),
\begin{equation}\label{UnInvRa}
(-\Del_{\m{x}})^{\frac{m-1}{2}} \, R^*_{m|0} \, R_{m|0}[\phi] = 2^{m}\pi^{m-1}  \phi.
\end{equation}
{Here, $R^*_{m|0}$ denotes the so-called} dual Radon transform, which to a continuous function $\psi(\m{w},p)$ on the space $\mathscr{P}^m$ of all hyperplanes in $\R^m$ associates the function
\[
R^*_{m|0} [\psi] (\m{x}) = \int_{\Sa^{m-1}} \psi(\m{w},\langle\m{x},\m{w}\rangle)\, dS_{\m{w}}.
\]
For $m$ even, the fractional power of the bosonic Laplacian $(-\Del_{\m{x}})^{\frac{m-1}{2}}$ is defined by means of the Fourier multiplier
\begin{equation}\label{DefLapFou}
\mathcal{F}\left[(-\Del_{\m{x}})^s \phi \right] = |\m{\xi}|^{2s}\mathcal{F}[\phi],{ \;\;\;\;\;\;  \phi\in\Sw(\R^m), \;\;\;\;\;\; 2s>-m}.
\end{equation}
%---------------------
%The operator $R^*_{m|0}$ is known as the dual Radon transform, which to a continuous function $\psi(\m{w},p)$ on the space $\mathscr{P}^m$ of all hyperplanes in $\R^m$ associates the function
%\[
%R^*_{m|0} [\psi] (\m{x}) = \int_{\Sa^{m-1}} \psi(\m{w},\langle\m{x},\m{w}\rangle)\, dS_{\m{w}}.
%\]

In this section, we prove a similar unified inversion formula for the super Radon transform for {all} values of the superdimension $M\in\Z$.  To that end, we shall first study in detail the fractional super Laplace operator and construct a fundamental solution for it.

\subsection{Fundamental solution of the fractional super Laplace operator.}
First, let us recall that the definition (\ref{DefLapFou}) of $(-\Del_{\m{x}})^s$ as a Fourier multiplier can be rewritten as (see \cite[Ch.1 - \S2.8]{MR754767})
\begin{equation*}%\label{LapAsRieszPot}
(-\Del_{\m{x}})^s[\phi] = I^{-2s}[\phi],\;\;\;\;\;\;  \phi\in\Sw(\R^m),
\end{equation*}
where $I^\gam$ is the Riesz potential,
\begin{align*}%\label{RiescPot}
I^\gam[\phi](\m{x}) %= H_m(\gam)^{-1} \left(r^{\gam-m} * \phi  \right)(\m{x}) 
&:= \frac{1}{H_m(\gam)} \int_{\R^m} |\m{x}-\m{y}|^{\gam-m}{\phi(\m{y})\, dV_{\m{y}}}, && {\mbox{with } \;\;\; H_m(\gam)= 2^\gam \pi^{\frac{m}{2}} \frac{\Gam\left(\frac{\gam}{2}\right)}{\Gam\left(\frac{m-\gam}{2}\right)}.}
\end{align*}
{Equivalently, we can write $I^\gam[\phi](\m{x}) = \left(K^m_\gam *  \phi  \right)(\m{x})$ where $K^m_\gam(\m{x})=H_m(\gam)^{-1} |\m{x}|^{\gam-m}$ is the corresponding Riesz potential. Note that the poles} of $|\m{x}|^{\gam-m}$ are cancelled by the poles of $\Gam\left(\frac{\gam}{2}\right)$, so that $I^\gam[\phi](\m{x})$ extends to a holomorphic function for all $\gam\in\C$ such that $\gam-m\notin 2\N_0$. Thus the power of the Laplacian  $(-\Del_{\m{x}})^s$ is well-defined if $-2s-m\notin2\N_0$.
%with $H_m(\gam)= 2^\gam \pi^{\frac{m}{2}} \frac{\Gam\left(\frac{\gam}{2}\right)}{\Gam\left(\frac{m-\gam}{2}\right)}$.
%In formula (\ref{RiescPot}), the poles of $r^{\gam-m}$ are cancelled by the poles of $\Gam\left(\frac{\gam}{2}\right)$, so that $I^\gam[\phi](\m{x})$ extends to a holomorphic function for all $\gam\in\C$ such that $\gam-m\notin 2\N_0$. Thus the the power of the Laplacian  $(-\Del_{\m{x}})^s$ is well-defined if $-2s-m\notin2\N_0$.

{An} important property of the Riesz kernels %$K^m_\gam(\m{x})=H_m(\gam)^{-1} |\m{x}|^{\gam-m}$ 
is that (see \cite[Ch.1 - \S1]{MR0350027})
\[
K^m_\al*K^m_\be=K^m_{\al+\be}, \;\;\;\;\;\;\;\mbox{ for }  \;\;\;\;\;\;\; \textup{Re}(\al+\be)<m,  \;\;\;\;\;\; %\al,\be\in \Sigma_m.
\al-m,\;\be-m\notin2\N_0.
\]
Moreover, from (\ref{ValSingP}) we have that 
\[
K^m_0(\m{x}) = H_m(\gam)^{-1} |\m{x}|^{\gam-m} \bigg|_{\gam=0} = \del(\m{x}).
\]
{In particular, these two properties imply that $K^m_\gam*K^m_{-\gam}=\del(\m{x})$ if $\pm\gam-m\notin 2\N_0$. It thus follows that a fundamental solution of $(-\Del_{\m{x}})^s$ is}
%From the last two properties it is seen that a fundamental solution of $(-\Del_{\m{x}})^s$ is given by 
\[
K_{2s}^m(\m{x}) = \frac{1}{H_m(2s)}\, |\m{x}|^{2s-m}, \;\;\;\;\;\;\; \;\;\; \pm2s-m\notin 2\N_0.
\]
We also recall that the action of the bosonic Laplacian on the generalized function ${{|\m{x}|}}^\lan$ is given by (see e.g.\ \cite[Ch.1 - \S3.9]{MR0166596}),
\[
\Del_{\m{x}}\left[\frac{{{|\m{x}|}}^{\lan+2}}{\Gam\left(\frac{\lan+m}{2}+1\right)}\right] = 2(\lan+2) \frac{{{|\m{x}|}}^\lan}{\Gam\left(\frac{\lan+m}{2}\right)},
\]
which is valid for the entire $\lan$-plane. It is hence easily seen that %$(-\Del_{\bf x}) K^m_\gam= K^m_{\gam-2}$ for $\gam-m\notin2\N_0$.
\begin{equation}\label{LapRiesz}
{(-\Del_{\m{x}})} K^m_\gam= K^m_{\gam-2}.
\end{equation}

% for all $\gam\in\Sigma_m$.

{If $m\neq 0$, we may define real powers of the super Laplace operator %$(-\Del_{\bf x})^s$ ($s\in\R$) 
by means of the following Taylor expansion (see (\ref{GenPow})),}
%We may now define real powers of the super Laplace operator. If $m\neq 0$, the operator $(-\Del_{\bf x})^s$ ($s\in\R$) is defined by means of the following Taylor expansion (see (\ref{GenPow})),
\begin{equation}\label{FracLapSupS}
(-\Del_{\bf x})^s = \sum_{j=0}^{n} \frac{\Del_{\m{x}\p}^j}{j!} \frac{\Gam(-s+j)}{\Gam(-s)} (-\Del_{\m{x}})^{s-j},  \;\;\;\;\;\;\; \;\;\; s\in\R, \;\; -2s-M\notin2\N_0.%-2s\in \Sigma_M.
\end{equation}
We now construct a fundamental solution for this operator. We will follow a procedure similar to the one used in \cite{MR2386499}, where fundamental solutions of the {natural powers of the} super Laplace operator %and its natural powers 
were computed.
\begin{teo}\label{FundSolLap}
Let $s\in\R$ be such that {$\pm2s -M\notin2\N_0$}. Then a fundamental solution of $(-\Del_{\bf x})^s$ is
\[
K_{2s}^M({\bf x}) = \frac{1}{H_M(2s)} |{\bf x}|^{2s-M} = \frac{1}{4^s\pi^{M/2}\Gam(s)} \sum_{{{j}}=0}^n \frac{\m{x}\p^{\, 2n-2{{j}}}}{(n-{{j}})!} \Gam\left(\frac{m}{2}-s-{{j}}\right) |\m{x}|^{2s+2{{j}}-m}.
 \]
\end{teo}
%\begin{remark}
%Observe that the conditions $2s -M\notin2\N_0$ and $-2s -M\notin2\N_0$ ensure that $(-\Del_{\bf x})^s$  and $K_{2s}^M({\bf x})$ are well-defined. 
%\end{remark}
\pf
{Note that the conditions $-2s -M\notin2\N_0$ and $2s -M\notin2\N_0$ ensure that {both} $(-\Del_{\bf x})^s$  and $K_{2s}^M({\bf x})$ are well-defined.} We begin by %proposing the following form for a 
{writing a fundamental solution of $(-\Del_{\bf x})^s$ in the following form}
\begin{equation}\label{PFS}
\rho = \sum_{{{j}}=0}^n a_{{j}} \, (-\Del_{\m{x}})^{n-{{j}}}[\phi] \, \m{x}\p^{\, 2n-2{{j}}},
\end{equation}
where $a_{{j}}\in\R$ and $\phi\in\Sw'(\R^m)$ are {(still)} to be determined. Combining (\ref{FracLapSupS}) and (\ref{PFS}) we obtain,
\[
(-\Del_{\bf x})^s [\rho]=\sum_{{{\el}},{{j}}=0}^n \frac{a_{{j}}}{{{\el}}!} \, \frac{\Gam(-s+{{\el}})}{\Gam(-s)}  \, (-\Del_{\m{x}})^{n-{{j}}+s-{{\el}}}[\phi] \, \Del_{\m{x}\p}^{{\el}}\left[\m{x}\p^{\, 2n-2{{j}}}\right].
\]
The action $\Del_{\m{x}\p}^{{\el}}\left[\m{x}\p^{\, 2n-2{{j}}}\right]$ {can be computed directly from (\ref{LapPowX})}
%has been explicitly computed in the literature (see e.g.\ \cite{MR2344451, MR2386499}) 
and it is given by
\[
\Del_{\m{x}\p}^{{\el}}\left[\m{x}\p^{\, 2n-2{{j}}}\right] = 4^{{\el}} \frac{(n-{{j}})!}{(n-{{j}}-{{\el}})!} \frac{({{j}}+{{\el}})!}{{{j}}!} \m{x}\p^{\, 2n-2{{j}}-2{{\el}}}, \;\;\;\;\;\;\; \mbox{ for } \;\;\;\;\; {{\el}}\leq n-{{j}}.
\]
Obviously, $\Del_{\m{x}\p}^{{\el}}\left[\m{x}\p^{\, 2n-2{{j}}}\right]=0$ if ${{\el}}> n-{{j}}$. Thus, we obtain
\begin{align*}
(-\Del_{\bf x})^s [\rho] &= \sum_{{{j}}=0}^n \sum_{{{\el}}=0}^{n-{{j}}} 4^{{\el}} \frac{a_{{j}}}{{{\el}}!} \,\frac{(n-{{j}})!}{(n-{{j}}-{{\el}})!} \frac{({{j}}+{{\el}})!}{{{j}}!}  \frac{\Gam(-s+{{\el}})}{\Gam(-s)}  \, (-\Del_{\m{x}})^{n-{{j}}+s-{{\el}}}[\phi] \,  \m{x}\p^{\, 2n-2{{j}}-2{{\el}}} \\
&= \sum_{{{j}}=0}^n \sum_{\el={{j}}}^{n} 4^{\el-{{j}}} \frac{a_{{j}}}{(\el-{{j}})!} \,\frac{(n-{{j}})!}{(n-\el)!} \frac{\el!}{{{j}}!}  \frac{\Gam(-s+\el-{{j}})}{\Gam(-s)}  \, (-\Del_{\m{x}})^{n+s-\el}[\phi] \,  \m{x}\p^{\, 2n-2\el},
\end{align*}
where we have replaced the index ${{\el}}$ by $\el-{{j}}$ in the second equality. Changing the order of summation we {get}, 
\begin{align*}
(-\Del_{\bf x})^s [\rho] &= \sum_{\el=0}^{n} \frac{\m{x}\p^{\, 2n-2\el}}{(n-\el)!} \, (-\Del_{\m{x}})^{n+s-\el}[\phi] \left(\sum_{{{j}}=0}^\el 4^{\el-{{j}}} \,a_{{j}} \,  (n-{{j}})! \binom{\el}{{{j}}}\frac{\Gam(-s+\el-{{j}})}{\Gam(-s)}\right)\\
&=\sum_{\el=0}^{n} \frac{\m{x}\p^{\, 2n-2\el}}{(n-\el)!} \, (-\Del_{\m{x}})^{n+s-\el}[\phi] \left(\sum_{j=0}^\el 4^{j} \,a_{\el-j} \,  (n-\el+j)! \binom{\el}{j}\frac{\Gam(-s+j)}{\Gam(-s)}\right),
\end{align*}
where we have now replaced the index ${{j}}$ by $\el-j$ in the last equality.

\noindent From our assumption $(-\Del_{\bf x})^s [\rho]=\del({\bf x})=\del(\underline{x}) \dfrac{\pi^n}{n!} \underline{x}\p^{\, 2n}$, we obtain for $\el=0$ that
\begin{equation}\label{53}
a_0  (-\Del_{\m{x}})^{n+s}[\phi] = \frac{\pi^n}{n!} \del(\underline{x}),
\end{equation}
and for $\el=1, \ldots, n$ that
\begin{equation}\label{54}
 \sum_{j=0}^\el 4^{j} \,a_{\el-j} \,  (n-\el+j)! \binom{\el}{j}\frac{\Gam(-s+j)}{\Gam(-s)} =0.
\end{equation}
From (\ref{53}) we immediately have that $a_0= \frac{\pi^n}{n!}$ and $\phi(\m{x})=K_{2s+2n}^m(\m{x})$. Indeed, %$\pm2s\in \Sigma_{m\pm2n}$ 
$\pm2s-M\notin2\N_0$ directly implies that %$\pm2(s+n)\in\Sigma_m$
$\pm2(s+n)-m\notin2\N_0$, and therefore, $K_{2s+2n}^m(\m{x})$ is a fundamental solution of $(-\Del_{\m{x}})^{n+s}$. On the other hand, if we use the substitution 
\[
a_{{j}}=\frac{\pi^n 4^{{j}}}{(n-{{j}})!}\frac{b_{{j}}}{\Gam(s)},
\]
then (\ref{54}) simplifies to
\[
 \sum_{j=0}^\el b_{\el-j} \,   \binom{\el}{j}\frac{\Gam(-s+j)}{\Gam(-s)\Gam(s)} =0,
\]
which has a solution given by $b_{{j}}=\Gam(s+{{j}})$, ${{j}}=0,\ldots, n$ (see the subsequent Lemma \ref{TechLem}). Hence we conclude that a solution for the system (\ref{53})-(\ref{54}) is 
\[
a_{{j}}=\pi^n \frac{ 4^{{j}}}{(n-{{j}})!}\frac{\Gam(s+{{j}})}{\Gam(s)}.
\]
Substituting this into (\ref{PFS}), together with $\phi=K_{2s+2n}^m$, we obtain {from (\ref{LapRiesz}) that}%the following fundamental solution of $(-\Del_{\bf x})^{s}$,
\begin{align*}
\rho &= \frac{\pi^n}{\Gam(s)} \sum_{{{j}}=0}^n  \frac{ 4^{{j}}\Gam(s+{{j}})}{(n-{{j}})!} \, (-\Del_{\m{x}})^{n-{{j}}}\left[K_{2s+2n}^m\right] \, \m{x}\p^{\, 2n-2{{j}}} \\
&=\frac{\pi^n}{\Gam(s)} \sum_{{{j}}=0}^n  \frac{ 4^{{j}}\Gam(s+{{j}})}{(n-{{j}})!} \, K_{2s+2{{j}}}^m \; \m{x}\p^{\, 2n-2{{j}}} \\
&=\frac{\pi^n}{\pi^{\frac{m}{2}} 4^s \Gam(s)} \sum_{{{j}}=0}^n \frac{ \Gam\left(\frac{m}{2}-{s}-{{j}}\right)}{(n-{{j}})!} \, |\m{x}|^{2s+2{{j}}-m} \, \m{x}\p^{\, 2n-2{{j}}} \\
&=\frac{1}{\pi^{\frac{M}{2}} 4^s \Gam(s)} \sum_{{{j}}=0}^n \frac{\m{x}\p^{\, 2{{j}}}}{{{j}}!} \Gam\left(\frac{M}{2}-s+{{j}}\right)\, |\m{x}|^{2s-M-2{{j}}},
\end{align*}
where we have replaced the index ${{j}}$ by $n-{{j}}$ in the last equality. Finally, using the definition of the generalized superfunction $|{\bf x}|^{2s-M}$ (see (\ref{PowSup})), we obtain 
\[
\rho = \frac{\Gam\left(\frac{M}{2}-s\right)}{\pi^{\frac{M}{2}} 4^s \Gam(s)} |{\bf x}|^{2s-M} ={K_{2s}^M({\bf x})}, %= \frac{1}{H_M(2s)} |{\bf x}|^{2s-M}
\]
%\\
%&= \frac{\Gam\left(\frac{M}{2}-s\right)}{\pi^{\frac{M}{2}} 4^s \Gam(s)} |{\bf x}|^{2s-M},
which proves the result. $\hfill\square$\\

We still need the following technical lemma regarding the solution of (\ref{54}). {We provide the proof due to the lack of reference.}
\begin{lem}\label{TechLem}
The following identity holds for all $s\in\C$ {and $\el\in\N$}
\[
 \sum_{j=0}^\el  \binom{\el}{j} \frac{\Gam(s+\el-j)}{\Gam(s)} \frac{\Gam(-s+j)}{\Gam(-s)} =0.
\]
\end{lem}
\pf
Recall that $(-1)^j \frac{\Gam\left(s+1\right)}{\Gam\left(s-j+1\right)} =  \frac{\Gam\left(-s+j\right)}{\Gam\left(-s\right)}$ {for all} $s\in\C$ and $j\in\N_0$. Then %the above sum equals to
\begin{align*}
 \sum_{j=0}^\el  \binom{\el}{j} \frac{\Gam(s+\el-j)}{\Gam(s)} \frac{\Gam(-s+j)}{\Gam(-s)} &=  \sum_{j=0}^\el  (-1)^j \binom{\el}{j} \frac{\Gam(s+\el-j)}{\Gam(s)} \frac{\Gam(s+1)}{\Gam(s-j+1)}\\
 &= s  \sum_{j=0}^\el  (-1)^j \binom{\el}{j} \frac{\Gam(s+\el-j)}{\Gam(s-j+1)}.
\end{align*}
Hence, it suffices to prove that the {following} polynomial {in the complex variable $s$ is identically zero,}
\begin{align*}
P(s) = \sum_{j=0}^\el  (-1)^j \binom{\el}{j} \frac{\Gam(s+\el-j)}{\Gam(s-j+1)} =  \sum_{j=0}^\el  (-1)^j \binom{\el}{j} (s-j+1)_{{\el-1}},%\ldots (s-j+\el)
\end{align*}
{where $(q)_\el:=\begin{cases} 1, & \el=0,\\ q(q+1)\cdots (q+\el-1), & \el>0, \end{cases}$ is the  rising Pochhammer symbol.}
Direct calculations show that we can rewrite $P(s)$ as the hypergeometric function
\begin{equation}\label{P(s)}
P(s) = \frac{\Gam(\el+s)}{\Gam({s}+1)}\; {}_2F_1(-\el,-s;-\el-s+1,1).
\end{equation}
From the Chu-Vandermonde identity for hypergeometric functions (see \cite[Corollary~2.2.3]{MR1688958}) it is known that
\[
{}_2F_1(-\el,b;c,1) = \frac{(c-b)_\el}{(c)_\el}, %\;\;\;\;\;\;\; -\el+b<c,
\;\;\;\;\;\;\;\;{ \mbox{ if }\;\;\;\;\; \textup{Re}(c)>\textup{Re}(-\el+b).}\]
%where $(q)_\el=\begin{cases} 1, & j=0,\\ q(q+1)\cdots (q+j-1), & j>0, \end{cases}$ 
%is the  rising Pochhammer symbol. 
Applying this result and the fact that $(-\el+1)_\el=(-\el+1)(-\el+2)\cdots(-\el+1+\el-1)=0$, we obtain from (\ref{P(s)}) that 
%Thus, using (\ref{P(s)}) and the fact that  $(-\el+1)_\el=(-\el+1)(-\el+2)\cdots(-\el+1+\el-1)=0$, we obtain
\[
P(s)=\frac{\Gam(\el+s)}{\Gam(s+1)} \frac{(-\el+1)_\el}{(-\el-s+1)_\el}=  0,
\]
which completes the proof. $\hfill\square$

\subsection{Inversion formula}
Given a superfunction $\psi({\bf w}, p)$ {of the supervector variable ${\bf w}$ and the commuting variable $p$,}
%with $p=\langle {\bf w}, {\bf x}\rangle$
 we define the dual of the super Radon transform by
\[
R^*_{m|2n} [\psi] ({\bf x}) = \int_{\Sa^{m-1,2n}} \psi({\bf w},\langle{\bf x},{\bf w}\rangle)\, dS_{{\bf w}}.
\]
We shall now establish the following unified inversion formulas for the super Radon transform, extending in this way formula (\ref{UnInvRa}) to the superspace setting.
\begin{teo}\label{InvExplRed}
Let %$M-1\notin -2\N_0$, $m>1$ 
{$m\neq 0$, $M=m-2n$,} and $\phi\in \Sw(\R^m)\otimes \mathfrak{G}_{2n}$. Then
\[
(-\Del_{\bf x})^{\frac{M-1}{2}} \, R^*_{m|2n} \, R_{m|2n} [\phi] ({\bf x}) = 2^M \pi^{{M-1}} \phi({\bf x}).
\]
\end{teo}
To begin with, observe that 
\begin{align}\label{adaf}
R^*_{m|2n} \, R_{m|2n} [\phi] ({\bf x}) &= \int_{\Sa^{m-1,2n}} R_{m|2n} [\phi] ({\bf w},\langle{\bf x},{\bf w}\rangle)\, dS_{{\bf w}} \nonumber\\
&= \int_{\Sa^{m-1,2n}}  \left(\int_{\R^{m|2n}_{{\bf y}}} \del\left(\langle{\bf y}- {\bf x},{\bf w}\rangle\right) \phi({\bf y}) \right) dS_{{\bf w}} \nonumber\\
&=\int_{\R^{m|2n}_{{\bf y}}}  \left( \int_{\Sa^{m-1,2n}} \del\left(\langle{\bf y}- {\bf x},{\bf w}\rangle\right) dS_{{\bf w}}   \right) \phi({\bf y}).
\end{align}
{Before proving} Theorem \ref{InvExplRed}, we {will} show first that the innermost integral in (\ref{adaf}) is {(up to a constant)} the plane wave decomposition of the super Riesz kernel {$K_{M-1}^M({\bf y}-{\bf x}) =H_M(M-1)^{-1} |{\bf y}-{\bf x}|^{-1}$}.
\begin{teo}\label{PWRK}
Let {$m\neq 0$ and $M=m-2n$}. Then
\[
\int_{\Sa^{m-1,2n}} \del\left(\langle{\bf x},{\bf w}\rangle\right) dS_{{\bf w}} = \sigma_{M-1} |{\bf x}|^{-1}, 
\]
{where we recall that $\sigma_{M-1} |{\bf x}|^{-1} =2\pi^{\frac{M-1}{2}} \dfrac{|{\bf x}|^\lan}{\Gam\left(\frac{M+\lan}{2}\right)}\Bigg|_{\lan=-1}$ is defined for all $M\in \Z$, see Theorem \ref{NorSup}.}
\end{teo}
\pf We begin by recalling that 
\begin{equation}\label{DelIntoL1}
\int_{\Sa^{m-1,2n}} \del\left(\langle{\bf x},{\bf w}\rangle\right) dS_{{\bf w}}=\sum_{j=0}^{2n}\frac{1}{j!}\left(\int_{\Sa^{m-1,2n}} \langle\m{x}\p,\m{w}\p\rangle^{j} \, \del^{(j)}(\langle\m{x},\m{w}\rangle)\, dS_{{\bf w}} \right)= \sum_{j=0}^{2n}\frac{1}{j!}\;  L_{j}[\del^{(j)}]({\bf x}),
\end{equation}
where we have introduced the notation 
\[
L_{j}[\del^{(j)}]({\bf x}) = \int_{\Sa^{m-1,2n}} \langle\m{x}\p,\m{w}\p\rangle^{j} \; \del^{(j)}(\langle\m{x},\m{w}\rangle)\, dS_{{\bf w}}, \;\;\;\;\; j=0,\ldots, {2n}.
\]
{The above integrals can be computed as follows (see the subsequent Lemma \ref{L3} in the Appendix section for a proof)
%Lemma \ref{L3} in the Appendix section shows that $L_{2j+1}[\del^{(2j+1)}]({\bf x})=0$ while 
%\[
%L_{2j}[\del^{(2j)}]({\bf x}) = 2(-1)^j \frac{\Gam\left(j+\frac{1}{2}\right)}{\pi^{n+\frac{1}{2}}} \; \m{x}\p^{\, 2j}\, \int_{\R^m} \del^{(n-j)}\left(1-|\m{w}|^2\right) \del^{(2j)}(\langle\m{x},\m{w}\rangle)\, dV_{{\m{w}}}.
%\]
\begin{align*}
&L_{2j+1}[\del^{(2j+1)}]({\bf x})=0, & j&=0, \ldots, n-1,\\
&L_{2j}[\del^{(2j)}]({\bf x}) = 2(-1)^j \frac{\Gam\left(j+\frac{1}{2}\right)}{\pi^{n+\frac{1}{2}}} \; \m{x}\p^{\, 2j}\, \int_{\R^m} \del^{(n-j)}\left(1-|\m{w}|^2\right) \del^{(2j)}(\langle\m{x},\m{w}\rangle)\, dV_{{\m{w}}} & j&=0, \ldots, n.
\end{align*}}
Then (\ref{DelIntoL1}) can be rewritten as 
\begin{equation}\label{DelIntoL}
\int_{\Sa^{m-1,2n}} \del\left(\langle{\bf x},{\bf w}\rangle\right) dS_{{\bf w}}= \frac{2}{\pi^{n+\frac{1}{2}}}\sum_{j=0}^{n}(-1)^j \frac{\Gam\left(j+\frac{1}{2}\right)}{(2j)!} \; \m{x}\p^{\, 2j}\,\int_{\R^m} \del^{(n-j)}\left(1-|\m{w}|^2\right) \del^{(2j)}(\langle\m{x},\m{w}\rangle)\, dV_{{\m{w}}}.
\end{equation}
Let us now compute the distribution $I_j:=\int_{\R^m} \del^{(n-j)}\left(1-|\m{w}|^2\right) \del^{(2j)}(\langle\m{x},\m{w}\rangle)\, dV_{{\m{w}}}$. Using spherical {coordinates  $\m{w}=t\m{\xi}$ with $t=|\m{w}|$ and $\m{\xi}\in\Sa^{m-1}$,} we obtain
\begin{align*}
I_j &= \int_0^\infty \int_{\Sa^{m-1}} \del^{(n-j)}(1-t^2) \; \del^{(2j)}(t\langle\m{x},\m{\xi}\rangle)\, t^{m-1}\; dS_{\m{\xi}} \; dt \\
&= \left(\int_0^\infty \del^{(n-j)}(1-t^2) t^{m-2j-2} \, dt\right) \left(\int_{\Sa^{m-1}} \del^{(2j)}(\langle\m{x},\m{\xi}\rangle)\; dS_{\m{\xi}} \right).
\end{align*}
To compute the first integral in the above formula, we take {$t=u^{1/2}$} and obtain
%The first integral in the above formula can be computed as follows by taking $u=t^{1/2}$, 
\begin{align*}
\int_0^\infty \del^{(n-j)}(1-t^2) t^{m-2j-2} \, dt &= \frac{1}{2} \int_0^\infty \del^{(n-j)}(1-u) \,u^{\frac{m-3}{2}-j} \, du\\
&=  \frac{1}{2}   \frac{d^{n-j}}{du^{n-j}} \left[u^{\frac{m-3}{2}-j}\right] \bigg|_{u=1} \\
&=  \frac{1}{2} \left(\frac{m-1}{2}-j-1\right) \left(\frac{m-1}{2}-j-2\right) \cdots \left(\frac{m-1}{2}-n\right) \\
&= \frac{1}{2} \frac{\Gam\left(\frac{m-1}{2}-j\right)}{\Gam\left(\frac{m-1}{2}-n\right)}.
\end{align*}
Hence, 
\[
I_j = \frac{1}{2} \frac{\Gam\left(\frac{m-1}{2}-j\right)}{\Gam\left(\frac{M-1}{2}\right)} \int_{\Sa^{m-1}} \del^{(2j)}(\langle\m{x},\m{\xi}\rangle)\; dS_{\m{\xi}}.
\]
An explicit formula for the  plane wave integral above can be obtained from {combining (\ref{PinValDim1}) with (\ref{PWRLan}). Indeed, substituting (\ref{PinValDim1}) into (\ref{PWRLan}) yields}
%combining (\ref{PWRLan}) with (\ref{ValSingP}). Indeed, for $\lan=-2j-1$ and $m=1$, formula (\ref{ValSingP}) yields
%\[
%\frac{|t|^\lan}{ \Gam\left(\frac{\lan+1}{2}\right)} \bigg|_{\lan=-2j-1} = (-1)^j \frac{j!}{(2j)!} \,\del^{(2j)}(t).
% \]
%Thus, formula (\ref{PWRLan}) reads for $\lan=-2j-1$ as
%Taking $t=\langle\m{x},\m{\xi}\rangle$, we can rewrite (\ref{PWRLan}) for  $\lan=-2j-1$ as
\begin{equation}\label{PWDelOS}
\frac{(-1)^j j!}{\pi^{\frac{m-1}{2}} (2j)!}  \int_{\Sa^{m-1}}\del^{(2j)}(\langle\m{x},\m{\xi}\rangle)\; dS_{\m{\xi}}= \frac{2{{|\m{x}|}}^\lan }{\Gam\left(\frac{\lan+m}{2}\right)} \bigg|_{\lan=-2j-1}.
\end{equation}
{Using formula (\ref{ValSingP}) we obtain
\[
\frac{2{{|\m{x}|}}^\lan }{\Gam\left(\frac{\lan+m}{2}\right)} \bigg|_{\lan=-2j-1} =
\begin{cases} \dfrac{2 {{|\m{x}|}}^{-2j-1}}{\Gam\left(\frac{m-1}{2}-j\right)}, & m-1-2j\notin -2\N_0,\\[+.4cm]
\dfrac{(-1)^{\frac{1-m}{2}+j}  \pi^{\frac{m}{2}}}{2^{-m+2j}\, \Gam\left(j+\frac{1}{2}\right)}\, \Del_{\m{x}}^{\frac{1-m}{2}+j} \del(\m{x}), & m-1-2j\in -2\N_0, \end{cases}
\]
which gives the following when substituted in (\ref{PWDelOS}) 
\begin{equation}\label{IntTwoFold}
\int_{\Sa^{m-1}}\del^{(2j)}(\langle\m{x},\m{\xi}\rangle)\; dS_{\m{\xi}}=
\begin{cases} 
2 (-1)^{j} \pi^{\frac{m-1}{2}} \dfrac{(2j)!}{j!}\dfrac{{{|\m{x}|}}^{-2j-1}}{\Gam\left(\frac{m-1}{2}-j\right)}, & m-1-2j\notin -2\N_0,\\[+.4cm]
\dfrac{(-1)^{\frac{1-m}{2}} \pi^{m-\frac{1}{2}}}{2^{-m+2j}\, \Gam\left(j+\frac{1}{2}\right)} \dfrac{(2j)!}{j!} \, \Del_{\m{x}}^{\frac{1-m}{2}+j} \del(\m{x}) & m-1-2j\in -2\N_0, \end{cases}
\end{equation}
Hence, if $M-1\notin -2\N_0$, we have that $m-1-2j\notin -2\N_0$ for all $j=0,\ldots, n$, and therefore 
\[
I_j = (-1)^j \frac{\pi^{\frac{m-1}{2}}  }{\Gam\left(\frac{M-1}{2}\right) }  \frac{(2j)!}{j!} \, {{|\m{x}|}}^{-2j-1}.
\]
Substituting this formula into (\ref{DelIntoL}) and in virtue of (\ref{PowSup}), we obtain 
\[
\int_{\Sa^{m-1,2n}} \del\left(\langle{\bf x},{\bf w}\rangle\right) dS_{{\bf w}} =   \sigma_{M-1} \sum_{j=0}^{n}  \frac{\m{x}\p^{\, 2j} }{j!}\ \frac{ \Gam\left(j+\frac{1}{2}\right)}{\Gam(\frac{1}{2})} \, {{|\m{x}|}}^{-1-2j} =  \sigma_{M-1} |{\bf x}|^{-1},
\]
which proves the theorem for this case.
}

{
\noindent On the other hand, if $M-1\in-2\N_0$, i.e.\ $m=2n-2k+1$ for some $k\in \N_0$ ($k\leq n$), we have that
\[
\dfrac{\Gam\left(\frac{m-1}{2}-j\right)}{\Gam\left(\frac{m-1}{2}-n\right)} =
\begin{cases} 
(-1)^{n-j} \dfrac{k!}{(k+j-n)!}, & k+j\geq n,\\[+.4cm]
0, & \mbox{otherwise}. 
\end{cases}
\]
Hence $I_j$ vanishes if $k+j<n$, while 
\[
I_j = \frac{(-1)^{n-j}}{2} \dfrac{k!}{(k+j-n)!}  \int_{\Sa^{m-1}}\del^{(2j)}(\langle\m{x},\m{\xi}\rangle)\; dS_{\m{\xi}}, \;\;\;\; \mbox{ if } k+j\geq n.
\]
The condition $k+j\geq n$ implies that $m-1-2j=2(n-k-j)\in-2\N_0$. Then, from (\ref{IntTwoFold}) we obtain
%\[
%\int_{\Sa^{m-1}}\del^{(2j)}(\langle\m{x},\m{\xi}\rangle)\; dS_{\m{\xi}} =  \dfrac{2 (-1)^{k-n} \pi^{m-\frac{1}{2}}}{2^{2k+2j-2n}\, \Gam\left(j+\frac{1}{2}\right)} \dfrac{(2j)!}{j!} \, \Del_{\m{x}}^{j+k-n} \del(\m{x})
%\]
%Hence,
\[
I_j = \dfrac{(-1)^{k-j}}{2^{2k+2j-2n}}  \dfrac{k!}{(k+j-n)!}  \dfrac{(2j)!}{j!} \dfrac{\pi^{m-\frac{1}{2}}}{\Gam\left(j+\frac{1}{2}\right)} \, \Del_{\m{x}}^{j+k-n} \del(\m{x}). 
\]
Substituting this into (\ref{DelIntoL}) yields
\begin{align*}
\int_{\Sa^{m-1,2n}} \del\left(\langle{\bf x},{\bf w}\rangle\right) dS_{{\bf w}} &= \frac{2(-1)^k}{\pi^{n+\frac{1}{2}}} \frac{k!\,  \pi^{m-\frac{1}{2}}}{2^{2k}}\sum_{j=n-k}^{n} \frac{\m{x}\p^{\,2j}}{j!} \, \frac{\Del_{\m{x}}^{j+k-n} \del(\m{x})}{2^{2j-2n}(k+j-n)!}.\\
&= \frac{2(-1)^k k!}{2^{2k}} \pi^{m-n-1} \sum_{{{j}}=0}^{k} \frac{4^{{j}} \, \m{x}\p^{\,2n-2{{j}}} }{(k-{{j}})!(n-{{j}})!} \Del_{\m{x}}^{k-{{j}}}\del(\m{x}),
\end{align*}
where we have replaced the index $j$ by $n-{{j}}$ in the last equality. Using Theorem \ref{NorSup} $i)$, we now obtain 
\begin{align}\label{PWRieszExcep}
\int_{\Sa^{m-1,2n}} \del\left(\langle{\bf x},{\bf w}\rangle\right) dS_{{\bf w}} = 2\pi^{M-1} \frac{(-1)^k}{2^{2k}} \,  \Del_{\bf x}^k \del({\bf x}).
\end{align}
Finally, since $M-1=-2k$, it is easily seen that 
\begin{align*}
\sigma_{M-1} |{\bf x}|^{-1} &=2\pi^{\frac{M-1}{2}} \dfrac{|{\bf x}|^\lan}{\Gam\left(\frac{M+\lan}{2}\right)}\Bigg|_{\lan=-1} = 2\pi^{\frac{M-1}{2}} \dfrac{|{\bf x}|^\lan}{\Gam\left(\frac{M+\lan}{2}\right)}\Bigg|_{\lan=-M-2k}.
\end{align*}
Then, in virtue of Theorem \ref{NorSup} $iii)$ we obtain 
\begin{align*}
\sigma_{M-1} |{\bf x}|^{-1} &=  2\pi^{\frac{M-1}{2}} \frac{(-1)^k \pi^{\frac{M}{2}}}{2^{2k} \Gam\left(\frac{M}{2}+k\right)} \Del_{\bf x}^k \del({\bf x}) = 2\pi^{M-1} \frac{(-1)^k}{2^{2k}} \,  \Del_{\bf x}^k \del({\bf x}),
\end{align*}
which completes the proof when compared with (\ref{PWRieszExcep}). $\hfill\square$\\
}
%Since $m-1-2n\notin -2\N_0$, we have that $m-1-2j\notin -2\N_0$ for all $j=0,\ldots, n$. Thus $\lan=-2j-1$ is not a singular point of $r^\lan$ and therefore, formula (\ref{PWDelOS}) yields 
%\[
%  \int_{\Sa^{m-1}}\del^{(2j)}(\langle\m{x},\m{\xi}\rangle)\; dS_{\m{\xi}} = 2\pi^{\frac{m-1}{2}} (-1)^j  \frac{(2j)! }{j!}  \frac{r^{-2j-1} }{\Gam\left(\frac{m-1}{2}-j\right)}.
%\]
%This implies that 
%\[
%I_j = (-1)^j \frac{\pi^{\frac{m-1}{2}}  }{\Gam\left(\frac{M-1}{2}\right) }  \frac{(2j)!}{j!} \, r^{-2j-1}.
%\]
%and hence,
%\[
%L_{2j}[\del^{(2j)}]({\bf x}) = \pi^{-\frac{1}{2}} \, \sigma_{M-1} \, \Gam\left(j+\frac{1}{2}\right) \frac{(2j)!}{j!} \, r^{-2j-1} \, \m{x}\p^{\, 2j}.
%\]
%Finally, substituting the later formula into (\ref{DelIntoL}) and in virtue of (\ref{PowSup}), we obtain 
%\[
%\int_{\Sa^{m-1,2n}} \del\left(\langle{\bf x},{\bf w}\rangle\right) dS_{{\bf w}} =   \sigma_{M-1} \sum_{j=0}^{n}  \frac{\m{x}\p^{\, 2j} }{j!}\ \frac{ \Gam\left(j+\frac{1}{2}\right)}{\Gam(\frac{1}{2})} \, r^{-1-2j} =  \sigma_{M-1} |{\bf x}|^{-1},
%\]
%which proves the theorem. $\hfill\square$\\

We now proceed with the proof of our main Theorem \ref{InvExplRed}. From Theorem \ref{PWRK},  we can rewrite (\ref{adaf}) as %follows
\begin{align}\label{adaacf}
R^*_{m|2n} \, R_{m|2n} [\phi] ({\bf x}) &= { 2\pi^{\frac{M-1}{2}}  \int_{\R^{m|2n}_{{\bf y}}} \dfrac{|{\bf y}-{\bf x}|^\lan}{\Gam\left(\frac{M+\lan}{2}\right)}\Bigg|_{\lan=-1}} \phi({\bf y}).
\end{align}
%\begin{align}\label{adaacf}
%R^*_{m|2n} \, R_{m|2n} [\phi] ({\bf x}) &= \sigma_{M-1} \int_{\R^{m|2n}} |{\bf y}-{\bf x}|^{-1}   \phi({\bf y})\, dV_{\bf y}.
%\end{align}
We may now take $s=\frac{M-1}{2}$ in Theorem \ref{FundSolLap}. Indeed, it is clear that %$\pm 2s=\pm (M-1)\in\Sigma_{m\pm2n}$.
$\pm2s-M=\pm(M-1)-M\notin2\N_0$ {for all $M\in\Z$}. Thus Theorem  \ref{FundSolLap} implies that 
\[
K_{M-1}^M({\bf x}) = \frac{1}{H_M(M-1)} |{\bf x}|^{-1} = \frac{1}{2^{M-1} \pi^{\frac{M-1}{2}} } \, {\dfrac{|{\bf x}|^\lan}{\Gam\left(\frac{M+\lan}{2}\right)}\Bigg|_{\lan=-1},}
\]
is a fundamental solution of $(-\Del_{\bf x})^{\frac{M-1}{2}}$. Finally, the action of $(-\Del_{\bf x})^{\frac{M-1}{2}}$ on both sides of (\ref{adaacf}) yields,
\begin{align*}
(-\Del_{\bf x})^{\frac{M-1}{2}} \, R^*_{m|2n} \, R_{m|2n} [\phi] ({\bf x}) &= {2\pi^{\frac{M-1}{2}}  \int_{\R^{m|2n}_{{\bf y}}}(-\Del_{\bf x})^{\frac{M-1}{2}}\, \dfrac{|{\bf y}-{\bf x}|^\lan}{\Gam\left(\frac{M+\lan}{2}\right)}\Bigg|_{\lan=-1}}   \phi({\bf y})\\
&= {2\pi^{\frac{M-1}{2}} \, 2^{M-1} \pi^{\frac{M-1}{2}}} \int_{\R^{m|2n}_{{\bf y}}} \del({\bf x} -{\bf y}) \phi({\bf y}) \\
&= 2^M \pi^{{M-1}} \phi({\bf x}),
\end{align*}
which completes the proof of Theorem  \ref{InvExplRed}.\\

{The explicit inversion formulas obtained in Theorem \ref{InvExpl} for $M+1\notin -2\N_0$ are particular cases of the general inversion formulas proved in Theorem  \ref{InvExplRed} for all $M\in\Z$. In a forthcoming paper, we shall study the plane wave decomposition of $\del({\bf x})$ when $M+1\in -2\N_0$, which will complete the list of explicit inversion formulas provided in Theorem \ref{InvExpl}.}
%Finally, we still need to prove the following technical Lemma used in the proof of   Theorem \ref{PWRK}.

%\section*{Acknowledgements}
%Alí Guzm\'an Ad\'an is supported by a BOF-post-doctoral grant from Ghent University. 

\appendix
\section{Alternative proof of Theorem \ref{ThmDelPW}}\label{App}

{In this appendix section, we derive from direct computations the plane wave decomposition formulas for $\del({\bf x})$ obtained in Theorem \ref{ThmDelPW}. To that end,} we need to gather a few preliminary results. We begin by the following particular case of the Funk-Hecke theorem in superspace, see e.g. \cite{MR2683546, MR2344451, CK_Ali}.
\begin{teo}\label{F-H_The}{\bf [Funk-Hecke]}
Let ${\bf x}, {\bf w}$ be independent super vector variables, $j,\el\in\N_0$, and let $H_\el$ be a harmonic polynomial homogeneous of degree $\el$, i.e.\ $\Del_{\bf x}[H_\el]=0$ and $\E[H_\el]=\el H_\el$. If $m\neq 0$, then
\[\int_{\Sa^{m-1,2n}} \langle {\bf x}, {\bf w} \rangle^j H_\el({\bf w})\, dS_{\bf w} = \al_{M,\el}[t^j] \, |{\bf x}|^{j-\el}\, H_\el({\bf x}),\]
where $\displaystyle \al_{M,\el}[t^j] = \begin{cases} \frac{j!}{(j-\el)!} \frac{2\pi^{\frac{M-1}{2}}}{2^\el} \frac{\Gam\left(\frac{j-\el+1}{2}\right)}{\Gam\left(\frac{M+j+\el}{2}\right)} & \mbox{ if } j+\el \mbox{ even and  } j\geq \el, \\ 0, & \mbox{ otherwise.}\end{cases}$\\[+.2cm]

\noindent Moreover, a similar result holds for the normalized integral defined in (\ref{NormInt}). If  $M=-2k$ (including the case $m=0$) and $j+\el\leq 2k+1$, then
\[\frac{1}{\sigma_{-2k}} \int_{\Sa^{m-1,2n}} \langle {\bf x}, {\bf w} \rangle^j H_\el({\bf w})\, dS_{\bf w} = \al_{M,\el}^*[t^j] \, {\bf x}^{j-\el}\, H_\el({\bf x}),\]
where $\displaystyle \al^*_{M,\el}[t^j] = \begin{cases}   \frac{(-1)^j \pi^{-1/2}}{2^\el} \frac{\left(k-\frac{j+\el}{2}\right)!}{k!} \frac{j!}{(j-\el)!} \Gam\left(\frac{j-\el+1}{2}\right) & \mbox{ if } j+\el \mbox{ even and  } j\geq \el, \\ 0, & \mbox{ otherwise.}\end{cases}$
\end{teo}

As a consequence of this result we obtain the following technical Lemma.
\begin{lem}\label{L3}
Given $g\in\Sw'(\R)$, consider the integrals $\displaystyle L_\el[g]({\bf x}):=\int_{\Sa^{m-1|2n}} \langle\m{x}\p,\m{w}\p\rangle^\el g(\langle\m{x},\m{w}\rangle)\, dS_{{\bf w}}$, with $\el=0, 1, \ldots, 2n$.  Then, for $\el$ odd we have $L_{2j+1}[g]({\bf x})=0$, while for $\el$ even 
\begin{equation}\label{L2jFin}
L_{2j}[g]({\bf x}) = 2(-1)^j \frac{\Gam\left(j+\frac{1}{2}\right)}{\pi^{n+\frac{1}{2}}} \, \m{x}\p^{\, 2j} \, \int_{\R^m} \del^{(n-j)}\left(1-|\m{w}|^2\right) g(\langle\m{x},\m{w}\rangle)\, dV_{{\m{w}}}.
%\int_{\Sa^{m-1|2n}} \langle\m{x}\p,\m{w}\p\rangle^j g(\langle\m{x},\m{w}\rangle) dS_{{\bf w}}.
\end{equation}
\end{lem}
\pf Formula (\ref{IntSupSph}) yields
\begin{align*}
L_\el[g]({\bf x}) &= 2 \int_{\R^m} \int_{B,{\m{w}\p}} \del(1+{\bf w}^2) \, \langle\m{x}\p,\m{w}\p\rangle^\el \,g(\langle\m{x},\m{w}\rangle) \, dV_{\m{w}} \\
&= 2 \sum_{{{p}}=0}^n   \left(\int_{B,{\m{w}\p}} \frac{\m{w}\p^{\, 2{{p}}}}{{{p}}!}\langle\m{x}\p,\m{w}\p\rangle^\el  \right) \left(\int_{\R^m} \del^{({{p}})}\left(1-|\m{w}|^2\right) g(\langle\m{x},\m{w}\rangle)\, dV_{{\m{w}}} \right).
\end{align*}
If $\el=2j+1$, it is clear that $\int_{B,{\m{w}\p}} \frac{\m{w}\p^{\, 2{{p}}}}{{{p}}!}\langle\m{x}\p,\m{w}\p\rangle^{2j+1} =0$, and therefore, $L_{2j+1}[g]({\bf x})=0$. On the other hand, if $\el=2j$, we obtain
\begin{align}\label{L1jInter}
L_{2j}[g]({\bf x}) &=  \frac{2}{(n-j)!} \left(\int_{B,{\m{w}\p}} \m{w}\p^{\, 2n-2j} \langle\m{x}\p,\m{w}\p\rangle^{2j}  \right) \left(\int_{\R^m} \del^{(n-j)}\left(1-|\m{w}|^2\right) g(\langle\m{x},\m{w}\rangle)\, dV_{{\m{w}}} \right)\nonumber \\
&=  \frac{2}{(n-j)!} \frac{\pi^{-n}}{4^n n!} \; \Del_{\m{w}\p}^n\left[ \m{w}\p^{\, 2n-2j} \langle\m{x}\p,\m{w}\p\rangle^{2j}\right] \left(\int_{\R^m} \del^{(n-j)}\left(1-|\m{w}|^2\right) g(\langle\m{x},\m{w}\rangle)\, dV_{{\m{w}}} \right).
\end{align}
Using {the} property (\ref{NormIntMod1}) of the normalized integral (\ref{NormInt}) we obtain
\[
\frac{(-1)^n}{4^n (n!)^2}\Del_{\m{w}\p}^n\left[ \m{w}\p^{\, 2n-2j} \langle\m{x}\p,\m{w}\p\rangle^{2j}\right] = \frac{1}{\sigma_{-2n}}  \int_{\Sa^{-1,2n}}\m{w}\p^{\, 2n-2j} \langle\m{x}\p,\m{w}\p\rangle^{2j}\, dS_{\m{w}\p} = (-1)^{n-j} \frac{1}{\sigma_{-2n}}  \int_{\Sa^{-1,2n}} \langle\m{x}\p,\m{w}\p\rangle^{2j}\, dS_{\m{w}\p}.
\]
Moreover, the Funk-Hecke Theorem \ref{F-H_The} for the normalized integral yields (see \cite{MR2344451, CK_Ali}), 
\[
\frac{1}{\sigma_{-2n}}  \int_{\Sa^{-1,2n}} \langle\m{x}\p,\m{w}\p\rangle^{2j}\, dS_{\m{w}\p} = \frac{(n-j)!}{n!} \frac{\Gam\left(j+\frac{1}{2}\right)}{\pi^{\frac{1}{2}}} \m{x}\p^{\, 2j}.
\]
Thus
\[
\Del_{\m{w}\p}^n\left[ \m{w}\p^{\, 2n-2j} \langle\m{x}\p,\m{w}\p\rangle^{2j}\right] = (-1)^j 4^n n! (n-j)! \frac{\Gam\left(j+\frac{1}{2}\right)}{\pi^{\frac{1}{2}}}  \m{x}\p^{\, 2j}.
\]
Finally, substituting the latter into (\ref{L1jInter}), we obtain (\ref{L2jFin}). $\hfill\square$\\

We now proceed to the proof of Theorem \ref{ThmDelPW}.

\paragraph{Case $M>1$, even.} In this case we must prove that 
\[
\del({\bf x}) =  \frac{(-1)^{\frac{M}{2}}(M-1)!}{(2\pi)^M} \int_{\Sa^{m-1,2n}}  \langle{\bf x},{\bf w}\rangle^{-M} \, dS_{\bf w}.
\]
We first recall that
\[
\langle{\bf x},{\bf w}\rangle^{-M} = \sum_{j=0}^{2n} (-1)^j \frac{\langle{\m{x}\p},\m{w}\p\rangle^j}{j!} \; \frac{(M+j-1)!}{(M-1)!} \,\langle{\m{x}},\m{w}\rangle^{-M-j}.
\]
Using Lemma \ref{L3} we have
\begin{align*}
\frac{(-1)^{\frac{M}{2}}(M-1)!}{(2\pi)^M}  & \int_{\Sa^{m-1,2n}}  \langle{\bf x},{\bf w}\rangle^{-M} \, dS_{\bf w}\nonumber\\
 &= \frac{(-1)^{\frac{M}{2}}}{(2\pi)^M} \sum_{j=0}^n \frac{(M+2j-1)!}{(2j)!}  \int_{\Sa^{m-1,2n}}  \langle{\m{x}\p},\m{w}\p\rangle^{2j}  \,\langle{\m{x}},\m{w}\rangle^{-M-2j} \, dS_{\bf w} \nonumber\\
&=  \frac{(-1)^{\frac{M}{2}}}{(2\pi)^M} \sum_{j=0}^n \frac{(M+2j-1)!}{(2j)!}  \frac{(-1)^j 2}{\pi^n} \frac{\Gam\left(j+\frac{1}{2}\right)}{{\pi^{\frac{1}{2}}}} \, \m{x}\p^{\, 2j} \, \int_{\R^m} \del^{(n-j)}\left(1-|\m{w}|^2\right) \langle{\m{x}},\m{w}\rangle^{-M-2j}\, dV_{{\m{w}}}. 
\end{align*}
From the identity $\dfrac{\Gam\left(j+\frac{1}{2}\right)}{\pi^{\frac{1}{2}}(2j)!} = \dfrac{1}{2^{2j} j!}$ we obtain
\begin{align}\label{A1}
\frac{(-1)^{\frac{M}{2}}(M-1)!}{(2\pi)^M} & \int_{\Sa^{m-1,2n}}  \langle{\bf x},{\bf w}\rangle^{-M} \, dS_{\bf w}\nonumber\\
&= \frac{2(-1)^{\frac{M}{2}}}{2^M\pi^{m-n}} \sum_{j=0}^n \frac{ (-1)^j (M+2j-1)!}{2^{2j} j!}  \, \m{x}\p^{\, 2j} \, \int_{\R^m} \del^{(n-j)}\left(1-|\m{w}|^2\right) \langle{\m{x}},\m{w}\rangle^{-M-2j}\, dV_{{\m{w}}} \nonumber\\
&= \frac{(-1)^{\frac{M}{2}} 2\pi^n}{2^M\pi^{m}} \sum_{j=0}^n \frac{ (-1)^{n-j} (m-2j-1)!}{2^{2n-2j} (n-j)!}  \, \m{x}\p^{\, 2n-2j} \, \int_{\R^m} \del^{(j)}\left(1-|\m{w}|^2\right) \langle{\m{x}},\m{w}\rangle^{2j-m}\, dV_{{\m{w}}}.
\end{align}
Let us now compute the integrals 
\[
I_j:=\int_{\R^m} \del^{(j)}\left(1-|\m{w}|^2\right) \langle{\m{x}},\m{w}\rangle^{2j-m}\, dV_{{\m{w}}}, \;\;\;\;\;\;\;\; j=0,\ldots, n.
\]
Using spherical coordinates, i.e. $\m{w}=r\m{\xi}$ with $r=|\m{w}|$ and $\m{\xi}\in\Sa^{m-1}$, we obtain
\begin{equation}\label{A2}
I_j= \left(\int_{\Sa^{m-1}} \langle{\m{x}},\m{\xi}\rangle^{2j-m}\, dS_{\m{\xi}} \right)\left(\int_{0}^{+\infty} \del^{(j)}\left(1-r^2\right) r^{2j-1}\, dr \right) = \begin{cases} \frac{1}{2} \int_{\Sa^{m-1}} \langle{\m{x}},\m{\xi}\rangle^{-m}\, dS_{\m{\xi}}, & j=0, \\[+.2cm] 0, & j\neq 0.\end{cases}
\end{equation}
Indeed, taking $t=r^2$ we get
\[
\int_{0}^{+\infty} \del^{(j)}\left(1-r^2\right) r^{2j-1}\, dr = \frac{1}{2} \int_{0}^{+\infty} \del^{(j)}\left(1-t\right) t^{j-1}\, dt = \begin{cases} \frac{1}{2}, & j=0, \\ 0, & j\neq 0.\end{cases}
\]
Finally, substituting (\ref{A2}) into (\ref{A1}) yields
\begin{align*}
\frac{(-1)^{\frac{M}{2}}(M-1)!}{(2\pi)^M}  \int_{\Sa^{m-1,2n}}  \langle{\bf x},{\bf w}\rangle^{-M} \, dS_{\bf w} &= \frac{(-1)^{\frac{m}{2}} (m-1)!}{2^m\pi^{m}} \left( \int_{\Sa^{m-1}} \langle{\m{x}},\m{\xi}\rangle^{-m}\, dS_{\m{\xi}} \right)\; \frac{\pi^n}{n!}  \, \m{x}\p^{\, 2n} \\
&= \del(\m{x})\; \frac{\pi^n}{n!}  \, \m{x}\p^{\, 2n},
\end{align*}
where we have used the plane wave decomposition (\ref{PWDelRm}) of the Dirac delta distribution in $\R^m$. % in the last equality.

\paragraph{Case $M>1$, odd.} We now must show that 
\[
\del({\bf x}) =  \frac{(-1)^{\frac{M-1}{2}}}{2 (2\pi)^{M-1}} \int_{\Sa^{m-1,2n}} \del^{(M-1)}(\langle{\bf x},{\bf w}\rangle) \, dS_{\bf w}.
\]
We recall that 
\[
\del^{(M-1)}(\langle{\bf x},{\bf w}\rangle) = \sum_{j=0}^{2n} \frac{\langle{\m{x}\p},\m{w}\p\rangle^{j}}{j!} \del^{(M+j-1)}(\langle{\m{x}},\m{w}\rangle).
\]
Lemma \ref{L3} now implies that 
\begin{align}\label{L4}
 \frac{(-1)^{\frac{M-1}{2}}}{2 (2\pi)^{M-1}} & \int_{\Sa^{m-1,2n}} \del^{(M-1)}(\langle{\bf x},{\bf w}\rangle) \, dS_{\bf w} \nonumber\\
 &=   \frac{(-1)^{\frac{M-1}{2}}}{2 (2\pi)^{M-1}}  \sum_{j=0}^n \frac{1}{(2j)!} \int_{\Sa^{m-1,2n}}  \langle{\m{x}\p},\m{w}\p\rangle^{2j} \, \del^{(M+2j-1)}(\langle{\m{x}},\m{w}\rangle) \, dV_{\m{w}} \nonumber\\
 &= \frac{(-1)^{\frac{M-1}{2}}}{(2\pi)^{M-1} \pi^n}  \sum_{j=0}^n  \frac{(-1)^j \m{x}\p^{\,2j}}{2^{2j} j!}  \int_{\R^m}  \del^{(n-j)}\left(1-|\m{w}|^2\right) \del^{(M+2j-1)}(\langle{\m{x}},\m{w}\rangle) \, dV_{\m{w}}\nonumber\\
&=  \frac{(-1)^{\frac{M-1}{2}}}{(2\pi)^{M-1}\, \pi^n}  \sum_{j=0}^n  \frac{(-1)^{n-j} \m{x}\p^{\,2n-2j}}{2^{2n-2j} (n-j)!}  \int_{\R^m}  \del^{(j)}\left(1-|\m{w}|^2\right) \del^{(m-2j-1)}(\langle{\m{x}},\m{w}\rangle) \, dV_{\m{w}}.
 \end{align}
 Similarly to the previous case, we now compute the integrals
 \[
I_j:=\int_{\R^m} \del^{(j)}\left(1-|\m{w}|^2\right) \del^{(m-2j-1)}(\langle{\m{x}},\m{w}\rangle)\, dV_{{\m{w}}}, \;\;\;\;\;\;\;\; j=0,\ldots, n,
\]
using spherical coordinates. We thus get
\begin{align*}
I_j &= \int_{0}^{+\infty} \int_{\Sa^{m-1}}  \del^{(j)}\left(1-r^2\right) r^{m-1} \frac{\del^{(m-2j-1)}(\langle{\m{x}},\m{\xi}\rangle)}{r^{m-2j}}\, dS_{\m{\xi}}  \, dr \\
&= \left( \int_{\Sa^{m-1}} \del^{(m-2j-1)}(\langle{\m{x}},\m{\xi}\rangle)\, dS_{\m{\xi}} \right) \left(\int_{0}^{+\infty} \del^{(j)}\left(1-r^2\right) r^{2j-1} \, dr\right)\\
&= \begin{cases} \frac{1}{2} \int_{\Sa^{m-1}} \del^{(m-1)}(\langle{\m{x}},\m{\xi}\rangle)\, dS_{\m{\xi}}, & j=0, \\[+.2cm] 0, & j\neq 0.\end{cases}
\end{align*}
Substituting this into (\ref{L4}), and using formula (\ref{PWDelRm}), we obtain 
\begin{align*}
 \frac{(-1)^{\frac{M-1}{2}}}{2 (2\pi)^{M-1}}  \int_{\Sa^{m-1,2n}} \del^{(M-1)}(\langle{\bf x},{\bf w}\rangle) \, dS_{\bf w} &= \frac{(-1)^{\frac{M-1}{2}}}{(2\pi)^{M-1} \, \pi^n}  \frac{(-1)^n \m{x}\p^{\, 2n}}{2^{2n+1} n!}  \int_{\Sa^{m-1}} \del^{(m-1)}(\langle{\m{x}},\m{\xi}\rangle)\, dS_{\m{\xi}} \\
 &=\frac{(-1)^{\frac{m-1}{2}}}{2(2\pi)^{m-1}} \left( \int_{\Sa^{m-1}} \del^{(m-1)}(\langle{\m{x}},\m{\xi}\rangle)\, dS_{\m{\xi}} \right)  \frac{\pi^n}{n!}  \, \m{x}\p^{\, 2n}\\
 &=\del(\m{x})\, \frac{\pi^n}{n!}  \, \m{x}\p^{\, 2n},
\end{align*}
which proves the stament.

Before proceeding to the next case, we recall that a cornerstone in the proof of this result in Theorem \ref{ThmDelPW} is formula (\ref{BVDel2}). This formula was obtained from the identity 
\[
\del^{(M-1)}(a) =\frac{-(M-1)!}{2\pi i} \, \textup{B.V.} \left[ z^{-M}\right], \;\;\;\;\;\;\;\; z=a+ib, \;\; a,b\in\R 
\]
by means of the identifications $a\mapsto \langle{\bf x},{\bf w}\rangle$, $b\mapsto x_0$, $i\mapsto \frac{-{\bf w}}{|{\bf w}|}$ and $\frac{1}{i} \mapsto  \frac{{\bf w}}{|{\bf w}|}$. We will now show that formula (\ref{BVDel2}) also follows from direct computations, {which justifies} %justifying %thus the correctness of 
the above identifications.
\begin{lem}
Let {$\el\in\N$}. Then
\[
\del^{(\el-1)}(\langle{\bf x},{\bf w}\rangle) = (-1)^\el \frac{(\el-1)!}{2\pi} \, \textup{B.V.} \left[\frac{{\bf w}}{|{\bf w}|} \left(\langle{\bf x},{\bf w}\rangle - x_0 \frac{{\bf w}}{|{\bf w}|}\right)^{-\el}  \right].
\]
\end{lem}
\pf
We begin by recalling the definition of $g\left(\langle{\bf x},{\bf w}\rangle - x_0 \frac{{\bf w}}{|{\bf w}|}\right)$, where $g(z)=g_1(a,b)+ig_2(a,b)$ is a complex holomorphic function in an open domain $\Om\subseteq \R^2\iso \C$ and $(\langle\m{x},\m{w}\rangle,x_0)\in\Om$, see (\ref{MonPWMod}). Using the Taylor expansion of the real and imaginary parts of $g$ we have,
\begin{align*}
g\left(\langle{\bf x},{\bf w}\rangle - x_0 \frac{{\bf w}}{|{\bf w}|}\right) &= g_1(\langle{\bf x},{\bf w}\rangle, x_0 ) - \frac{{\bf w}}{|{\bf w}|} g_2(\langle{\bf x},{\bf w}\rangle, x_0 )\\
&=\sum_{j=0}^{2n} \frac{\langle{\m{x}\p},\m{w}\p\rangle^{j}}{j!} \left[\pa_a^j g_1(\langle\m{x},\m{y}\rangle,x_0)-\frac{{\bf w}}{|{\bf w}|} \pa_a^j g_2(\langle\m{x},\m{y}\rangle,x_0)\right] \\
&=\sum_{j=0}^{2n} \frac{\langle{\m{x}\p},\m{w}\p\rangle^{j}}{j!} \; g^{(j)}\left(\langle{\m{x}},{\m{w}}\rangle - x_0 \frac{{\bf w}}{|{\bf w}|}\right).
\end{align*}
Thus, taking $g(z)=z^{-\el}$ we obtain
\begin{equation}\label{MinM}
\left(\langle{\bf x},{\bf w}\rangle - x_0 \frac{{\bf w}}{|{\bf w}|}\right)^{-\el} = \sum_{j=0}^{2n} (-1)^j \frac{\langle{\m{x}\p},\m{w}\p\rangle^{j}}{j!} \; \frac{(\el+j-1)!}{(\el-1)!} \left(\langle{\m{x}},{\m{w}}\rangle - x_0 \frac{{\bf w}}{|{\bf w}|}\right)^{-\el-j}.
% g^{(j)}\left(\langle{\bf x},{\bf w}\rangle - x_0 \frac{{\bf w}}{|{\bf w}|}\right)
\end{equation}
We now recall that $\del^{(\el-1)}(a) =\dfrac{(-1)^{\el} (\el-1)!}{2\pi i} \, \textup{B.V.} \left[ z^{-\el}\right]$ for all $\el\in\N$. 
%\[
%\del^{(\el)}(a) =\frac{(-1)^{\el+1} \el!}{2\pi i} \, \textup{B.V.} \left[ z^{-\el+1}\right]. %\;\;\;\;\;\;\;\; z=a+ib, \;\; a,b\in\R 
%\]
Clearly, we can identify the complex imaginary unit $i$ with $\dfrac{-{\bf w}}{|{\bf w}|}$ in this identity. Moreover, we can make the identifications $a\mapsto \langle{\m{x}},{\m{w}}\rangle$ and $b\mapsto x_0$, {which pose no problem because both {are substitutions among real variables}}. Therefore we obtain
\[
\del^{(\el-1)}(\langle{\m{x}},{\m{w}}\rangle) =\dfrac{(-1)^{\el} (\el-1)!}{2\pi} \, \textup{B.V.} \left[ \left(\langle{\m{x}},{\m{w}}\rangle - x_0 \frac{{\bf w}}{|{\bf w}|}\right)^{-\el}  \frac{{\bf w}}{|{\bf w}|}\right].
\]
Substituting this identity into the definition of $\del^{(M-1)}(\langle{\bf x},{\bf w}\rangle)$ as a Taylor expansion (see Definition \ref{ConDelSS}), we get
\begin{align*}
\del^{(\el-1)}(\langle{\bf x},{\bf w}\rangle) &= \sum_{j=0}^{2n} \frac{\langle{\m{x}\p},\m{w}\p\rangle^{j}}{j!} \; \del^{(\el-1+j)}(\langle{\m{x}},{\m{w}}\rangle) \\
&=\sum_{j=0}^{2n} (-1)^{\el+j} \frac{\langle{\m{x}\p},\m{w}\p\rangle^{j}}{j!} \; \frac{(\el+j-1)!}{2\pi}\, \textup{B.V.} \left[ \left(\langle{\m{x}},{\m{w}}\rangle - x_0 \frac{{\bf w}}{|{\bf w}|}\right)^{-\el-j}  \frac{{\bf w}}{|{\bf w}|}\right] \\
&= (-1)^{\el} \frac{(\el-1)!}{2\pi} \frac{{\bf w}}{|{\bf w}|} \; \textup{B.V.} \left[ \sum_{j=0}^{2n}  (-1)^j \frac{\langle{\m{x}\p},\m{w}\p\rangle^{j}}{j!} \; \frac{(\el+j-1)!}{(\el-1)!} \left(\langle{\m{x}},{\m{w}}\rangle - x_0 \frac{{\bf w}}{|{\bf w}|}\right)^{-\el-j} \right],
\end{align*}
which proves the result when combined with (\ref{MinM}). $\hfill\square$

\paragraph{Case $M=-2k$, $m\neq0$.}
In this case we need to show that 
\[
\del({\bf x})= \frac{1}{ 2^{2k-1} (k!)^2 \;\sigma_{-2k+1}}\, \Del_{\bf w}^k \left[\langle{\bf x},{\bf w}\rangle^{2k}\right] -  (-1)^k (4\pi^2)^k  \int_{\Sa^{m-1,2n}}  G_{2k}\left(|\langle{\bf x},{\bf w}\rangle| \right)\, dS_{\bf w}.
\]
To that end, we first denote the right hand side of our statement by $S$, i.e.\
\[
S:=\frac{1}{ 2^{2k-1} (k!)^2 \;\sigma_{-2k+1}}\, \Del_{\bf w}^k \left[\langle{\bf x},{\bf w}\rangle^{2k}\right] -  (-1)^k (4\pi^2)^k  \int_{\Sa^{m-1,2n}}  G_{2k}\left(|\langle{\bf x},{\bf w}\rangle| \right)\, dS_{\bf w}.
\]
The first term in $S$ can be computed using the Funk-Hecke Theorem \ref{F-H_The} for the normalized integral. Indeed,
\[
\Del_{\bf w}^k \left[\langle{\bf x},{\bf w}\rangle^{2k}\right]  = (-1)^k 2^{2k} (k!)^2 \; \frac{1}{\sigma_{-2k}} \int_{\Sa^{m-1,2n}} \langle {\bf x}, {\bf w} \rangle^{2k} \, dS_{\bf w} = (-1)^k 2^{2k} k! \, \frac{\Gam\left(k+\frac{1}{2}\right)}{\pi^{\frac{1}{2}}} \, {\bf x}^{2k}.
\] 
Thus, using the fact that $\displaystyle\frac{\Gam(-k+\frac{1}{2})}{\pi^{\frac{1}{2}}} \frac{\Gam(k+\frac{1}{2})}{\pi^{\frac{1}{2}}}=(-1)^k$, we obtain
\[
\frac{1}{ 2^{2k-1} (k!)^2 \;\sigma_{-2k+1}}\, \Del_{\bf w}^k \left[\langle{\bf x},{\bf w}\rangle^{2k}\right] = \frac{(-1)^k}{\pi^{-k} k! } \frac{\Gam(-k+\frac{1}{2})}{\pi^{\frac{1}{2}}} \frac{\Gam(k+\frac{1}{2})}{\pi^{\frac{1}{2}}} \, {\bf x}^{2k} = \frac{\pi^k}{k! } \, {\bf x}^{2k},
\]
and therefore
\begin{equation}\label{SFundEq}
S=\frac{\pi^k}{k! } \, {\bf x}^{2k}  -  (-1)^k (4\pi^2)^k  \int_{\Sa^{m-1,2n}}  G_{2k}\left(|\langle{\bf x},{\bf w}\rangle| \right)\, dS_{\bf w}.
\end{equation}
Let us now compute the integral $\int_{\Sa^{m-1,2n}}  G_{2k}\left(|\langle{\bf x},{\bf w}\rangle| \right)\, dS_{\bf w}$. To that end, we define the functions $F_\el(x)=G_\el(|x|)$, which are analytic functions in $\R\setminus\{0\}$. Then the superfunction $G_{2k}\left(|\langle{\bf x},{\bf w}\rangle| \right)$ may be written as
\[
G_{2k}\left(|\langle{\bf x},{\bf w}\rangle| \right) = \sum_{j=0}^{2n} \frac{\langle\m{x}\p, \m{w}\p\rangle^j}{j!} F_{2k}^{(j)}(\langle\m{x}, \m{w}\rangle).
\]
{By} virtue of Lemma \ref{L3} we obtain
\begin{align}\label{G_2kInt}
\int_{\Sa^{m-1,2n}} G_{2k}\left(|\langle{\bf x},{\bf w}\rangle| \right)\, dS_{\bf w} &= \sum_{j=0}^{n} \frac{1}{(2j)!} \int_{\Sa^{m-1,2n}}  \langle\m{x}\p, \m{w}\p\rangle^{2j}  F_{2k}^{(2j)}(\langle\m{x}, \m{w}\rangle) \, dS_{\bf w} \nonumber \\
 &= \sum_{j=0}^{n} \frac{2(-1)^j}{(2j)! \pi^n} \frac{\Gam\left(j+\frac{1}{2}\right)}{\pi^{\frac{1}{2}}} \, \m{x}\p^{\,2j}  \int_{\R^m}  \del^{(n-j)}\left(1-|\m{w}|^2\right) F_{2k}^{(2j)}(\langle{\m{x}},\m{w}\rangle) \, dV_{\m{w}} \nonumber \\
%&= \frac{2}{\pi^n} \sum_{j=0}^{n}  \frac{(-1)^j \m{x}\p^{\,2j} }{2^{2j} j!}  \int_{\R^m}  \del^{(n-j)}\left(1-|\m{w}|^2\right) F^{(2j)}(\langle{\m{x}},\m{w}\rangle) \, dV_{\m{w}},\\
&= \frac{2}{\pi^n} \sum_{j=0}^{n}  \frac{(-1)^{n-j} \m{x}\p^{\,2n-2j} }{2^{2n-2j} (n-j)!}  \int_{\R^m}  \del^{(j)}\left(1-|\m{w}|^2\right) F_{2k}^{(2n-2j)}(\langle{\m{x}},\m{w}\rangle) \, dV_{\m{w}},
\end{align}
where we {have again used} the identity $\dfrac{\Gam\left(j+\frac{1}{2}\right)}{\pi^{\frac{1}{2}}(2j)!} = \dfrac{1}{2^{2j} j!}$ and reversed the order of summation in the last equality.

\noindent We now recall that the functions $G_\el$ satisfy that  $G'_{\el+1}=G_\el$ and $G_0(z)=\ln(z)$. We thus obtain
\begin{align*}
F_{2k}^{(2j)}(x) &= F_{2k-2j}(x)=G_{2k-2j}(|x|), & j&\leq k, \\
%F_{2k}^{(2k)}(x) &=\ln(|x|),\\
F_{2k}^{(2k+j)}(x) &= (-1)^{j-1} (j-1)! \; x^{-j},  & j&\in\N.
\end{align*}
Substituting these formulas into (\ref{G_2kInt}), and using the fact that $2n-2j=m+2k-2j$, we get 
\begin{align}\label{G2kInt}
\int_{\Sa^{m-1,2n}} G_{2k}\left(|\langle{\bf x},{\bf w}\rangle| \right)\, dS_{\bf w} %&= \phantom{+} \frac{2}{\pi^n} \sum_{j=0}^{k}  \frac{(-1)^j \m{x}\p^{\,2j} }{2^{2j} j!}  \int_{\R^m}  \del^{(n-j)}\left(1-|\m{w}|^2\right) G_{2k-2j}(|\langle{\m{x}},\m{w}\rangle|) \, dV_{\m{w}} \\
%&\phantom{=}+  \frac{2}{\pi^n} \sum_{j=k+1}^{n}  \frac{(-1)^j \m{x}\p^{\,2j} }{2^{2j} j!}  \int_{\R^m}  \del^{(n-j)}\left(1-|\m{w}|^2\right)  F_{2k}^{(2n-2j)}(\langle{\m{x}},\m{w}\rangle) \, dV_{\m{w}}\\
&=\phantom{+}  \frac{2}{\pi^n} \sum_{j=0}^{\frac{m}{2}-1}  \frac{(-1)^{n-j} \m{x}\p^{\,2n-2j} }{2^{2n-2j} (n-j)!}  \int_{\R^m}  \del^{(j)}\left(1-|\m{w}|^2\right)  F_{2k}^{(2n-2j)}(\langle{\m{x}},\m{w}\rangle) \, dV_{\m{w}}\nonumber\\
&\phantom{=} + \frac{2}{\pi^n} \sum_{j=\frac{m}{2}}^{n}  \frac{(-1)^{n-j} \m{x}\p^{\,2n-2j} }{2^{2n-2j} (n-j)!}  \int_{\R^m}  \del^{(j)}\left(1-|\m{w}|^2\right)  G_{2j-m}(|\langle{\m{x}},\m{w}\rangle|) \, dV_{\m{w}}.
\end{align}
Let us now compute the integrals appearing in the above sum, which will be denoted by 
\begin{align*}
I_j&:= \int_{\R^m}  \del^{(j)}\left(1-|\m{w}|^2\right)  F_{2k}^{(2n-2j)}(\langle{\m{x}},\m{w}\rangle) \, dV_{\m{w}}, &j&=0,\ldots, \frac{m}{2}-1,
\end{align*}
and 
\begin{align*}
J_j&:=\int_{\R^m}  \del^{(j)}\left(1-|\m{w}|^2\right)  G_{2j-m}(|\langle{\m{x}},\m{w}\rangle|) \, dV_{\m{w}}, &j&=\frac{m}{2},\ldots, n,
\end{align*}
respectively.

\noindent To compute $I_j$, we first observe that 
$
F_{2k}^{(2n-2j)}(\langle{\m{x}},\m{w}\rangle) = F_{2k}^{(2k+m-2j)}(\langle{\m{x}},\m{w}\rangle)= -(m-2j-1)! \langle{\m{x}},\m{w}\rangle^{-m+2j}.
$
Using again spherical coordinates, i.e.\ $\m{w}=r\m{\xi}$ with $r=|\m{w}|$ and $\m{\xi}\in\Sa^{m-1}$, we obtain
\begin{align}\label{I_jComp}
I_j&= -(m-2j-1)! \left( \int_{\Sa^{m-1}}\langle{\m{x}},\m{\xi}\rangle^{-m+2j}\, dS_{\m{\xi}} \right) \left(\int_{0}^{+\infty} \del^{(j)}\left(1-r^2\right) r^{2j-1} \, dr\right) \nonumber \\[+.2cm]
&= \begin{cases} \frac{-(m-1)!}{2} \int_{\Sa^{m-1}} \langle{\m{x}},\m{\xi}\rangle^{-m}\, dS_{\m{\xi}}, & j=0, \\[+.2cm] 0, & j\neq 0.\end{cases}
\end{align}
To compute $J_j$ we also use spherical coordinates, which yields
\[
J_j=\int_{\Sa^{m-1}}\left(\int_{0}^{+\infty}  \del^{(j)}\left(1-r^2\right) G_{2j-m}(r |\langle{\m{x}},\m{\xi}\rangle|)\,  r^{m-1} \, dr  \right) \, dS_{\m{\xi}}.
\]
By the definition (\ref{RecFunc}) of the functions $G_\el$, we have
\begin{align*}
G_{2j-m}(r |\langle{\m{x}},\m{\xi}\rangle|) &= r^{2j-m} \left[\frac{\langle{\m{x}},\m{\xi}\rangle^{2j-m}}{(2j-m)!} \left(\ln(|\langle{\m{x}},\m{\xi}\rangle|)+\ln(r)\right)-a_{2j-m}\langle{\m{x}},\m{\xi}\rangle^{2j-m}\right]\\
&= r^{2j-m}G_{2j-m}(|\langle{\m{x}},\m{\xi}\rangle|) + \frac{\langle{\m{x}},\m{\xi}\rangle^{2j-m}}{(2j-m)!} \, r^{2j-m}\ln(r).
\end{align*}
Hence,
\begin{multline*}
J_j=  \left( \int_{\Sa^{m-1}}G_{2j-m}(|\langle{\m{x}},\m{\xi}\rangle|)\, dS_{\m{\xi}} \right) \left(\int_{0}^{+\infty} \del^{(j)}\left(1-r^2\right) r^{2j-1} \, dr\right) \\  + \left( \int_{\Sa^{m-1}} \frac{\langle{\m{x}},\m{\xi}\rangle^{2j-m}}{(2j-m)!}\, dS_{\m{\xi}} \right) \left(\int_{0}^{+\infty} \del^{(j)}\left(1-r^2\right) \ln(r) r^{2j-1} \, dr\right).
\end{multline*}
We now recall that $\int_{0}^{+\infty} \del^{(j)}\left(1-r^2\right) r^{2j-1} \, dr =0$ for $j=\frac{m}{2},\ldots, n$. Then 
\[
J_j=\left( \int_{\Sa^{m-1}} \frac{\langle{\m{x}},\m{\xi}\rangle^{2j-m}}{(2j-m)!}\, dS_{\m{\xi}} \right) \left(\int_{0}^{+\infty} \del^{(j)}\left(1-r^2\right) \ln(r) r^{2j-1} \, dr\right).
\]
Considering the change of coordinates $r=t^{\frac{1}{2}}$, we obtain 
\[\int_{0}^{+\infty} \del^{(j)}\left(1-r^2\right) \ln(r) r^{2j-1} \, dr = \frac{1}{4} \int_{0}^{+\infty} \del^{(j)}\left(1-t\right) \ln(t) t^{j-1} \, dt= \frac{1}{4} \frac{d^j}{dt^j} \left[t^{j-1}\ln(t)\right]\bigg|_{t=1}. %= \frac{(j-1)!}{4} 
\] 
By induction, it can be proven that \[\frac{d^j}{dt^j} \left[t^{j-1}\ln(t)\right]\bigg|_{t=1}=(j-1)!.\] Indeed, for $j=1$ one has $\frac{d}{dt} \left[\ln(t)\right]\big|_{t=1}=1$, while the induction step follows from the identity
\begin{align*}
\frac{d^{j+1}}{dt^{j+1}} \left[t^{j}\ln(t)\right] %&= \frac{d^{j}}{dt^{j}} \frac{d}{dt} \left[t \, t^{j-1}\ln(t)\right] \\
&= \frac{d^{j}}{dt^{j}} \left[ t^{j-1}\ln(t)+ t \left(t^{j-1}\ln(t)\right)'\right] \\
&= \frac{d^{j}}{dt^{j}} \left[ t^{j-1}\ln(t)\right] + \frac{d^{j}}{dt^{j}} \left[ (j-1)t^{j-1}\ln(t)+t^{j-1}\right]\\
&=j \frac{d^{j}}{dt^{j}} \left[ t^{j-1}\ln(t)\right].
\end{align*}
Thus, $\displaystyle\int_{0}^{+\infty} \del^{(j)}\left(1-r^2\right) \ln(r) r^{2j-1} \, dr=\frac{(j-1)!}{4}$ and therefore,
\[
J_j=\frac{(j-1)!}{4(2j-m)!} \int_{\Sa^{m-1}} \langle{\m{x}},\m{\xi}\rangle^{2j-m}\, dS_{\m{\xi}}.
\]
The Funk-Hecke Theorem \ref{F-H_The} in the purely bosonic case (i.e.\ $n=0$) now yields
\begin{equation}\label{J_jComp}
J_j=\frac{\pi^{\frac{m-1}{2}}}{2(2j-m)!} \Gam\left(j-\frac{m}{2}+\frac{1}{2}\right) |\m{x}|^{2j-m}= \frac{\pi^{\frac{m}{2}}}{2^{2j-m+1}\left(j-\frac{m}{2}\right)!} |\m{x}|^{2j-m}.
\end{equation}
Substituting (\ref{I_jComp}) and (\ref{J_jComp}) into (\ref{G2kInt}) we obtain
\begin{align*}\label{G2kInt}
\int_{\Sa^{m-1,2n}} G_{2k}&\left(|\langle{\bf x},{\bf w}\rangle| \right)\, dS_{\bf w} \\
&=  \frac{(-1)^{n+1}(m-1)!}{2^{2n}\, n! \,  \pi^n} \m{x}\p^{\, 2n} \int_{\Sa^{m-1}} \langle{\m{x}},\m{\xi}\rangle^{-m}\, dS_{\m{\xi}} 
\;+\; 2^{-2k} \pi^{-k}  \sum_{j=\frac{m}{2}}^{n} (-1)^{n-j}  \frac{ \m{x}\p^{\,2n-2j} }{(n-j)!} \frac{|\m{x}|^{2j-m}}{\left(j-\frac{m}{2}\right)!}\\
&=  \frac{(-1)^{n+1}(m-1)!}{2^{2n}\, n! \,  \pi^n} \m{x}\p^{\, 2n} \int_{\Sa^{m-1}} \langle{\m{x}},\m{\xi}\rangle^{-m}\, dS_{\m{\xi}} 
\;+\; (-1)^k 2^{-2k} \pi^{-k}  \sum_{j=0}^{k}   \frac{ \m{x}\p^{\,2k-2j} }{(k-j)!} \frac{\m{x}^{2j}}{j!}\\
&=\frac{(-1)^{n+1}(m-1)!}{2^{2n}\, n! \,  \pi^n} \m{x}\p^{\, 2n} \int_{\Sa^{m-1}} \langle{\m{x}},\m{\xi}\rangle^{-m}\, dS_{\m{\xi}} 
\;+\; \frac{(-1)^k  2^{-2k} \pi^{-k} }{k!} {\bf x}^{2k}.
\end{align*}
Finally, substituting this last expression into (\ref{SFundEq}) gives
\begin{align*}
S&=\frac{\pi^k}{k! } \, {\bf x}^{2k}  -  (-1)^k (4\pi^2)^k  \left[ \frac{(-1)^{n+1}(m-1)!}{2^{2n}\, n! \,  \pi^n} \m{x}\p^{\, 2n} \int_{\Sa^{m-1}} \langle{\m{x}},\m{\xi}\rangle^{-m}\, dS_{\m{\xi}} 
\;+\; \frac{(-1)^k  2^{-2k} \pi^{-k} }{k!} {\bf x}^{2k}\right]\\
&= \frac{(-1)^{\frac{m}{2}} (m-1)!}{2^m \pi^m} \left(\int_{\Sa^{m-1}} \langle{\m{x}},\m{\xi}\rangle^{-m}\, dS_{\m{\xi}} \right) \frac{\pi^n}{n!} \m{x}\p^{\, 2n},
\end{align*}
which yields our assertion when combined with the plane wave decomposition (\ref{PWDelRm}).

\paragraph{Case $M =-2n$, i.e.\ $m= 0$.} In this case we must show that 
\begin{align*}
\del(\m{x}\p)&=\frac{1}{ 2^{2n-1} (n!)^2 \;\sigma_{-2n+1}}\, \Del_{\m{w}\p}^n \left[\langle\m{x}\p,\m{w}\p\rangle^{2n}\right].
\end{align*}
This easily follows from the Funk-Hecke Theorem \ref{F-H_The} for the nomalized integral in the purely fermionic case. Indeed,
\[
\Del_{\m{w}\p}^n \left[\langle{\m{x}\p},{\m{w}\p}\rangle^{2n}\right]  = (-1)^n 2^{2n} (n!)^2 \; \frac{1}{\sigma_{-2n}} \int_{\Sa^{-1,2n}} \langle {\m{x}\p}, {\m{w}\p} \rangle^{2n} \, dS_{\m{w}\p} = (-1)^n 2^{2n} n! \, \frac{\Gam\left(n+\frac{1}{2}\right)}{\pi^{\frac{1}{2}}} \, {\m{x}\p}^{2n}.
\] 
Then
\begin{align*}
\frac{1}{ 2^{2n-1} (n!)^2 \;\sigma_{-2n+1}}\, \Del_{\m{w}\p}^n \left[\langle\m{x}\p,\m{w}\p\rangle^{2n}\right] &= \frac{(-1)^n}{\pi^{-n} n!} \frac{\Gam(-n+\frac{1}{2})}{\pi^{\frac{1}{2}}} \frac{\Gam(n+\frac{1}{2})}{\pi^{\frac{1}{2}}} \, {\m{x}\p}^{2n} = \frac{\pi^{n}}{ n!}\, {\m{x}\p}^{2n} = \del(\m{x}\p),
\end{align*}
which proves the result.

\bibliographystyle{abbrv}
%\bibliography{0References}

\end{document}